%% file: main.tex
\documentclass[
    reprint,
    superscriptaddress,
    amsmath, amssymb,
    aps,
    prx,
    floatfix,
]{revtex4-2}

\input{setup.tex}

\begin{document}

\title{Geometric encoding of turbulence for end-to-end quantum simulation}

\author{Zhaoyuan Meng}
    \affiliation{State Key Laboratory for Turbulence and Complex Systems, School of Mechanics and Engineering Science, Peking University, Beijing 100871, China}
    \affiliation{State Key Laboratory of Nonlinear Mechanics, Institute of Mechanics, Chinese Academy of Sciences, Beijing 100190, China}
\author{Xiao-Ming Zhang}
    \affiliation{School of Physics, South China Normal University, Guangzhou, China}
\author{Xiao Yuan}
    \affiliation{Center on Frontiers of Computing Studies, Peking University, Beijing 100871, China}
    \affiliation{School of Computer Science, Peking University, Beijing 100871, China}
\author{Yue Yang}
    \email{yyg@pku.edu.cn}
    \affiliation{State Key Laboratory for Turbulence and Complex Systems, School of Mechanics and Engineering Science, Peking University, Beijing 100871, China}
    \affiliation{HEDPS-CAPT, Peking University, Beijing 100871, China}
    
\date{\today}

\begin{abstract}
\noindent
Simulating multiscale turbulence in engineering remains challenging. Any quantum advantage for such simulations requires efficient initial-state preparation. Yet exact loading of an arbitrary classical snapshot requires exponential quantum resources. We introduce ``turbuloscope'', a physics-informed geometric method that generates statistically representative turbulent initial states in a gate-based register from a parameterized Pauli-spinor multiscale distribution. Scale invariance renders conditional rotations approximately affine in Gray-code space, enabling linear-depth, ancilla-free preparation. The generalized Madelung transform and Hopf fibration map the encoded Pauli spinor to fluid observables and vortex tubes. The qubit count scales logarithmically with the Reynolds number, yielding an exponential reduction in preparation overhead relative to full-grid classical spectral synthesis. We classically emulate a 30-qubit preparation circuit and generate a turbulent field over one billion grid points at a scale-inferred Reynolds number of 35,000, exhibiting Kolmogorov's five-thirds energy spectrum, tangled vortex tubes, and intermittency. This method provides scalable initial states for quantum simulation of turbulence and multiscale systems.
\end{abstract}

\maketitle

\noindent Fluid turbulence represents a canonical grand challenge in mechanical engineering and computational science, characterized by nonlinear interactions that span a vast range of scales---from macroscopic eddies to microscopic dissipation~\cite{Feynman2015_The}. Simulating this complete energy cascade~\cite{Alexakis2018_Cascades} poses a formidable barrier for classical computers, as the number of spatial grid points $N_d$ required to resolve a $d$-dimensional flow scales polynomially with the Reynolds number ($\Rey$), following the lower bound $N_d=\Omega(\Rey^{5d/4-3/2})$. 
Consequently, quantum computing, which offers an exponential speedup for certain applications~\cite{Shor1994_Algorithms, Weinstein2001_Implementation, Harrow2009_Quantum, Liu2021_Efficient}, has emerged as a highly promising alternative~\cite{Gourianov2022_A, Succi2023_Quantum, Tennie2025_Quantum, Wang2026_Simulating}. By utilizing amplitude encoding, the state of a velocity field can be represented such that the number of qubits, $n$, relates to the grid size by $2^n=dN_d$. In this paradigm, the required number of qubits grows merely logarithmically with the Reynolds number as $n=\Omega(\log_2 \Rey)$. 
This logarithmic scaling suggests that a linear increase in qubit count could encode turbulent fields with exponentially larger $\Rey$, theoretically unlocking unprecedented capacity for the representation of multiscale physics~\cite{Stanley1999_Scaling}.

However, translating this vast representational capacity into a practical, end-to-end quantum advantage for solving partial differential equations (PDEs)~\cite{Childs2021_High, An2023_Linear, Jaksch2023_Variational, Jin2024_Quantum} is hindered by the state-preparation bottleneck~\cite{Preskill2018_Quantum}.
A precise distinction exists between two state-preparation approaches.
The first is instance-specific loading, where a prescribed classical array of $2^n$ amplitudes is transferred into an $n$-qubit register.
For arbitrary input data, this process demands exponential resources within the standard gate model~\cite{Mottonen2005_Transformation, Kempe2006_The, Nielsen2006_Geometry, Zhang2022_Quantum, Sun2023_Asymptotically, Ben-Dov2024_Approximate}.
Uniformly controlled rotations provide a general ancilla-free construction but retain an exponential gate count~\cite{Mottonen2005_Transformation}, whereas linear-depth preparation of an arbitrary state requires exponentially many ancillary qubits~\cite{Zhang2022_Quantum}.
The second is physics-informed generative initialization, which synthesizes a quantum state from the prescribed scaling law of a target multiscale distribution rather than loading it from a full-resolution classical snapshot, analogous to synthetic-turbulence initialization in classical simulations~\cite{Shen2024_Designing}.
In this work, we focus on the latter approach.
Specifically, our approach takes as input the power-law exponent of the Pauli-spinor spectral distribution, the relative component population, and spatial resolution of a turbulent ensemble, rather than the explicit values of a pre-existing flow field at every grid point.
The circuit parameters are compiled directly from this analytic input by retaining only the effective spectral support, thereby bypassing the construction of a $2^n$-component probability array.

Overcoming this initialization bottleneck is exceptionally difficult for turbulent flows. Historically, efficient quantum state preparation relies on exploiting the structural regularity of specific problems. Established techniques, such as adiabatic evolution~\cite{Farhi2000_Quantum} or variational methods~\cite{Cerezo2021_Variational}, succeed primarily in structured quantum systems governed by local interactions or distinct symmetries, such as those in quantum chemistry~\cite{McArdle2020_Quantum} or lattice gauge theory~\cite{Martinez2016_Real, Zhang2022_Synthesizing}. 
Fully developed fluid turbulence, in stark contrast, is governed by strong nonlinearity, cross-scale coupling, and chaotic dynamics. It exhibits pronounced intermittency and multifractal statistics, stubbornly resisting the algorithmic shortcuts used for well-ordered systems. 
Consequently, preparing a chaotic, highly resolved turbulent state using a shallow quantum circuit without the use of resource-heavy ancillary qubits has remained a formidable open challenge. 

\begin{figure*}[t!]
    \centering
    \includegraphics{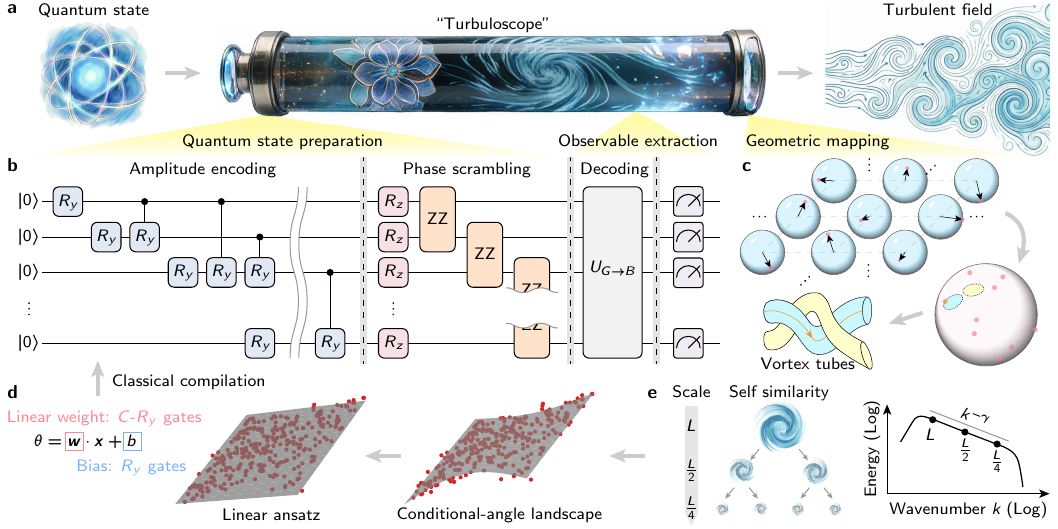}
    \caption{Physics-informed generative preparation of a turbulent quantum state.
    Classical compilation in \textbf{d} configures the state-preparation circuit in \textbf{b}, followed by geometric decoding in \textbf{c} to generate turbulent fields with the multiscale scaling in \textbf{e}.
    \textbf{a}, Conceptual ``turbuloscope'', a kaleidoscope-like generator of emergent complexity, maps a quantum-circuit-encoded perturbed state onto macroscopic turbulent vortices. 
    \textbf{b}, The amplitude encoding block encodes the prescribed Pauli-spinor modal distribution by directly mapping the regression bias $b$ and weights $\vec{w}$ onto a base $R_y$ rotation and a cascade of controlled-$R_y$ gates. Subsequently, single-qubit $R_z$ rotations and pairwise $\mathrm{ZZ}$ interactions generate a correlated phase landscape across all $2^n$ computational basis states using $O(n^2)$ circuit parameters, without the need to load each mode's phase individually. Finally, a basis transformation sequence converts the encoded state from the Gray code basis, originally utilized to preserve locality, back to the standard binary computational basis, producing the $n$-qubit register state $|\hat{\varPsi}_s\rangle$ for one physical Pauli-spinor component. 
    This ansatz achieves full state preparation in approximately $10n$ layers, excluding routing and fault-tolerant overheads.
    \textbf{c}, The geometric decoding stage of the ``turbuloscope''.
    On a grid in $\mathbb{R}^3$, the spin vector $\vec{s}$, a physical unit vector field defined at each spatial point, transforms into vorticity via Eq.~\eqref{eq:s_to_vor}. The entire spin vector field maps to the unit sphere $\mathbb{S}^2$, where a patch or point on the sphere corresponds to a vortex tube or line in $\mathbb{R}^3$, respectively. While geometrically equivalent to a Bloch sphere, this manifold represents the base space of $\vec{s}$ rather than the state space of an individual register qubit.
    \textbf{d}, Motivated by the self-similarity in \textbf{e}, we approximate the conditional rotation angles $\theta$ (which dictate the local probability split based on predecessor states) by an affine landscape in the Boolean feature space under Gray-code ordering. Such smooth geometry permits the angle distribution to be effectively approximated by a linear hyperplane within the logarithmic qubit space. 
    \textbf{e}, Multiscale fields characterized by a power-law energy spectrum $E(k)\propto k^{-\gamma}$ exhibit intrinsic scale invariance. For classical turbulence ($\gamma=5/3$), this invariance manifests as geometric self-similarity among vortices within the inertial range. This fractal-like hierarchy reflects the Richardson energy cascade, whereby kinetic energy is sequentially transferred from larger eddies to smaller ones.}
    \label{fig:encode}
\end{figure*}

To address this bottleneck, we introduce a conceptual and algorithmic framework termed the ``turbuloscope'' (Fig.~\ref{fig:encode}a). 
Analogous to a kaleidoscope, which uses mirrors to generate complex patterns from compact internal elements, the ``turbuloscope'' generates turbulent realizations from compact circuit parameters, bypassing the need to load pre-existing flow snapshots. 
As illustrated in Figs.~\ref{fig:encode}b and c, we utilize the Hopf fibration~\cite{Irvine2008_Linked} and a generalized Madelung transform~\cite{Meng2023_Quantum, Meng2024_Quantum} to mathematically map the inherent topological structures of a perturbed quantum state directly onto the macroscopic vortex tubes of a fluid flow. 
By structurally embedding a compact generative description into a quantum circuit, we successfully encode the multiscale characteristics and intermittency of fully developed turbulence using an approximately linear-depth quantum circuit. 
This approach circumvents instance-specific classical-to-quantum data loading, which typically entails exponential data-access and state-preparation overheads.
By constructing a turbulent state directly in spectral space within a compact register, this geometric encoding provides a route toward quantum investigations of classical turbulence~\cite{Preskill2018_NISQ, Xu2018_Emulating, Katabarwa2024_Early, Meng2025_Challenges, Gourianov2025_Tensor} and toward end-to-end quantum simulation of complex multiscale systems.

\firtitle{Results}
\sndtitle{Method overview}
We propose a three-stage workflow for generative initialization and geometric decoding of an instantaneous turbulent field in particular, and a multiscale field in general (see Fig.~\ref{fig:encode}).
The classical inputs comprise the qubit count $n$, the Pauli-spinor amplitude exponent $\beta$, the spectral cutoff $k_d$, the zero-mode probability $P_0$, and the relative component population $\varTheta$, and the random phase parameters $\vec{\xi}$. 
The normalized modal target of each component is compiled from $\beta$, $k_d$, and $P_0$. 
The population $\varTheta$ specifies the relative physical norms of the two components, while the phase parameters select a realization. 
The decoded velocity statistics depend on their combination through the nonlinear geometric transform.
Before quantum execution, the classical compiler evaluates the conditional-angle landscape from the prescribed spectrum and applies weighted regression to obtain the parameters for quantum circuit, as illustrated in Figs.~\ref{fig:encode}d and \ref{fig:encode}e. 
These parameters configure the amplitude-encoding gates in Fig.~\ref{fig:encode}b, after which phase scrambling and the Gray-to-binary conversion are executed on the quantum register.

First, we construct the quantum circuit to prepare the Fourier coefficients of each Pauli-spinor component.
A linear ansatz~\cite{Sarma2024_Quantum} determines the amplitude-encoding rotations, while a random-phase-scrambling layer generates correlated phases across the encoded spectral amplitudes.
Subsequently, a Gray-to-binary transformation yields the final register state.
This generative state-preparation stage achieves a circuit depth that scales approximately linearly with the qubit count.
Here and below, preparation depth refers to an ideal gate model with elementary arbitrary-angle rotations, all-to-all connectivity, and parallel scheduling. 
These estimates exclude device-specific routing and fault-tolerant overheads. 
The qubit count refers to data-register qubits, and the ancilla-free statement applies only to state preparation.

Second, measurement-specific convolution operators act on the prepared register state to extract the spectral coefficients of the fluid density, momentum, and density-weighted spin-vector fields.
These operators introduce the quadratic mode coupling inherent to the generalized Madelung transform.

Third, observable deconvolution recovers the velocity and spin-vector fields.
The Hopf fibration subsequently maps the normalized spin-vector field to vortex lines and vortex tubes in Euclidean space~\cite{Irvine2008_Linked,Yang2023_Applications}.
Consequently, the convolution and deconvolution stages decode physical information from the prepared state rather than loading additional classical flow data into the register.

The resulting workflow yields two complementary outputs.
The circuit-prepared register states encode the Fourier coefficients of the physical Pauli-spinor components, jointly providing the initial condition for Hamiltonian simulation via the generalized Madelung transform~\cite{Meng2023_Quantum, Meng2024_Quantum, Meng2024_Simulating}.
In parallel, targeted measurements and post-processing extract the corresponding physical flow observables, whereas full-field reconstruction is employed in this numerical demonstration solely to validate the statistical and geometric fidelity of the prepared state.
The displayed velocity fields are obtained from classical statevector emulation and full-field Fourier reconstruction, without quantum measurements.
By synthesizing the initial state directly within the quantum register rather than loading a prescribed classical flow field point by point, this approach circumvents the data-loading bottleneck, facilitating the subsequent simulation of multiscale vortex dynamics.

The multiscale organization of turbulence motivates the structured generative construction in Fig.~\ref{fig:encode}.
State preparation acts on the prescribed spinor modal distribution, while the velocity spectrum is evaluated after nonlinear decoding.
This physical hierarchy maps naturally onto the quantum register, where binary degrees of freedom collectively encode wavenumber modes spanning from macroscopic scales to fine-grained fluctuations.
Crucially, we employ Gray-code ordering~\cite{Caruana1988_Representation}, to ensure that physical proximity in wavenumber space is translated into Hamming locality, thereby preventing the artificial discontinuities introduced by standard binary encoding from disrupting this scale-invariant structure.

\vspace{1em}
\sndtitle{Geometric quantum encoding in spectral space}
For a flow field in $d$ spatial dimensions, each dimension $\alpha\in\{0,1,\cdots,d-1\}$ in spectral space of the wavenumber vector $\vec{k}=(k_0,k_1,\cdots,k_{d-1})$ is encoded using $n_\alpha$ qubits, with a total of $n=\sum_{\alpha=0}^{d-1} n_\alpha$ qubits.
The state vector
\begin{equation}\label{eq:Psi_hat_s_binary}
    \ket{\hat{\varPsi}_s} = \frac{1}{\|\hat\psi_s\|_2}\prod_{\alpha=0}^{d-1} \sum_{j_\alpha=0}^{2^{n_\alpha}-1} \hat{\psi}_s(\vec{k}) \bigotimes_{i=0}^{d-1}\ket{j_i}
\end{equation}
encodes the Fourier coefficients of the Pauli spinor field $\ket{\psi(\vec{x})}=[\psi_+(\vec{x}), \psi_-(\vec{x})]\T$.
Here, the two-component complex physical field $\ket{\psi(\vec{x})}$, introduced in the generalized Madelung representation~\cite{Meng2024_Quantum}, is distinct from a computational qubit.
Each spinor component $\psi_s$ ($s=\pm$) is separately amplitude-encoded within an $n$-qubit register.
Accordingly, $s$ labels the internal component of the Pauli spinor field, whereas $j_{\alpha}$ indexes a computational basis state of the qubit register.
The relative population of the two components is quantified by $\varTheta=\|\hat{\psi}_+\|_2^2/(\|\hat{\psi}_+\|_2^2 + \|\hat{\psi}_-\|_2^2)$.
The wavenumber $k_{\alpha}=0,1,\ldots,2^{n_{\alpha}-1}-1,-2^{n_{\alpha}-1},\ldots,-1$ follows the standard ordering of the discrete Fourier transform.
For each dimension $\alpha$, indices $j_\alpha$ and $k_\alpha$ are bijectively related by $k_\alpha = \mathrm{mod}(j_\alpha+2^{n_\alpha-1}, 2^{n_\alpha}) - 2^{n_\alpha-1}$ and its inverse $j_\alpha = k_\alpha + 2^{n_\alpha} [1 - H(k_\alpha)]$, where $\mathrm{mod}(\cdot, \cdot)$ is the modulo operation and $H(\cdot)$ is the Heaviside step function.

\begin{figure}[t!]
    \centering
    \includegraphics{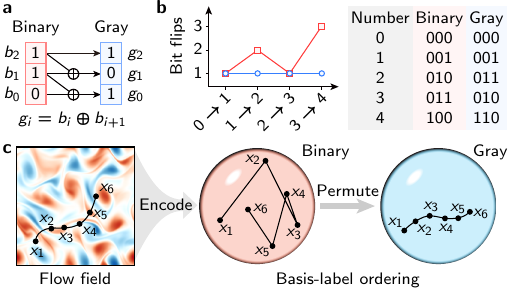}
    \caption{Adjacency-preserving Gray-code ordering for field encoding.
    \textbf{a}, The conversion from canonical binary code $b_i$ to Gray code $g_i$ is implemented via the bitwise XOR operation $g_i = b_i \oplus b_{i+1}$, which reorders quantum basis states to align physical adjacency with bitwise adjacency.
    \textbf{b}, Comparison of the Hamming distance (number of bit flips) required to transition between adjacent physical grid points.
    Standard binary encoding (red) exhibits periodic discontinuities characterized by simultaneous multi-bit flips (e.g., $011 \to 100$).
    In contrast, Gray code (blue) maintains a constant Hamming distance of 1 between consecutive indices.
    \textbf{c}, Schematic comparison of binary and Gray label orderings for sampled field coordinates. Connecting lines indicate succession of basis labels, rather than physical trajectories or distances between quantum states.}
    \label{fig:Gray_encode}
\end{figure}

First, the Fourier coefficients $\hat{\psi}_s(\boldsymbol{k})$ are synthesized via a shallow quantum circuit configured according to the nonzero-mode amplitude profile $|\hat{\psi}_s(\boldsymbol{k})|\propto k^{-\beta}\mathrm{e}^{-(k/k_d)^{10}}$, with the zero-mode probability $P_0$ specified separately.
For the turbulence examples reported here, we use $\beta=0$, which introduces no additional algebraic decay into the nonzero-mode amplitudes. This baseline choice separates the imposed modal profile from the velocity scaling obtained through component mixing and nonlinear decoding.
The register size $n$ determines the spatial resolution and the available Fourier bandwidth.
Consequently, no grid-resolved velocity field is required as input.
The modes of this target distribution are subsequently ordered according to the Gray code (see ``Methods'' and Fig.~\ref{fig:Gray_encode}).
Unlike standard binary encoding, which divides continuous physical fields into separate quantum states, this mapping ensures that the topology of the physical wavenumber space is preserved within the computational Hilbert space of the $n$-qubit register.
As shown in Fig.~\ref{fig:encode}d, this topological preservation suppresses the high-frequency fluctuations in the conditional rotation angles induced by Hamming discontinuities, yielding a near-affine landscape in the Boolean feature space.
For the prescribed spinor targets, the conditional-angle landscape admits an accurate affine approximation over the weighted configurations.
The approximate affinity is quantitatively defined and validated in SI~\cite{SI}.

Leveraging this geometric regularity, we introduce a linear ansatz that substantially reduces the complexity of state preparation.
Rather than fitting the rotation angles to a high-order polynomial, we find that the discrete conditional-angle landscape in Fig.~\ref{fig:encode}d is accurately described by an affine hyperplane over the probability-weighted region of the Boolean feature space.
Physically, this approximation implies that the intricate long-range dependencies of the turbulent field can be decomposed into simple pairwise correlations between qubits.
Subsequently, amplitude-weighted ridge regression determines the biases $b_j$ and coupling weights $w_{j,m}$ via a deterministic closed-form solution (see ``Methods'').
These parameters are obtained solely from the analytically prescribed amplitude distribution, requiring neither extraction from a deconvolved turbulent snapshot nor fitting to existing flow-field data.
The biases $b_j$ are implemented via single-qubit $R_y(b_j)$ rotations, while the weights $w_{j,m}$ parameterize the angles of the controlled-$R_y(w_{j,m})$ gates (see Figs.~\ref{fig:encode}b and \ref{fig:encode}d).

\begin{figure}[t!]
    \centering
    \includegraphics{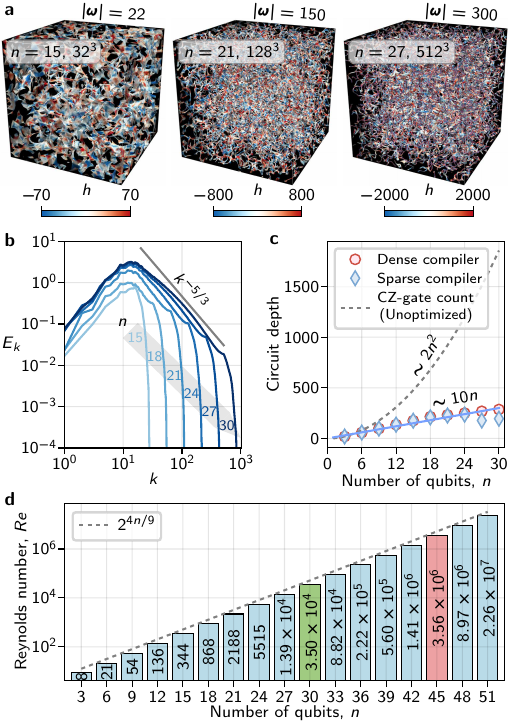}
    \caption{Scalable quantum encoding of high-$\Rey$ turbulence.
    \textbf{a}, Isosurfaces of vorticity magnitude, color-coded by helicity density, representing the quantum-encoded turbulent field with different numbers $n$ of qubits. Increasing $n$ reveals increasingly fine-grained vortices.
    \textbf{b}, As $n$ increases, the range of inertial energy-spectrum scaling broadens.
    \textbf{c}, The all-pair interaction ansatz has an unoptimized CZ-gate count that scales quadratically, whereas parallel scheduling under all-to-all connectivity yields a transpiled circuit depth that grows approximately linearly with $n$.
    The sparse compiler further reduces the transpiled depth relative to the dense compiler for larger registers, particularly at $n=27$ and $30$.
    This linear scaling demonstrates the efficiency of the proposed method for encoding highly turbulent regimes on near-term quantum processors. 
    \textbf{d}, Scale-inferred Reynolds number increases according to the scaling relation $\Rey \sim 2^{4n/9}$, with $\Rey$ values indicated for qubit counts ranging from $n=3$ to $51$. The green and red bars denote the computational limits of a desktop computer (current work) and a supercomputer~\cite{Yeung2025_GPU}, respectively.}
    \label{fig:different_nq}
\end{figure}

Following amplitude encoding, $R_z$ rotations and pairwise ZZ interactions parameterized by the random angles $\boldsymbol{\xi}$ generate a randomized but correlated phase distribution.
Notably, this circuit generates a restricted quadratic phase family rather than assigning independent phases to each Fourier mode.
In the implementation, $\beta=0$ and $k_d=15$ are held fixed, while $\varTheta$ is prescribed as a function of the qubit count $n$, and the phase parameters $\boldsymbol{\xi}$ are sampled randomly.
The two components use separately sampled phase vectors.
For a given $n$ and a sampled $\boldsymbol{\xi}$, the two-component registers $\ket{\hat{\varPsi}_{\pm}}$ jointly define a deterministic and reproducible Pauli-spinor field corresponding to a single turbulence realization.
Resampling $\boldsymbol{\xi}$ generates distinct realizations that retain the same multiscale power-law scaling for the velocity field.
This generative procedure requires neither full-resolution flow-field snapshots nor instance-specific training data, while the complete state-preparation circuit has a depth of $\Theta(n)$ (see Fig.~\ref{fig:different_nq}c).

\begin{figure}[tp!]
    \centering
    \includegraphics{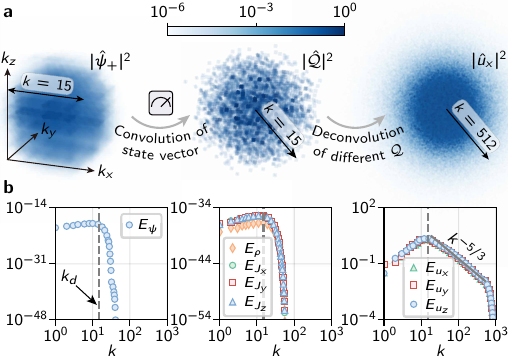}
    \caption{Three-stage geometric quantum encoding of a turbulent field in spectral space.
    \textbf{a}, Volume rendering of the spectral field magnitudes $|\hat{\psi}_+|^2$, $|\hat{\mathcal{Q}}|^2$, and $|\hat{u}_x|^2$, normalized to their respective maximum values.
    The size of the spectral support for the corresponding quantities is marked by the wavenumber $k=k_d$.
    \textbf{b}, Corresponding energy spectra $E_f(k)\equiv \sum_{\vec{k}'}|\hat{f}|^2 \delta(|\vec{k}'|-k)$.}
    \label{fig:three_stage}
\end{figure}

\begin{figure*}[t!]
    \centering
    \includegraphics{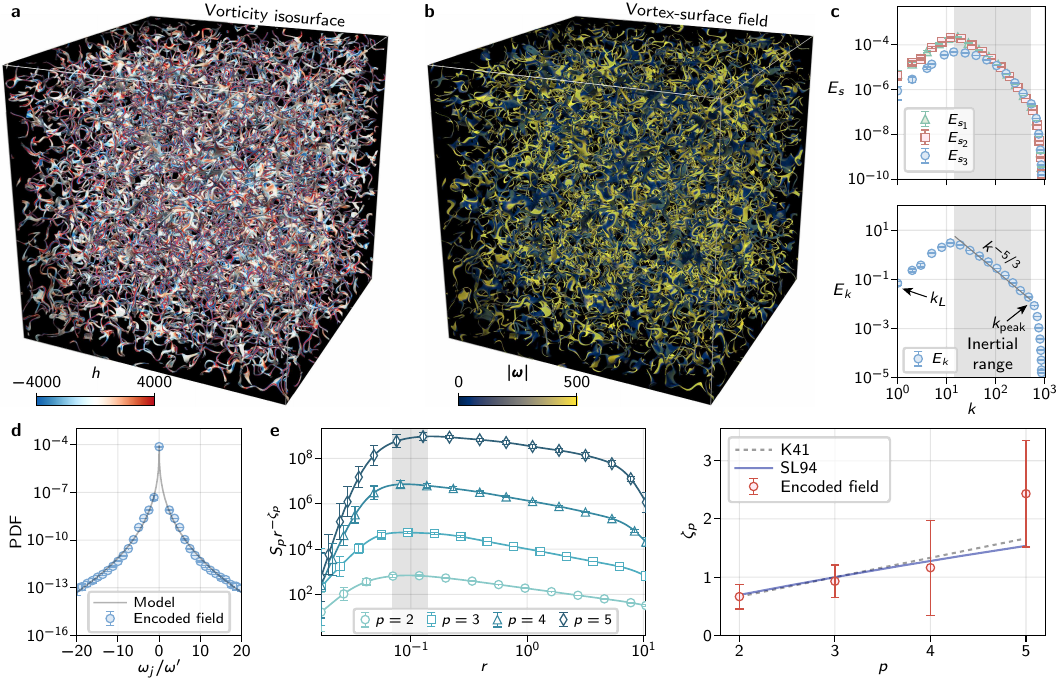}
    \caption{Geometric encoding of homogeneous isotropic turbulence using $n=30$ qubits on a $1024^3$ uniform grid.
    \textbf{a}, Isosurface of vorticity magnitude $|\vec{\omega}| = 400$ of one realization, colored by the helicity density $h \equiv \vec{u} \cdot \vec{\omega}$.
    \textbf{b}, Isosurface of vortex-surface field (VSF) $s_1 = 0.3$ of one realization, colored by $|\vec{\omega}|$. These visualizations reveal a dense, entangled network of coiled vortex tubes and convoluted sheet-like structures, highlighting the intense topological complexity and fine-scale intermittency characteristic of high-$\Rey$ flows.
    \textbf{c}, In the shaded inertial range $k\in[k_d, k_{\mathrm{peak}}]$, the energy spectrum $E_k$ displays Kolmogorov's $k^{-5/3}$ scaling, implying self-similarity of the quantum encoded turbulent field.
    Across 10 independent realizations, the mean fitted spectral exponents, with one standard deviation, are $-1.668\pm 0.004$ for the velocity spectrum and $-2.268\pm 0.004$, $-2.268\pm 0.005$, and $-1.864\pm 0.008$ for the three VSF spectra, respectively. The VSF spectra exhibit a steeper decay at high wavenumbers than the velocity spectrum.
    \textbf{d}, Probability density function of the vorticity components, normalized by $\omega'\equiv (\|\vec{\omega}\|_2^2/ 3)^{1/2}$. The data follow a stretched exponential distribution $P(|\vec{\omega}|) \sim \exp(-(|\vec{\omega}|/\sigma)^{\beta_\omega})$ with characteristic width $\sigma \approx 0.007$ and stretching exponent $\beta_\omega \approx 0.38$, which provides a statistical signature of intermittent vorticity fluctuations.
    \textbf{e}, Velocity structure functions $S_p(r) =\langle |[\vec{u}(\vec{x}+\vec{r}) - \vec{u}(\vec{x})]\cdot \vec{r}/r|^p \rangle$ for orders $p=2,3,4,5$, computed via Monte-Carlo sampling and compensated by the scaling exponents $\zeta_p$ from the She-Lévêque (SL94) model~\cite{She1994_Universal}.
    The gray shaded area indicates the self-similar range.
    The low-order scaling exponents $\zeta_p$ of the encoded field, extracted from linear fits, are compared with the K41~\cite{Kolmogorov1991_The} and SL94 models.
    The resulting exponents exhibit a nonlinear, concave dependence on the order $p$ that departs from the K41 prediction $\zeta_p=p/3$, providing quantitative evidence of anomalous scaling and intermittency.
    For low orders ($p \le 4$), the consistency between the computed $\zeta_p$ values and the SL94 model $\zeta_p=p/9 + 2[1-(2/3)^{p/3}]$ demonstrates that the generated field captures the multifractal characteristics of turbulence. Markers denote means over 10 independent realizations, and error bars indicate one standard deviation.}
    \label{fig:GeoEncodeTurb_nq=30}
\end{figure*}

Second, measurement operators evaluate the quadratic convolution among the Fourier coefficients required by the generalized Madelung transform, thereby capturing its intrinsic cross-mode coupling.
The generalized Madelung transform~\cite{Meng2023_Quantum, Meng2024_Quantum} is adopted to define the fluid density $\rho \equiv \langle\psi|\psi\rangle_s$ and momentum $\vec{J} \equiv \real\langle\psi|\hat{\vec{p}}|\psi\rangle_s$.
Here, $\hat{\vec{p}}$ is the momentum operator, and $\langle\cdot|\cdot\rangle_s$ denotes a local contraction over the two Pauli-spinor components. 
Additionally, for general compressible flows, we define the unit spin vector
\begin{equation}\label{eq:spin_vector}
    \vec{s} \equiv \frac{\langle \psi |\vec{\sigma} | \psi\rangle_s}{\langle\psi|\psi\rangle_s},
\end{equation}
with the Pauli vector $\vec{\sigma}$. 
The vorticity $\vec{\omega}\equiv\bn\times \vec{u}$ is then given in terms of the spin vector~\cite{Meng2024_Lagrangian} by
\begin{equation}\label{eq:s_to_vor}
    \vec{\omega} = \frac{1}{4}\varepsilon_{ijk}s_i\bn s_j\times \bn s_k.
\end{equation}
Consequently, the spin component $s_i=C$ is a vortex-surface field (VSF)~\cite{Yang2010_On, Yang2023_Applications}. 
Since the VSF satisfies $\vec{\omega}\cdot\bn s_i=0$, its isosurfaces can visualize the topological conservation and dynamical evolution of the vorticity field \cite{Meng2024_Lagrangian}. 
Furthermore, mapping $\vec{s}$ to a unit sphere establishes a correspondence, where patches on the sphere represent vortex tubes in $\mathbb{R}^3$. 
The area of each element is proportional to the local circulation~\cite{Chern2016_Schrödinger}, as shown in Fig.~\ref{fig:encode}c. 
Thus, this encoding method inherently incorporates the fundamental laws of vorticity dynamics.

The generalized Madelung transform maps the prescribed spinor amplitudes, relative component population, and phases to the multiscale velocity and VSF fields through quadratic convolution and density inversion (detailed in SI~\cite{SI}).
The Fourier coefficients of a fluid quantity $\mathcal{Q}$ (e.g., $\rho$, $\vec{J}$, or $\rho\vec{s}$) are expressed as the convolution
\begin{equation}\label{eq:Q}
    \hat{\mathcal{Q}}(\vec{k}) = \sum_{s,s'=\pm}\sum_{\vec{k}'\in\mathbb{Z}^d} \hat{c}_{s,s'}(\vec{k}, \vec{k}') \hat{\psi}_s(\vec{k}'+\vec{k}) \hat{\psi}_{s'}^*(\vec{k}'),
\end{equation}
where the kernel $\hat{c}_{s,s'}(\vec{k}, \vec{k}')$ depends on the physical quantity considered, as summarized in Tab.~\ref{tab:c_s}. 

\begin{table}[ht!]
    \renewcommand\arraystretch{1.3}
    \caption{Summary of the kernel $\hat{c}_{s,s'}$ for the fluid observables $\rho$, $\vec{J}$, and $\rho\vec{s}$.}
    \label{tab:c_s}
    \begin{ruledtabular}
        \begin{tabular}{cccccc}
	    $\mathcal{Q}$ & $\rho$ & $\vec{J}=\rho\vec{u}$ & $\rho s_1$ & $\rho s_2$ & $\rho s_3$ \\
            \colrule
            $\hat{c}_{+,+}$ & 1 & $\frac{1}{2}\vec{k}+\vec{k}'$ & 0 & 0 & 1 \\
            $\hat{c}_{+,-}$ & 0 & 0 & 1 & $-\ii$ & 0 \\
            $\hat{c}_{-,+}$ & 0 & 0 & 1 & $\ii$ & 0 \\
            $\hat{c}_{-,-}$ & 1 & $\frac{1}{2}\vec{k}+\vec{k}'$ & 0 & 0 & $-1$
	\end{tabular}
    \end{ruledtabular}
\end{table}

The convolution in Eq.~\eqref{eq:Q} is quadratic in $\hat{\psi}_s$, thereby defining a linear measurement operator \cite{Fano1957_Description} acting on the density matrix $\varrho = \ket{\hat{\varPsi}_s}\bra{\hat{\varPsi}_s}$. 
This operator for $\rho$ or $\vec{J}$, with kernel $\hat{c}_{s,s}$, is constructed as 
\begin{equation}\label{eq:observable_fluid}
    \hat{\mathcal{O}}_{\mathcal{Q},s}(\vec{j}) = \sum_{\vec{j}'\in \mathcal{J}(\vec{j})} \mathcal{C}(\vec{j}, \vec{j}')\ket{n(\vec{j}')}\bra{m(\vec{j}, \vec{j}')},
\end{equation}
where the detailed derivation and the definitions of the state indices $m(\vec{j}, \vec{j}')$ and $n(\vec{j}')$, the interaction coefficient $\mathcal{C}$, and the Cartesian product of the admissible per-axis index sets $\mathcal{J}(\vec{j})$ are provided in SI~\cite{SI}. 
Subsequently, the Fourier coefficients are obtained via $\hat{\mathcal{Q}}(\vec{k}(\vec{j}))=\sum_{s=\pm} \|\hat\psi_s\|_2^2 \mathrm{Tr}\big(\hat{\mathcal{O}}_{\mathcal{Q},s}(\vec{j})\varrho_s\big)$.
The convolution in Eq.~\eqref{eq:Q} transforms the cubic spectral support of $\psi_s$ into a larger spherical support of $\mathcal{Q}$ in Fig.~\ref{fig:three_stage}a, while the angle-averaged spectrum is nearly unaffected in Fig.~\ref{fig:three_stage}b. 

Finally, the velocity field in spectral space is obtained by the deconvolution $\hat{\vec{u}}=\sum_{\vec{k}'\in\mathbb{Z}^d} \hat{\rho}^{-1}(\vec{k}-\vec{k}')\hat{\vec{J}}(\vec{k}')$, where  $\hat{\rho}$ and $\hat{\vec{J}}$ are computed by convolutions of the state vector with Eq.~\eqref{eq:Q}, and $\hat{\rho}^{-1}$ denotes the formal inverse of the convolution kernel $\hat{\rho}$. 
For efficient implementation, $\hat{\vec{u}}$ is calculated by the Fourier transform of $\vec{u}=\vec{J}/\rho$, and the spin vector field $\hat{\vec{s}}$ in spectral space is determined analogously. 
The nonlinear deconvolution introduces broad cross-scale coupling in the decoded velocity and spin-vector fields, producing the approximately isotropic multiscale distributions shown in Fig.~\ref{fig:three_stage}a.
Consequently, applying this three-stage procedure to the state vector prepared by a shallow circuit constructs a turbulent field characterized by both multiscale power-law scaling and coherent vortex structures.

We analyze the computational complexity of the encoding method. 
The notations $\Omega$, $O$, and $\Theta$ are employed to denote the lower, upper, and tight asymptotic bounds, respectively.
The first stage employs a $10n$-depth quantum circuit with comparatively low overhead under all-to-all connectivity. 
The cost of the second stage increases linearly with the number of spectral measurements $N_M$. 
Note that the measurement-resource analysis below is restricted to density and momentum, because the transverse spin observables require coherent cross-component interference, as discussed in SI~\cite{SI}.
Each measurement incorporates a pre-conditioning circuit of depth $O(\poly(n))$, a post-selection step with a success probability of at least $1/2$, and $O(\varLambda/\varepsilon)$ quantum-amplitude-estimation queries to achieve an absolute precision $\varepsilon$, where $\varLambda$ bounds the coefficient norm of the linear-combination-of-unitaries representation of the measured observable (see ``Methods'').
Hence, the total complexity of the quantum component is $O(N_M \varLambda \poly(n) / \varepsilon)$.
The third stage performs classical post-processing with complexity $O(N_M\log N_M)$. 
This yields an overall time complexity of $O(N_M (\varLambda\poly(n)/\varepsilon + \log N_M))$.
For a fixed or polynomial number of target observables, $N_M=O(\operatorname{poly}(n))$, the selected-coefficient measurement stage scales polynomially with the number of qubits if $\varLambda=O(\operatorname{poly}(n))$.
Conversely, full reconstruction of all $2^n$ amplitudes reintroduces an exponential scaling in output cost and is unnecessary for the targeted quantum-simulation workflow~\cite{Meng2024_Simulating}.
The readout estimates apply to selected quadratic convolution coefficients and do not establish resource for nonlinear velocity reconstruction or subsequent time integration.

An analogous distinction applies at the input and circuit-compilation stages.
Specifically, a generic normalized $n$-qubit pure state is parameterized by $2^{n+1}-2$ independent real degrees of freedom.
Its exact preparation therefore demands $\Omega(2^n)$ gates and a circuit depth of $\Omega(2^n/n)$ under optimal parallelization~\cite{Zhang2022_Quantum,Sun2023_Asymptotically}.
Similarly, reading an arbitrary classically stored snapshot requires $\Omega(2^n)$ data-access operations, whereas quantum random-access memory merely shifts this exponential cost to memory construction and initialization rather than eliminating it~\cite{Giovannetti2008_Quantum}.

The present method instead addresses generative preparation from an analytically prescribed multiscale distribution and a compact set of phase parameters.
For this structured input, the circuit parameters are compiled without constructing or loading a full $2^n$-entry field, yielding both a circuit-compilation cost and a state-preparation circuit depth of $\Theta(n)$.
The assumptions, classical compilation algorithm, complexity bounds, error analysis, and numerical validation against the dense reference implementation are detailed in SI~\cite{SI}.

\vspace{1em}
\sndtitle{Circuit-generated turbulent initial state}
To demonstrate the turbuloscope method, we prepare a statistically homogeneous isotropic turbulent velocity field on a $1024^3$ uniform grid within a periodic box using $n=30$ qubits. 
This state is generated from Pauli-spinor target $|\hat{\psi}_s(\boldsymbol{k})|\propto k^{-\beta}\mathrm{e}^{-(k/k_d)^{10}}$ with $\beta=0$, $k_d=15$, and $P_0=0.1$, together with sampled random phase parameters $\vec{\xi}$, without relying on pre-existing direct numerical simulations or experimental snapshots.
We set $\varTheta=0.002$, which controls the relative component densities and hence their local weighting of phase gradients in the reconstructed velocity (detailed in SI~\cite{SI}).
The velocity spectrum exponent is then measured from the decoded field.
Solely for validation purposes, we classically simulate the prepared register state and reconstruct the corresponding physical fields~\cite{code}.
To the best of our knowledge, this represents the largest size achieved for a quantum-encoded multiscale field to date.
The uniform grid in each direction implies $n_\alpha=n/3=10$.
The generated instantaneous field with a nonuniform density distribution (see Fig.~S1 in SI~\cite{SI}) exhibits tangled vortex tubes in Figs.~\ref{fig:GeoEncodeTurb_nq=30}a and b, forming ``sinews'' of turbulence~\cite{Shen2024_Designing, Zhu2025_Quantum}. 
Moreover, isosurfaces of the vorticity magnitude in Fig.~\ref{fig:GeoEncodeTurb_nq=30}a and Fig.~S2 in SI~\cite{SI} demonstrate that the tubular structures are robust over a wide range of thresholds. 
This organized morphology is consistent with the coherent vortex structures observed in experiments and direct numerical simulations of high-$\Rey$ turbulence~\cite{Ishihara2009_Study}.

The generated field exhibits several characteristic instantaneous signatures of classical turbulence. To characterize its scale range, we adopt $k_L=1$ and use the enstrophy-spectrum peak $k_{\mathrm{peak}}\approx512$ as a characteristic spectral scale. The modeling convention $k_\eta\approx5k_{\mathrm{peak}}$ assigns $k_\eta\approx2560$ and a scale-inferred Reynolds number $\Rey\approx(k_\eta/k_L)^{4/3}\approx35000$. 
This inferred $k_\eta$ lies beyond the $1024^3$ Fourier grid, so the generated field does not resolve the inferred Kolmogorov scale. 
Under the same convention, the inferred Taylor-scale quantities are $\lambda/L\approx0.0207$ and $\Rey_\lambda\approx725$, with $L=k_L^{-1}$.
The velocity energy spectrum $E_k$ exhibits Kolmogorov's $k^{-5/3}$ scaling law over a broad inertial range in Fig.~\ref{fig:GeoEncodeTurb_nq=30}c.
Independently, the ensemble-averaged instantaneous advective energy flux~\cite{Fujino2023_Hierarchy} is predominantly positive, indicating a forward nonlinear transfer from large to small scales (Fig.~S3 in SI~\cite{SI}).
The VSF spectra in Fig.~\ref{fig:GeoEncodeTurb_nq=30}c decay more steeply than the $k^{-5/3}$ velocity spectrum, while the VSF isosurfaces reveal the coiling and bending of vortex tubes.
The coexistence of these statistical and geometric signatures demonstrates that the prepared instantaneous field exhibits several key signatures of turbulence.

To quantitatively assess the statistical isotropy of the quantum-encoded field, we evaluate the Reynolds stress anisotropy tensor $\varLambda_{ij} = \langle u_i' u_j' \rangle / \langle u_k' u_k' \rangle - \delta_{ij}/3$, where $u_i' = u_i - \langle u_i \rangle$ denotes the fluctuating velocity component.
This tensor provides a normalized measure of deviation from an ideal isotropic state.
The small magnitude of these quantities, which are of order $10^{-3}$ (see SI~\cite{SI} for data), indicates that the quantum-encoded turbulent field is statistically isotropic, consistent with ideal isotropic turbulence~\cite{Ishihara2009_Study}.

Additionally, we assess the statistics and structures of the encoded flow field. 
The probability density function (PDF) of the vorticity magnitude is strongly non-Gaussian, exhibiting heavy tails that follow a stretched exponential distribution in Fig.~\ref{fig:GeoEncodeTurb_nq=30}d. 
This signature identifies intermittent vorticity fluctuations concentrated within sparse, intense, tube-like structures~\cite{Vincent1991_The, He1998_Statistics, Ishihara2009_Study, Shen2024_Designing}.
Furthermore, the anomalous structure-function scaling in Fig.~\ref{fig:GeoEncodeTurb_nq=30}e indicates multifractal velocity statistics.
The simultaneous agreement of these spectral, statistical, and geometric diagnostics demonstrates that the output constitutes a statistically representative and geometrically structured turbulent realization, rather than a random field matched solely to a prescribed power-law spectrum.

\vspace{1em}
\sndtitle{Dependence of Reynolds number on number of qubits}
To assess the representational scaling, we encode turbulent fields using varying numbers of qubits.
Along the family specified by $\varTheta=\varTheta(n)$, increasing $n$ extends the available Fourier bandwidth and produces finer vortex structures and a broader range of inertial scaling (see Figs.~\ref{fig:different_nq}a and \ref{fig:different_nq}b), corresponding to higher scale-inferred Reynolds numbers.
The values of $\varTheta(n)$ are reported in SI~\cite{SI}.
Quantitatively, we observe that the peak wavenumber scales as $k_{\mathrm{peak}} = 2^{n/3-1}$, with the corresponding Reynolds number following $\Rey \sim 2^{4n/9}$ (see Fig.~\ref{fig:different_nq}d). 
This exponential scaling implies that a linear increase in the qubit count facilitates the representation of turbulence regimes with Reynolds numbers orders of magnitude larger than those tractable by classical means. 
Under this scaling, quantum registers of $70$ and $200$ qubits suffice to represent flows with $\Rey\sim10^{9}$ in aerospace engineering and $\Rey\sim10^{26}$ in astrophysics~\cite{Fujita2003_Nonthermal}, respectively, which correspond to the highest estimated Reynolds numbers in these domains.

The scalability of this approach is underpinned by the algorithmic efficiency of the state preparation protocol.
By deriving circuit parameters analytically via a closed-form solution (see ``Methods''), our algorithm bypasses the computational bottlenecks associated with iterative optimization.
Crucially, the circuit transpiled under an all-to-all connectivity model exhibits a depth that scales linearly with $n$ (see Fig.~\ref{fig:different_nq}c).
Consequently, the linear ansatz provides a scalable route to generative initialization of high-$\Rey$ turbulent states on near-term quantum hardware.

\vspace{1em}
\sndtitle{Applications to quantum simulation of multiscale systems}
The geometric quantum encoding addresses the generative initialization bottleneck in a prospective end-to-end quantum simulation of turbulence. 
By preparing a statistically representative, high-$\Rey$ turbulent state with a circuit depth of $\Theta(n)$, the method provides a native initialization module for time-dependent quantum simulations.
The prepared state subsequently supplies the initial condition for Hamiltonian simulation based on the generalized Madelung transform~\cite{Meng2023_Quantum,Meng2024_Quantum,Meng2024_Simulating}.
Following the quantum evolution, targeted physical observables are extracted via measurement procedures (see ``Methods'').

This framework is tailored for simulations where the initial conditions are specified by an input multiscale power-law scaling.
Different selections of $\boldsymbol{\xi}$ generate statistically identical yet microscopically distinct initial states, thereby facilitating the estimation of rare-event probabilities and non-self-averaging global observables.
Notably, the method is not designed for data assimilation, restarting from a prescribed numerical state, or the trajectory-level reproduction of experimentally measured flows, which conventionally demand the instance-specific loading of complete classical datasets.
Within this scope, the combination of generative state preparation, quantum evolution, and targeted measurements defines a prospective end-to-end workflow that circumvents the need for full-field classical inputs and reconstructions.
The complete quantum evolution-and-readout sequence is not implemented in the present numerical demonstration.

Beyond fluid turbulence, this geometry-based encoding framework provides a versatile and scalable approach for initializing broad classes of multiscale and chaotic systems.
The core algorithmic innovations, specifically the Gray ordering and the encoding of self-similar distributions via a linear ansatz, are not intrinsically restricted to fluid turbulence.
Consequently, this representation may extend to other systems whose modal coefficients retain a compact, structured multiscale distribution, including magnetohydrodynamic turbulence in fusion plasmas, the scale-invariant distribution of dark matter in the cosmic web, and nonlinear reaction-diffusion processes.
Thus, the present method serves as a foundational algorithmic primitive for deploying near-term quantum devices to model complex natural systems.

\firtitle{Discussion}
We introduce the ``turbuloscope'', a physics-informed framework for the generative preparation of turbulent quantum states.
This approach addresses a distinct objective from the exact loading of arbitrary, pre-existing flow fields.
Building on the established hydrodynamic Schrödinger equation, spinor, and geometric mappings~\cite{Meng2023_Quantum, Meng2024_Quantum}, the present work contributes a Gray-ordered conditional-angle construction, a sparse classical compiler with explicit error control, and an ancilla-free linear-depth circuit for the multiscale state family considered here. 
Together, these elements enable the direct generative preparation of a turbulent state within a quantum register, for subsequent Hamiltonian simulation and targeted-observable measurements.
The physical consistency of this generative engine is demonstrated by our 30-qubit encoding of a homogeneous isotropic turbulent field with a scale-inferred Reynolds number of approximately 35,000. 
The synthesized field exhibits tangled vortex networks, Kolmogorov $k^{-5/3}$ scaling over a broad inertial range, a predominantly positive ensemble-averaged advective energy flux, a strongly non-Gaussian vorticity PDF, and anomalous structure-function scaling.
These signatures confirm that our geometry-based encoding naturally captures the elusive small-scale intermittency and multifractal statistics of high-$\Rey$ turbulence, matching the fine-scale features observed in direct numerical simulations and physical experiments of turbulence.

Importantly, the turbuloscope method recasts initial preparation of turbulent states as a physics-informed generative task.
While classical spectral synthesis addresses the same generative objective, it explicitly constructs all $2^n$ grid-resolved values, requiring at least $\Omega(2^n)$ classical operations.
In contrast, the proposed method prepares the Pauli-spinor state directly within a gate-based quantum register with a circuit depth of $\Theta(n)$, thereby achieving an exponential reduction in state-generation complexity with respect to $n$.
This scaling does not violate the lower bound for arbitrary state preparation because the prepared states belong to a restricted family specified by the component modal targets and the compact phase circuit, rather than being required to match every individual amplitude of a specific field.
An end-to-end advantage additionally requires that subsequent quantum evolution and targeted measurements remain efficient at the required accuracy, whereas full-field reconstruction reintroduces an exponential output cost.
Requiring $n=3[\log_2(\Rey^{3/4}/5)+1]$ qubits, this representation achieves asymptotic optimality in qubit count for the generative task considered here, as detailed in the SI~\cite{SI}.

Despite this logarithmic qubit-count scaling within the mode-resolving encoding model, significant practical challenges remain for near-term experimental realization. The current circuit architecture necessitates all-to-all qubit connectivity and entails a transpiled circuit depth exceeding 200 layers. On existing noisy intermediate-scale quantum devices, particularly superconducting processors with restricted coupling topologies, the substantial overhead from SWAP gates and limited qubit coherence times render direct execution prohibitive. Consequently, bridging the gap between this qubit-efficient ansatz and the topological constraints of near-term hardware, perhaps through hardware-efficient routing or advanced error mitigation, remains a critical next step toward practical demonstration.
From an engineering perspective, this gap also defines a clear co-design problem: quantum-circuit architectures, measurement strategies, and flow observables should be optimized together for quantities of practical interest, such as drag, lift, mixing efficiency, and heat-transfer rates.

Nevertheless, the turbuloscope lays the groundwork for large-scale, end-to-end quantum simulations of turbulence for mechanical-engineering analysis. 
As a statistically representative initial condition characterized by prescribed multiscale and geometric properties, the generated state is directly compatible with Hamiltonian-simulation protocols for subsequent time evolution~\cite{Meng2023_Quantum, Meng2024_Simulating, Sato2024_Hamiltonian}.
Although assessed here using synthetic initial fields with selected statistics of homogeneous isotropic turbulence, the proposed state-preparation procedure does not require flow stationarity.
Instead, the method remains applicable whenever the initial state admits a compact, structured generative representation, irrespective of statistical stationarity, thereby potentially extending its scope to transitional flows~\cite{Avila2011_The}.
Furthermore, for nonperiodic domains, the Fourier basis can be replaced by boundary-adapted orthogonal basis functions that satisfy the prescribed boundary conditions while maintaining the structured modal-encoding strategy.
Looking forward, scaling this framework to merely $n > 50$ qubits may support the initialization of extreme turbulent fields at $\Rey > 10^7$~\cite{Meng2025_Challenges}, a regime entirely beyond the reach of classical supercomputers~\cite{Yeung2025_GPU}. 
Such capability would be particularly relevant to aerodynamic and hydrodynamic flows, turbomachinery, and combustion devices, where full-resolution classical simulation is often replaced by empirical closure models in mechanical engineering. 
More broadly, because this geometric encoding employs measurement convolutions to efficiently capture self-similar multiscale interactions, the underlying principles extend far beyond fluid dynamics. This universal approach provides a novel pathway for the quantum simulation of a wide array of complex systems characterized by power-law scaling and fractal geometries, including nonlinear reaction-diffusion systems~\cite{Kardar1986_Dynamic, Kondo2010_Reaction}, cosmological structure formation~\cite{Geller1989_Mapping, Bond1996_How}, and biological sequences~\cite{Voss1992_Evolution, Arneodo1995_Characterizing}.

\firtitle{Methods}
\sndtitle{Adjacency-preserving encoding via Gray codes}
We use Gray-code ordering in Eq.~\eqref{eq:Psi_hat_s_binary} to preserve one-bit adjacency between consecutive coordinate indices. 
Conventional binary encoding exhibits Hamming cliffs~\cite{Caruana1988_Representation}, where neighboring indices differ in several bits, as illustrated by the transition from 3 to 4 in Fig.~\ref{fig:Gray_encode}b. 
For the prescribed multiscale distribution, the nonzero Fourier modes of the spinor follow a smooth amplitude envelope across spectral scales. 
Together with their phases, these amplitudes determine the fluid density and momentum through spectral convolutions, motivating an encoding that facilitates accurate approximation of their modal organization. 
Figure~\ref{fig:Gray_encode}c schematically contrasts the ordering of sampled field coordinates in binary and Gray codes. 
The benefit of Gray ordering concerns the conditional-angle approximation used for compact state preparation, whose probability-weighted fitting errors are quantified in SI~\cite{SI}. 

To avoid the multi-bit changes between consecutive indices, we implement Gray-code ordering.
For a spatial index $x$ with binary representation $b_i$, the Gray code sequence $g_i$ is generated via the transformation $g_i=b_i\oplus b_{i+1}$, as shown in Fig.~\ref{fig:Gray_encode}a.
Accordingly, the quantum state corresponding to the spatial coordinate $x$ is defined as $\ket{x}_G=\bigotimes_i\ket{g_i}$.
The adjacency property of this encoding is quantified by the constant Hamming distance 
\begin{equation}
    d_H(\ket{x}_G, \ket{x+1}_G) = 1
\end{equation}
between adjacent grid points, which strictly enforces that local translations correspond to local, single-qubit rotations in the Hilbert space, avoiding $O(n)$ bit flips of binary encoding.

To relate the Gray-encoded state to the standard binary representation in Eq.~\eqref{eq:Psi_hat_s_binary}, we define a bijective mapping $\mathcal{G}:\mathbb{Z}\to \{0,1\}^n$ that converts the state index into its Gray code bit string.
The Gray-encoded quantum state
\begin{equation}
    |\hat{\varPsi}_{G,s}\rangle =\frac{1}{\|\hat\psi_s\|_2} \sum_{\boldsymbol{k}}\hat\psi_s(\boldsymbol{k})\bigotimes_{\alpha=0}^{d-1}|G(k_\alpha)\rangle
\end{equation}
preserves the amplitude distribution $\hat{\psi}_s$ while permuting the computational basis.
Subsequently, the standard binary basis $\ket{\hat{\varPsi}_s}$ can be recovered via the unitary transformation
\begin{equation}
    \ket{\hat{\varPsi}_s} = \bigg(\bigotimes_{\alpha=0}^{d-1} U_{G\to B}^{(\alpha)} \bigg)\ket{\hat{\varPsi}_{G,s}},
\end{equation}
where the Gray-to-binary operator $U_{G\to B} = \prod_{j=0}^{n-2} \mathrm{CX}^{(j+1, j)}$ is efficiently implemented using a cascade of CX gates.
This unitary basis permutation allows mathematically exact retrieval of the prepared state in standard binary ordering without additional approximation error.

For the target distributions examined here, Gray ordering improves the accumulated probability-weighted affine-fit error relative to binary ordering, as quantified in SI~\cite{SI}. 
This benefit concerns the chosen approximation and does not imply removal of physical spectral components or a general reduction in entanglement.

\vspace{1em}
\sndtitle{Encoding self-similar distribution}
We propose a resource-efficient protocol to prepare the $n$-qubit register state $|\hat{\varPsi}_s\rangle$, which encodes the Fourier coefficients of the physical spinor component $\psi_s$.
For nonzero modes, the target amplitude follows $|\hat{\psi}_s(\boldsymbol{k})|\propto k^{-\beta}\mathrm{e}^{-(k/k_d)^{10}}$, while the zero-mode probability $P_0$ is specified independently.
The procedure begins by mapping the discretized wavevector grid onto $n$ qubits using Gray code ordering. 
To encode the desired amplitude distribution, the quantum state is constructed sequentially layer by layer, with $|\hat{\psi}_s(\boldsymbol{k})|$ of each mode determined by the rotation angle $\theta_j$ applied to the corresponding qubit.
In a standard state-preparation routine, the rotation angle $\theta_j$ for the $j$-th qubit depends on the specific configuration of all preceding qubits $\vec{c}_j$, which typically requires an exponential number of controlled operations (detailed in SI~\cite{SI}).
However, by exploiting the smoothness of power-law spectra together with the locality of the Gray code, we find that these angles can be approximated by a linear ansatz
\begin{equation}\label{eq:linear_ansatz}
    \theta_j(\boldsymbol{c}_j)
    \approx
    \tilde{\theta}_j(\boldsymbol{c}_j)
    =
    b_j
    +
    \sum_{m=0}^{j-1}
    w_{j,m}q_m,
\end{equation}
where $b_j$ denotes a bias term and $w_{j,m}$ represents the coupling weight from the $m$-th precursor qubit.
This ansatz reduces the parameter-space complexity from $O(2^n)$ to $O(n^2)$.

Crucially, Eq.~\eqref{eq:linear_ansatz} admits a direct mapping onto a hardware-efficient quantum circuit architecture.
The bias $b_j$ is implemented as a local rotation $R_y(b_j)$, whereas the weights $w_{j,m}$ are realized through controlled-rotation gates $C$-$R_y(w_{j,m})$.
To determine the optimal circuit parameters $\{\vec{w}_j, b_j\}$, we employ an amplitude-weighted ridge regression that prioritizes accurate fitting of low-wavenumber modes to the target amplitude while suppressing numerical noise in the high-wavenumber region.
The optimal parameter vector is determined through a deterministic closed-form solution, thereby avoiding iterative optimization.
A scalable implementation for determining these regression parameters without constructing the full probability array is detailed and validated in SI~\cite{SI}.

Following the encoding of the amplitude distribution, we apply a phase-scrambling layer $U_p(\boldsymbol{\xi})$ to construct the final complex-valued state.
Since the amplitude-encoding circuit yields a real-valued state, $U_p(\boldsymbol{\xi})$ introduces the phase fluctuations and correlations necessary to represent turbulent realizations.
Specifically, we define the operator
\begin{equation}
    U_p(\boldsymbol{\xi})
    =
    \prod_{i=1}^{n}
    \prod_{j=i+1}^{n}
    \mathrm{ZZ}^{(i,j)}(\gamma_{ij})
    \bigotimes_{j=1}^{n}
    R_z^{(j)}(\phi_j),
    \label{eq:phase_scrambling}
\end{equation}
with the set of rotation angles $\boldsymbol{\xi}=\left(\{\phi_j\},\{\gamma_{ij}\}_{i<j}\right)$.
These diagonal gates preserve the encoded probability distribution while introducing phase fluctuations and correlations.
While a fixed configuration of $\boldsymbol{\xi}$ yields a deterministic, reproducible state, resampling $\boldsymbol{\xi}$ generates a distinct realization governed by the same input power-law amplitude scaling.
The resulting circuit depth scales linearly with $n$ on connectivity-rich quantum architectures.

\vspace{1em}
\sndtitle{Measurement of mean momentum}
We address the measurement of the mean momentum components $\bar{J}_x, \bar{J}_y$, and $\bar{J}_z$. 
The corresponding mean momentum observables $\hat{p}_x, \hat{p}_y$, and $\hat{p}_z$ are diagonal in the computational basis.
For each normalized component state $|\hat\varPsi_s\rangle$, we estimate $\bar J_{s,x}=\langle\hat\varPsi_s|\hat p_x|\hat\varPsi_s\rangle$ from $T$ independent measurement outcomes using $\tilde J_{s,x}=\frac{1}{T}\sum_{t=1}^{T}\hat p_x(j_{s,t},j_{s,t})$. 
Its variance satisfies $\operatorname{Var}(\tilde J_{s,x})\leq(2^{n/3}-1)^2/(4T)$. 
The physical mean momentum is recovered as $\bar J_x=\sum_{s=\pm}\|\hat\psi_s\|_2^2\bar J_{s,x}$, with independent component estimates combined using the same weights. 
The corresponding variance and the analogous procedures for the other directions are detailed in SI~\cite{SI}.

\vspace{1em}
\sndtitle{Measurement of spectral observables}
To evaluate physical observables governed by quadratic convolutions in spectral space, such as the density $\rho$ and momentum $\vec{J}$, we implement a structured measurement protocol based on the linear combination of unitaries~\cite{Childs2012_Hamiltonian}.
To measure the generally non-Hermitian target convolution operator in Eq.~\eqref{eq:observable_fluid}, we decompose it into a linear combination of unitary operators, expressed as $\hat{\mathcal{Q}} = \sum_{\ell} \frac{c_\ell}{2} (\hat{U}_{\ell, X} - i\hat{U}_{\ell, Y}) - \sum_{\ell} c_\ell \hat{P}_\ell^\perp$, where $\hat{U}_{\ell, X}$ and $\hat{U}_{\ell, Y}$ are unitary extensions involving Pauli gates acting on the subspace spanned by the basis states and $\hat{P}_\ell^\perp = I - \ket{m_\ell}\bra{m_\ell} - \ket{n_\ell}\bra{n_\ell}$ is an projection operator.
By constructing efficient preparation and selection oracles to encode the coefficients $c_\ell$ and address the coupled basis states, we estimate the expectation values via an ancilla-assisted Hadamard test (detailed in SI~\cite{SI}).
This procedure circumvents full state tomography, achieving a single block-encoding query that has polynomial depth, whereas the complete measurement cost additionally scales with $\varLambda/\varepsilon$.

\firtitle{Data availability}
The data sets generated during the current study are available at https://github.com/YYgroup/QEncodeTurb.

\firtitle{Code availability}
The simulation source code is available in the QEncodeTurb repository at https://github.com/YYgroup/QEncodeTurb.

\firtitle{Acknowledgements}
The authors thank G.-W. He and C. Song for helpful discussions. 
This work was supported by the National Natural Science Foundation of China (Grant Nos.~12525201, 12432010, 12602344, and 12361161602), the NSFC Excellence Research Group Program for multiscale problems in nonlinear mechanics (Grant No.~12588201), and the New Cornerstone Science Foundation through the Xplorer Prize.

\firtitle{Author contributions}
Y.Y. and Z.M. conceived the theoretical ideas. Z.M. designed the quantum circuit and carried out the numerical simulation under the supervision of Y.Y. All authors contributed to data analysis, discussion of the results, and writing of the paper.

\firtitle{Competing interests}
The authors declare no competing interests.

\bibliographystyle{apsrev4-2.bst}
\bibliography{main.bib}
\end{document}


\title{Supplementary Information for \\ ``Geometric encoding of turbulence for end-to-end quantum simulation''}

\author{Zhaoyuan Meng}
    \affiliation{State Key Laboratory for Turbulence and Complex Systems, School of Mechanics and Engineering Science, Peking University, Beijing 100871, China}
    \affiliation{State Key Laboratory of Nonlinear Mechanics, Institute of Mechanics, Chinese Academy of Sciences, Beijing 100190, China}
\author{Xiao-Ming Zhang}
    \affiliation{School of Physics, South China Normal University, Guangzhou, China}
\author{Xiao Yuan}
    \affiliation{Center on Frontiers of Computing Studies, Peking University, Beijing 100871, China}
    \affiliation{School of Computer Science, Peking University, Beijing 100871, China}
\author{Yue Yang}
    \email{yyg@pku.edu.cn}
    \affiliation{State Key Laboratory for Turbulence and Complex Systems, School of Mechanics and Engineering Science, Peking University, Beijing 100871, China}
    \affiliation{HEDPS-CAPT, Peking University, Beijing 100871, China}
    
\date{\today}

\maketitle

\beginsupplement
\renewcommand{\thepage}{S\arabic{page}}
\renewcommand{\citenumfont}[1]{S#1}
\renewcommand{\bibnumfmt}[1]{[S#1]}

\makeatletter
\renewcommand{\l@subsubsection}[2]{}
\makeatother
\tableofcontents

\clearpage
\section{Supplementary figures and data}

\begin{table}[ht!]
    \centering
    \setlength{\tabcolsep}{10pt}
    \renewcommand{\arraystretch}{1.2}
    \begin{tabular}{@{}cc@{\qquad}cc@{}}
        \hline\hline
        $b_{11}$ & $-0.022422 \pm 0.001350$
        & $b_{12}$ & $-0.000204 \pm 0.001252$ \\
        $b_{22}$ & $-0.001215 \pm 0.001559$
        & $b_{13}$ & $-0.000823 \pm 0.001008$ \\
        $b_{33}$ & $0.023638 \pm 0.001376$
        & $b_{23}$ & $-0.000264 \pm 0.001079$ \\
        \hline\hline
    \end{tabular}
    \caption{Assessment of the statistical isotropy of the quantum-encoded turbulent field with $n=30$ qubits.
    The six independent components of the Reynolds stress anisotropy tensor $\varLambda_{ij}=\langle u_i'u_j'\rangle / \langle u_k'u_k'\rangle - \delta_{ij}/3$ are tabulated, where $u_i'=u_i-\langle u_i\rangle$ represents the velocity fluctuation.
    The reported values are the means and standard deviations obtained from 10 independent realizations.
    The second invariant of this tensor is found to be $I_2(\varLambda)=-\varLambda_{ij}\varLambda_{ji}/2\approx -0.000538 \pm 0.000054$.
    The small magnitude of these quantities indicates that the quantum-encoded turbulent field is consistent with second-order statistical isotropy expected in ideal homogeneous isotropic turbulence~\cite{Ishihara2009_Study}.}
    \label{tab:anisotropy}
\end{table}

\begin{table}[ht!]
    \centering
    \setlength{\tabcolsep}{15pt}
    \renewcommand{\arraystretch}{1.35}
    \begin{tabular}{ccccc}
        \hline\hline
        $n$ & $k_{\max}$ & $\Rey$ & $\lambda/L$ & $\Rey_\lambda$ \\ \hline
        15 & 16 & 345 & 0.209 & 72 \\
        18 & 32 & 869 & 0.131 & 114 \\
        21 & 64 & 2189 & 0.0828 & 181 \\
        24 & 128 & 5515 & 0.0521 & 288 \\
        27 & 256 & 13898 & 0.0329 & 457 \\
        30 & 512 & 35020 & 0.0207 & 725 \\
        \hline\hline
    \end{tabular}
    \caption{Scale-inferred characteristic quantities for turbulent states encoded with various qubit counts $n$.
    The qubits are equally distributed across the three spatial dimensions, giving the coordinate Nyquist number $k_{\max}=2^{n/3-1}$.
    We adopt $k_L=1$, $k_\eta=5k_{\max}$, and $\Rey=(k_\eta/k_L)^{4/3}$ as a scale-assignment convention, where the factor 5 specifies the assumed relation between the characteristic spectral peak and the inferred Kolmogorov wavenumber.
    The corresponding Taylor microscale and Taylor Reynolds number follow the standard relations for homogeneous isotropic turbulence, $\lambda/L=(15/\Rey)^{1/2}$ and $\Rey_\lambda=(15\Rey)^{1/2}$, respectively. 
    They are not independent measurements of velocity gradients or viscous dissipation. Since $k_{\max}\eta=0.2$, the inferred Kolmogorov scale lies outside the represented Fourier grid in every case. The quoted Reynolds numbers are therefore model-inferred scale labels rather than independently determined values of $UL/\nu$.}
    \label{tab:Rey_lam}
\end{table}

\begin{figure}[ht!]
    \centering
    \includegraphics{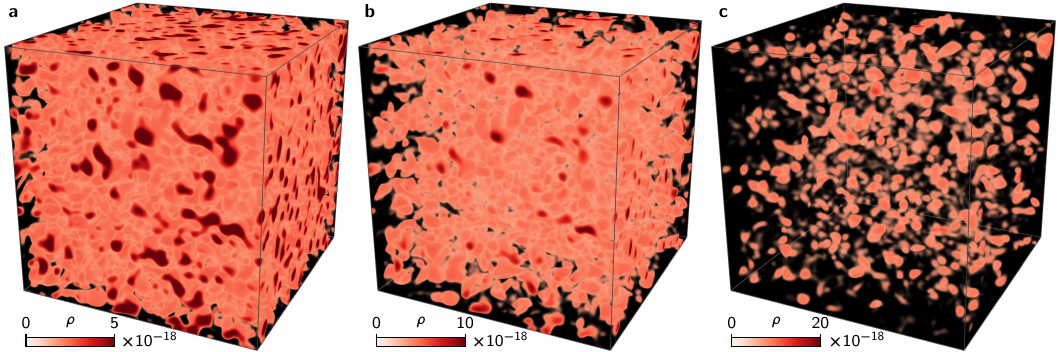}
    \caption{Volume rendering of the density for the geometrically quantum-encoded turbulent field with $n=30$ qubits. The density field exhibits blob-like structures.}
    \label{fig:rho_render}
\end{figure}

\begin{figure*}[ht!]
    \centering
    \includegraphics{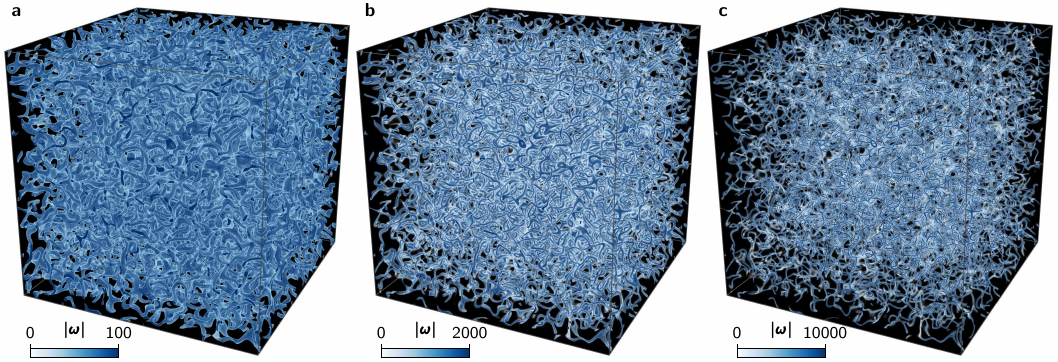}
    \caption{Volume rendering of the vorticity magnitude for the geometrically quantum-encoded turbulent field with $n=30$ qubits. The robustness of the filamentary structures across a wide range of thresholds indicates that vorticity is highly concentrated within well-defined tubular structures.}
    \label{fig:vor_render}
\end{figure*}

\begin{figure}[ht!]
    \centering
    \includegraphics{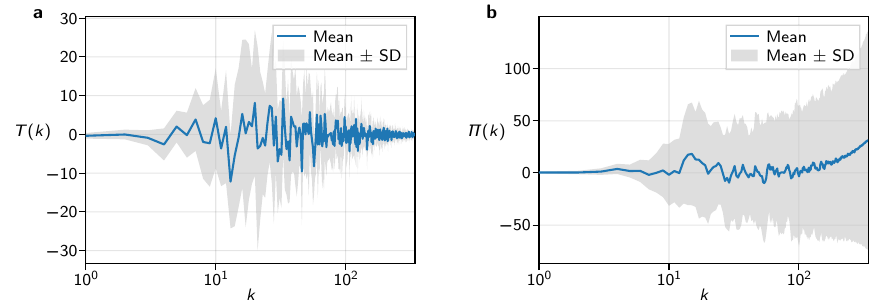}
    \caption{Ensemble statistics of scale-dependent nonlinear energy transfer. 
    Results for ten turbulent-field realizations generated with $n=30$ qubits on a $1024^3$ grid.
    \textbf{a}, Shell-wise advective energy-transfer rate $T(k)=-\sum_{\boldsymbol{q}\in S_k}\operatorname{Re}[\hat{u}_i^{*}(\boldsymbol{q})\widehat{u_j\partial_j u_i}(\boldsymbol{q})]$, with $S_k=\{\boldsymbol{q}\mid k-1/2<|\boldsymbol{q}|\leq k+1/2\}$ \cite{Goto2017_Hierarchy,Fujino2023_Hierarchy}.
    Positive and negative values of $T(k)$ represent net energy gain and loss for shell $k$, respectively.
    The alternating signs of $T(k)$ reflect the fluctuating, shell-to-shell redistribution of energy within the instantaneous field.
    \textbf{b}, Cumulative energy flux $\varPi(k)=-\sum_{k'\leq k}T(k')$, where $\varPi(k)>0$ indicates forward transfer from low to high wavenumbers.
    The cumulative flux remains positive at most scales, demonstrating an instantaneous, nonlinear tendency to transfer energy from larger to smaller scales.
    The persistence of a non-vanishing $\varPi(k)$ at the highest resolved wavenumbers is consistent with the scope of the present framework, which encodes a self-similar inertial range without resolving the terminal dissipation range.
    Combined with the Kolmogorov-like energy spectrum, the positive spectral flux serves as a complementary diagnostic of the multiscale turbulent structure of the prepared state.
    The advective nonlinearity is computed pseudospectrally using the standard $2/3$ dealiasing rule.}
    \label{fig:energy_flux}
\end{figure}

\begin{figure}[ht!]
    \centering
    \includegraphics{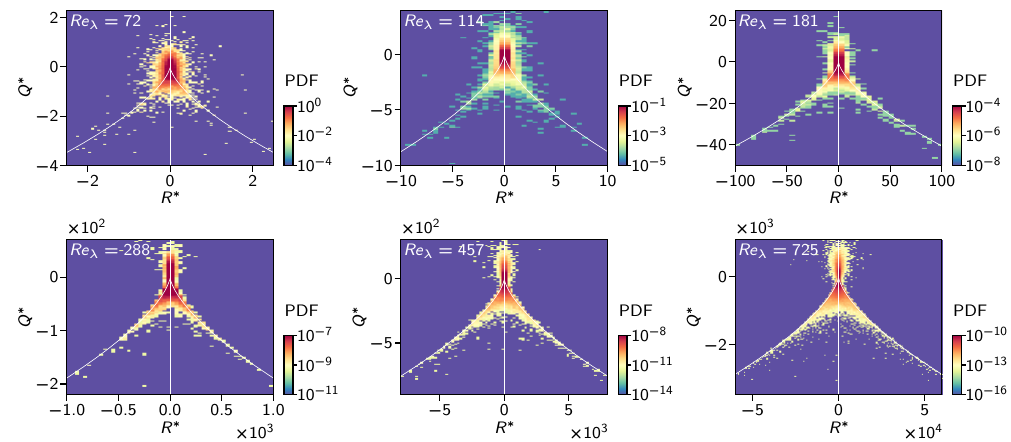}
    \caption{Normalized joint probability distributions of the velocity-gradient invariants for quantum-encoded turbulent fields across $\Rey_\lambda=72$, $114$, $181$, $288$, $457$, and $725$.
    To enable comparison with incompressible homogeneous isotropic turbulence, the reconstructed velocity field is subjected to the Helmholtz projection $\hat{u}_i^\perp(\boldsymbol{k})=(\delta_{ij}-k_i k_j/|\boldsymbol{k}|^2)\hat{u}_j(\boldsymbol{k})$ for $\boldsymbol{k}\neq\boldsymbol{0}$.
    This projection is applied exclusively for the present diagnostic.
    The projected velocity-gradient tensor is defined as $A_{ij}=\partial u_i^\perp/\partial x_j$ and satisfies $\operatorname{tr}(\tensor{A})=0$.
    The corresponding invariants are $Q=-\operatorname{tr}(\tensor{A}^2)/2$ and $R=-\operatorname{tr}(\tensor{A}^3)/3$.
    With the characteristic gradient magnitude defined as $\sigma_A=\langle A_{ij}A_{ij}\rangle^{1/2}$, we define dimensionless invariants $Q^*=Q/\sigma_A^2$ and $R^*=R/\sigma_A^3$.
    Across the investigated scale-inferred Reynolds numbers, the joint distributions exhibit a cusp-shaped core and extended strain-dominated tails that align closely with the Vieillefosse line $27(R^*)^2+4(Q^*)^3=0$ (white lines), reproducing the characteristic support of the teardrop-shaped distribution observed in Navier-Stokes turbulence.
    Unlike dynamically evolved turbulence, however, the distributions prepared here remain approximately symmetric about $R^*=0$.
    This symmetry indicates that while the prepared fields capture the Reynolds-number-robust velocity-gradient topology, they do not reproduce the signed third-order correlations generated by strain self-amplification and vortex stretching.}
    \label{fig:RQ_Reynolds}
\end{figure}

\begin{figure}[ht!]
    \centering
    \includegraphics{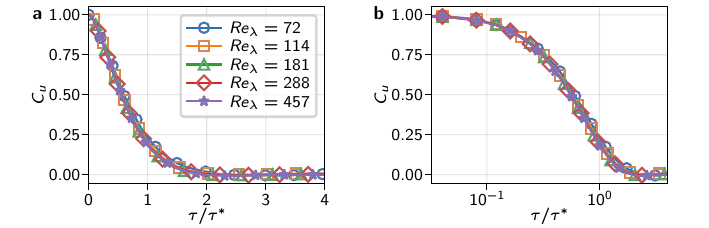}
    \caption{Two-time velocity correlations of quantum-encoded turbulent fields under two-component Schr\"odinger evolution $\ii\p_t\ket{\psi_s} = \hat{\vec{p}}^2 \ket{\psi_s}/2$, where $s=\pm$ and $\hat{\vec{p}}$ is the momentum operator.
    We evaluate the normalized single-point, two-time correlation of the longitudinal velocity component $C_u(\tau) = \langle u_x'(\boldsymbol{x},0)u_x'(\boldsymbol{x},\tau) \rangle_{\boldsymbol{x}} / \{ \langle |u_x'(\boldsymbol{x},0)|^2\rangle_{\boldsymbol{x}} \langle |u_x'(\boldsymbol{x},\tau)|^2\rangle_{\boldsymbol{x}} \}^{1/2}$, where $u_x'(\boldsymbol{x},\tau) = u_x(\boldsymbol{x},\tau) - \langle u_x(\boldsymbol{x},\tau)\rangle_{\boldsymbol{x}}$.
    For homogeneous isotropic turbulence, the spatial average $\langle\cdot\rangle_{\boldsymbol{x}}$ serves as an ensemble estimate of the two-time correlation measured at an arbitrary fixed position.
    The characteristic decorrelation time is defined as $\tau^*= 2[ \langle k^4\rangle_{\psi} - \langle k^2\rangle_{\psi}^2 ]^{-1/2}$, where $\langle f\rangle_{\psi} = \sum_{\boldsymbol{k}}W(\boldsymbol{k})f(\boldsymbol{k}) / \sum_{\boldsymbol{k}}W(\boldsymbol{k})$ with $W(\boldsymbol{k}) = \sum_{s=\pm}|\hat{\psi}_s(\boldsymbol{k},0)|^2$.
    The encoded states yield $\tau^*\approx0.0292$ across the register sizes considered, since the fixed spectral cutoff preserves the normalized distribution of occupied spinor frequencies, whereas additional high-wavenumber modes introduced at larger $n$ remain exponentially suppressed.
    The corresponding correlations for $\Rey_\lambda=72$, $114$, $181$, $288$, and $457$ are presented using \textbf{a}, linear and \textbf{b}, logarithmic coordinates.
    These curves exhibit a near-collapse, decaying from $C_u(0)=1$ to approximately zero over the range $\tau/\tau^*=O(1)$.
    This decay reflects the progressive loss of memory of the initial velocity field as the occupied spectral modes acquire distinct dynamical phases.
    The collapse is a consequence of the spinor-frequency distribution imposed during state preparation and does not imply any universal, Reynolds-number independence of physical Navier-Stokes temporal correlations.}
    \label{fig:time_correlation}
\end{figure}

\begin{figure}[ht!]
    \centering
    \includegraphics{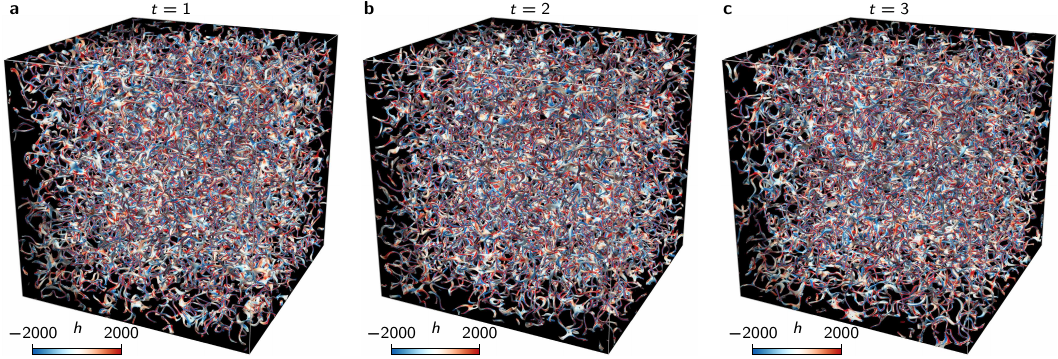}
    \caption{Temporal evolution of the vortical structures using $n=27$ qubits under two-component Schr\"odinger evolution $\ii\p_t \ket{\psi_s} = \hat{\vec{p}}^2 \ket{\psi_s}/2$, where $s=\pm$ and $\hat{\vec{p}}$ is the momentum operator.
    The vorticity isosurface $|\vec{\omega}|=300$ is visualized at $t=1$, $2$, and $3$. The persistence of the densely entangled, multiscale vortex-tube network suggests that the evolution continuously reorganizes individual vortex lines without disrupting the entire vortical structure.}
    \label{fig:iso-vor_evo_n=27}
\end{figure}

\begin{figure}[ht!]
    \centering
    \includegraphics{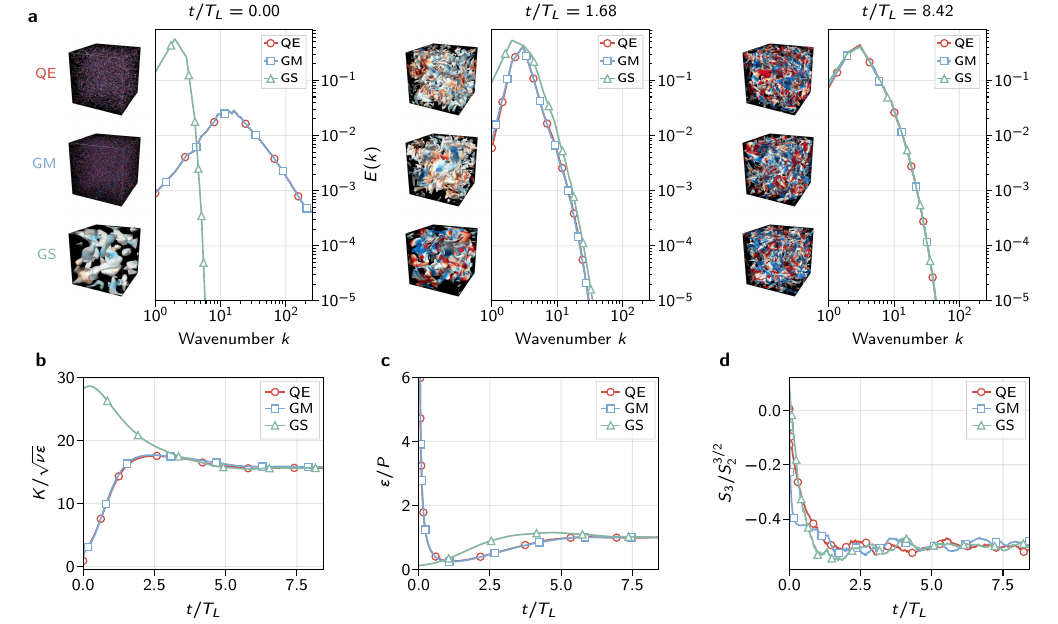}
    \caption{Relaxation to statistically stationary turbulence from quantum-encoded and Gaussian initial fields. Forced homogeneous isotropic turbulence with a kinematic viscosity of $\nu=0.001$ is simulated by solving the Navier-Stokes equations with three different initial conditions: a quantum-encoded field (QE), a Gaussian random field matching the initial energy spectrum of QE (GM), and a Gaussian random field with a conventional model spectrum $E_k\propto k^4\exp[-2(k/k_p)^2]$, where $k_p=4$ (GS). The simulations use the same initial kinetic energy, viscosity, forcing, and numerical settings~\cite{Alvelius1999_Random}. 
    \textbf{a}, Isosurfaces of the normalized vorticity magnitude, $|\vec{\omega}|/\|\vec{\omega}\|_2 = 1.35$, colored by the helicity density over the range $[-15, 15]$, together with the corresponding energy spectra at $t/T_L = 0$, $1.68$, and $8.42$.
    \textbf{b}--\textbf{d}, Evolution of the total kinetic energy, dissipation rate, and longitudinal velocity-gradient skewness, respectively. 
    Energy, dissipation, and time are normalized by the common reference scales $\sqrt{\nu\varepsilon}$, the prescribed kinetic-energy injection rate $P=1$, and the mean eddy turnover time $T_L\approx 0.594$. All three initializations approach consistent stationary turbulence statistics. QE reaches this regime within approximately five eddy turnover times, on a timescale comparable to those of GM and GS.}
    \label{fig:three_cases_evolution}
\end{figure}

\clearpage
\section{Generalized Madelung transform in spectral space}
In physical space, the generalized Madelung transform~\cite{Meng2024_Quantum} relates the Pauli spinor $\ket{\psi}=[\psi_+, \psi_-]\T$ to the density $\rho \equiv \langle\psi | \psi\rangle_s$ and momentum $\vec{J} \equiv \real\langle \psi | \hat{\vec{p}} | \psi\rangle_s$. 
Here, $\hat{\vec{p}}$ is the momentum operator, and $\langle\cdot|\cdot\rangle_s$ denotes a local contraction over the two Pauli-spinor components.
The velocity field is subsequently defined by $\vec{u}\equiv \vec{J}/\rho$.
For simplicity, we employ natural units $\hbar=1$ and $m=1$. 

We now derive the generalized Madelung transform in spectral space.
Expanding the wave function for a $d$-dimensional flow field in a Fourier basis as $\psi_s=\sum_{\vec{k}\in\mathbb{Z}^d} \hat{\psi}_s(\vec{k}) \ee^{\ii\vec{k}\cdot\vec{x}}$ yields
\begin{equation}
    \rho = \langle\psi | \psi\rangle_s
    = \sum_{s=\pm} \psi_s \psi_s^*
    = \sum_{s=\pm}\sum_{\vec{k}\in\mathbb{Z}^d} \sum_{\vec{k}'\in\mathbb{Z}^d}\hat{\psi}_s(\vec{k})\hat{\psi}_s^*(\vec{k}')\ee^{\ii(\vec{k}-\vec{k}')\cdot\vec{x}}
    = \sum_{\vec{k}\in\mathbb{Z}^d} \bigg(\sum_{s=\pm} \sum_{\vec{k}'\in\mathbb{Z}^d} \hat{\psi}_s(\vec{k}+\vec{k}')\hat{\psi}_s^*(\vec{k}') \bigg) \ee^{\ii\vec{k}\cdot\vec{x}}.
\end{equation}
Consequently, the Fourier coefficient of $\rho$ is given by a convolution
\begin{equation}
    \hat{\rho} = \sum_{s=\pm} \sum_{\vec{k}'\in\mathbb{Z}^d} \hat{\psi}_s(\vec{k}+\vec{k}')\hat{\psi}_s^*(\vec{k}').
\end{equation}
Similarly, we derive
\begin{equation}
    \begin{aligned}
        \vec{J} = \real\langle \psi | \hat{\vec{p}} | \psi\rangle_s
        = \frac{\ii}{2}\sum_{s=\pm} (\psi_s\bn\psi_s^* - \psi_s^*\bn\psi_{s})
        &= \frac{1}{2}\sum_{s=\pm}\sum_{\vec{k},\vec{k}'\in\mathbb{Z}^d} (\vec{k}+\vec{k}')\hat{\psi}_{s}(\vec{k})\hat{\psi}_{s}^*(\vec{k'})\ee^{\ii(\vec{k}-\vec{k}')\cdot\vec{x}}
        \\
        &= \sum_{\vec{k}\in\mathbb{Z}^d} \bigg(\frac12\sum_{s=\pm}\sum_{\vec{k}'\in\mathbb{Z}^d} (\vec{k}+2\vec{k}')\hat{\psi}_s(\vec{k}+\vec{k}')\hat{\psi}_s^*(\vec{k}') \bigg)\ee^{\ii\vec{k}\cdot\vec{x}},
    \end{aligned}
\end{equation}
which yields
\begin{equation}
    \hat{\vec{J}} = \frac12\sum_{s=\pm}\sum_{\vec{k}'\in\mathbb{Z}^d} (\vec{k}+2\vec{k}')\hat{\psi}_s(\vec{k}+\vec{k}')\hat{\psi}_s^*(\vec{k}').
\end{equation}
Similarly, the Fourier coefficients for the components of the spin vector field are calculated as
\begin{equation}
    \langle\psi|\sigma_i|\psi\rangle_s
    = \sum_{s,s'=\pm} \hat{c}_{s,s'}\psi_s\psi_{s'}^*
    = \sum_{\vec{k}\in\mathbb{Z}^d} \bigg(\sum_{s,s'=\pm}\sum_{\vec{k}'\in\mathbb{Z}^d} \hat{c}_{s,s'}(\vec{k}, \vec{k}') \hat{\psi}_s(\vec{k}'+\vec{k}) \hat{\psi}_{s'}^*(\vec{k}') \bigg) \ee^{\ii\vec{k}\cdot\vec{x}}, \quad i=1,2,3.
\end{equation}
The kernel $\hat{c}_{s,s'}$ is summarized in Tab.~I.
Consequently, the Fourier coefficients of $\rho$, $\vec{J}$, and $\rho s_i$ are expressed in the unified form of a convolution
\begin{equation}\label{S-eq:Q}
    \hat{\mathcal{Q}}(\vec{k}) = \sum_{s,s'=\pm}\sum_{\vec{k}'\in\mathbb{Z}^d} \hat{c}_{s,s'}(\vec{k}, \vec{k}') \hat{\psi}_s(\vec{k}'+\vec{k}) \hat{\psi}_{s'}^*(\vec{k}').
\end{equation}
Here and below, $\hat{\mathcal{Q}}(\boldsymbol{k})$ denotes a physical Fourier coefficient. The corresponding operator acting on the data register is denoted by $\hat{\mathcal{O}}_{\mathcal{Q},s}(\boldsymbol{j})$.

The generalized Madelung transform also provides a dynamical interface between the prepared state and the subsequent fluid evolution.
Evolution of the circuit-prepared spinor $\ket{\psi(0)}$ under the hydrodynamic Schr\"odinger-Pauli equation~\cite{Meng2024_Quantum} yields a mapped density and velocity that obey the Navier-Stokes equations, allowing $\ket{\psi(0)}$ to directly provide the initial condition without requiring full-field reconstruction.
The reduced two-component Schr\"odinger evolution, as illustrated in Figs.~\ref{fig:iso-vor_evo_n=27} and \ref{fig:time_correlation}, demonstrates this dynamical compatibility.

\section{Details of state-preparation circuit construction}

\subsection{Control of fluid density homogeneity via the zero-mode probability}

\begin{figure}[tp!]
    \centering
    \includegraphics{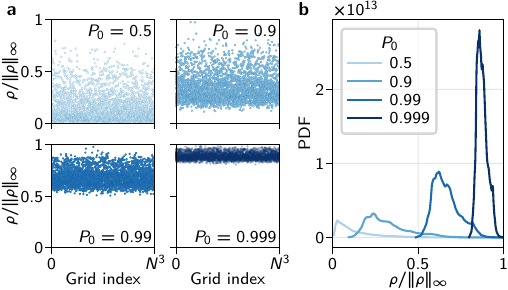}
    \caption{Effect of the ground-state probability $P_0$ on fluid density uniformity. We explicitly regulate the density homogeneity of the generated field by tuning the probability of $\ket{0}^{\otimes n}$, denoted as $P_0$.
    In general, the closer $P_0$ is to 1, the more uniform the density of the flow field, and the entire system can therefore be viewed as a perturbative system with respect to density. 
    \textbf{a}, Scatter plots of the normalized density (down-sampled to 3000 points) for $P_0=0.5$, 0.9, 0.99, and 0.999.
    The variance of density fluctuations decreases as $P_0$ approaches 1.
    \textbf{b}, The probability density function of the density converges to the Dirac delta function as $P_0$ approaches 1.
    To retain pronounced density fluctuations in the decoding step, we set $P_0=0.1$. 
    This choice controls density contrast but does not impose a divergence-free velocity field.}
    \label{fig:rho_pdf}
\end{figure}

In the proposed quantum encoding framework, the spatial distribution of the fluid density $\rho(\vec{x})$ is governed by the probability amplitudes of the quantum state $|\hat{\varPsi}_s\rangle$.
A critical control parameter is the occupancy of the zero-frequency mode, $P_{0} = |\langle 0 |^{\otimes n} |\hat{\varPsi}_s \rangle|^2$, which regulates the homogeneity of the generated fluid field.
This parameter controls the density contrast between background-dominated and strongly fluctuating regimes.

Here, we derive the relationship between the spectral probability distribution encoded in the quantum state and the spatial homogeneity of the resulting density field.
We explicitly demonstrate how the population of the zero-momentum mode $P_0$ modulates the real-space density contrast via convolution.
For simplicity without loss of generality, we consider a single-component wave function.

The $\vec{k}$-th Fourier component of the density is given by the convolution of the state vector amplitudes
\begin{equation}\label{eq:convolution}
    \hat{\rho}({\vec{k}}) = \sum_{\vec{k}' \in \mathbb{Z}^d} \hat{\psi}(\vec{k}+\vec{k}') \hat{\psi}^*(\vec{k}').
\end{equation}
Equation \eqref{eq:convolution} indicates that the spatial structure of the density field $\rho(\vec{x})$ arises from quantum interference between momentum modes separated by a wavevector $\vec{k}$.
We decompose the spectral amplitudes into a macroscopic mean field and turbulent fluctuations as
\begin{equation}\label{eq:decomposition_psi_spec}
    \hat{\psi} (\vec{k}) = \sqrt{P_0} \ee^{\ii\phi_0} \delta_{\vec{k}, \vec{0}} + \delta\hat{\psi}(\vec{k}),
\end{equation}
where $P_0 \in [0, 1]$ is the zero-mode probability, $\phi_0$ is an arbitrary global phase, and $\delta\hat{\psi}(\vec{k})$ represents the fluctuations.
The normalization constraint implies $\sum_{\vec{k}\neq\vec{0}} |\delta\hat{\psi}(\vec{k})|^2 = 1 - P_0$.

Substituting Eq.~\eqref{eq:decomposition_psi_spec} into the real-space expression yields
\begin{equation}
    \psi(\vec{x}) = \sqrt{P_0} \ee^{\ii\phi_0} + \delta\psi(\vec{x}),
\end{equation}
where $\delta\psi(\vec{x}) = \sum_{\vec{k}\neq\vec{0}} \delta\hat{\psi}(\vec{k}) \ee^{\ii\vec{k}\cdot\vec{x}}$ is the spatially varying fluctuation field.
The density field subsequently expands as
\begin{equation}
    \begin{aligned}
        \rho(\vec{x}) 
        &= \left( \sqrt{P_0} \ee^{-\ii\phi_0} + \delta\psi^*(\vec{x}) \right) \left( \sqrt{P_0} \ee^{\ii\phi_0} + \delta\psi(\vec{x}) \right) \\
        &= P_0 + |\delta\psi(\vec{x})|^2 + \sqrt{P_0} \left( \ee^{-\ii\phi_0}\delta\psi(\vec{x}) + \ee^{\ii\phi_0}\delta\psi^*(\vec{x}) \right) \\
        &= P_0 + |\delta\psi(\vec{x})|^2 + 2\sqrt{P_0} \operatorname{Re}\left[ \ee^{-\ii\phi_0}\delta\psi(\vec{x}) \right].
    \end{aligned}
    \label{eq:density_expansion}
\end{equation}
 
To quantify the uniformity, we analyze the relative density fluctuation.
Assuming the fluctuating part $\delta\psi(\vec{x})$ has a zero spatial mean, the average density approximates $\langle \rho \rangle \approx P_0 + \langle |\delta\psi|^2 \rangle = 1$.
The spatial variance of the density, $\sigma_{\rho}^2 = \langle (\rho(\vec{x}) - \langle \rho \rangle)^2 \rangle$, is primarily determined by the interference terms.
In the background-dominated limit $P_0\to1$, the interference term on the right-hand side of Eq.~\eqref{eq:density_expansion} dominates the quadratic fluctuation term.
Consequently, the variance scales as
\begin{equation}
    \sigma_{\rho}^2 \approx \left\langle \left( 2\sqrt{P_0} \operatorname{Re}[\ee^{-\ii\phi_0}\delta\psi] \right)^2 \right\rangle 
    \propto P_0 \cdot \langle |\delta\psi|^2 \rangle = P_0 (1 - P_0).
\end{equation}
Accordingly, the relative fluctuation intensity scales as
\begin{equation}
    \frac{\sigma_{\rho}}{\langle \rho \rangle} \approx \sqrt{P_0 (1 - P_0)}.
\end{equation}
When considering the amplitude of the fluctuations relative to the background as $P_0 \to 1$, the fluctuation field scales as $\sqrt{1-P_0}$.
Thus, the ratio of the fluctuating component to the constant background in Eq.~\eqref{eq:density_expansion} scales as $\sqrt{P_0(1-P_0)}/P_0 = \sqrt{(1-P_0)/P_0}$.

Therefore, the parameter $P_0$ serves as a tunable regulator of density uniformity within the encoded turbulent field.
In the limit $P_0 \to 0$, the density $\rho(\vec{x})$ is governed exclusively by the higher-order correlation $|\delta\psi(\vec{x})|^2$, resulting in pronounced spatial inhomogeneity and the formation of voids where $\rho \to 0$.
Conversely, a nonzero $P_0$ introduces a uniform background component (see Fig.~\ref{fig:rho_pdf}). The perturbative scaling applies as $P_0\to1$, whereas the density and velocity-dilatation statistics at $P_0=0.1$ are evaluated directly in the numerical diagnostics.

In the implementation with $n=30$ qubits, we sample 10 independent random phase vectors and evaluate the dimensionless velocity-dilatation measure
\begin{equation}
    \vartheta=
    \frac{\left\langle
    (\boldsymbol{\nabla}\cdot\boldsymbol{u})^2
    \right\rangle}
    {\left\langle
    (\boldsymbol{\nabla}\cdot\boldsymbol{u})^2
    \right\rangle+
    \left\langle
    |\boldsymbol{\nabla}\times\boldsymbol{u}|^2
    \right\rangle}.
\end{equation}
At the target $P_0=0.1$, we obtain $\vartheta=0.3333\pm0.0032$, reported as the mean $\pm$ sample standard deviation across the 10 phase realizations.
These values indicate appreciable velocity dilatation.

\subsubsection*{Numerical conditioning of fluid decoding}
The velocity and spin fields involve division by the local density. 
For small reconstruction errors, the velocity perturbation satisfies $\delta\boldsymbol{u}\simeq(\delta\boldsymbol{J}-\boldsymbol{u}\,\delta\rho)/\rho$, indicating potential error amplification in low-density regions. 
A density zero does not necessarily imply a singular physical velocity because the momentum may vanish simultaneously. 
Nevertheless, accurate numerical evaluation requires consistent reconstruction of both quantities. The original decoding uses direct division without a density floor and serves as the reference for the following tests.

At the prescribed $P_0=0.1$, we characterize the normalized density $q=\rho/\langle\rho\rangle$ over 10 independent random phase realizations on grids of $32^3$, $64^3$, $128^3$, $256^3$, $512^3$, and $1024^3$. 
For each realization, we calculate the minimum and the $0.01\%$, $0.1\%$, and $1\%$ quantiles of $q$.
Table~\ref{tab:norm_rho} reports their means and sample standard deviations across phase realizations.
The smallest normalized density observed across all tested grids and realizations is $5.378\times10^{-7}$, and no grid points have nonfinite reconstructed velocity or spin.
Normalization by the spatial mean separates local density contrast from the overall normalization of the encoded spinor.

\begin{table}[ht!]
    \centering
    \setlength{\tabcolsep}{10pt}
    \renewcommand{\arraystretch}{1.35}
    \begin{tabular}{ccccc}
        \hline\hline
        Grid & Minimum $q$ & $0.01\%$ quantile & $0.1\%$ quantile & $1\%$ quantile \\
        \hline
        $32^3$ & $(2.905 \pm 1.655)\times10^{-3}$ & $(7.764 \pm 2.240)\times10^{-3}$ & $0.02231 \pm 0.00210$ & $0.07438 \pm 0.00163$  \\
        $64^3$ & $(1.219 \pm 0.732)\times10^{-3}$ & $(7.262 \pm 1.083)\times10^{-3}$ & $0.02287 \pm 0.00086$ & $0.07485 \pm 0.00092$  \\
        $128^3$ & $(2.888 \pm 1.118)\times10^{-4}$ & $(3.857 \pm 0.146)\times10^{-3}$ & $0.01242 \pm 0.00013$ & $0.04199 \pm 0.00032$  \\
        $256^3$ & $(3.905 \pm 2.012)\times10^{-5}$ & $(2.011 \pm 0.053)\times10^{-3}$ & $(6.613 \pm 0.073)\times10^{-3}$ & $0.02377 \pm 0.00016$  \\
        $512^3$ & $(7.137 \pm 3.630)\times10^{-6}$ & $(1.122 \pm 0.013)\times10^{-3}$ & $(3.807 \pm 0.041)\times10^{-3}$ & $0.01549 \pm 0.00015$  \\
        $1024^3$ & $(1.416 \pm 0.725)\times10^{-6}$ & $(6.637 \pm 0.054)\times10^{-4}$ & $(2.391 \pm 0.016)\times10^{-3}$ & $0.01205 \pm 0.00011$  \\
        \hline\hline
    \end{tabular}
    \caption{Minimum and low-quantile normalized densities across grids and phase realizations. For each realization, $q=\rho/\langle\rho\rangle$ is evaluated on the full grid. Density statistics are reported as mean ± standard deviation across phase realizations.}
    \label{tab:norm_rho}
\end{table}

To assess sensitivity to low-density regions, we introduce the diagnostic regularization $\boldsymbol{u}_{\epsilon}=\boldsymbol{J}/(\rho+\epsilon\langle\rho\rangle)$.
The density and momentum are held fixed while $\epsilon$ takes the values $10^{-6}$, $10^{-5}$, $10^{-4}$, and $10^{-3}$.
The corresponding low-density volume fraction is $f_{\epsilon}=\langle \mathbbm{1}_{q<\epsilon}\rangle$, where $\mathbbm{1}$ denotes the indicator function.
Vorticity is recomputed from each regularized velocity using the same discrete derivative operator as for the unregularized reference.
For $F=\boldsymbol{u}$ or $\boldsymbol{\omega}$, the relative change is measured by $\Delta_F(\epsilon)=\|F_{\epsilon}-F_0\|_2/\|F_0\|_2$, where $F_0$ denotes the unregularized result.
Table~\ref{tab:decoding_regularization} reports the low-density volume fractions, relative field changes, and velocity-spectrum slope changes as means and sample standard deviations across 10 independent phase realizations at each resolution.
Spectral slopes are fitted over $[15,N/2]$.
The $32^3$ grid contains only two shells in this interval, so its slopes remain undefined.
For the $1024^3$ grid, $\epsilon=10^{-6}$ produces mean relative changes of $0.00620$ in velocity and $0.0144$ in vorticity, with a mean spectral-slope change of $-0.00403$.
At $\epsilon=10^{-3}$, these changes increase to $0.191$, $0.417$, and $-0.656$, respectively, despite a low-density volume fraction of only $2.151\times10^{-4}$.
These results show that stronger regularization substantially alters the reconstructed fields and spectrum, even when the volume fraction below the regularization threshold is small.

\begin{table}[ht!]
    \centering
    \setlength{\tabcolsep}{8pt}
    \renewcommand{\arraystretch}{1.35}
    \begin{tabular}{cccccc}
        \hline\hline
        $n$ & $\epsilon$ & $f_\epsilon$ & $\Delta_u$ & $\Delta_\omega$
        & Velocity-spectrum slope change \\
        \hline

        \multirow{4}{*}{15}
        & $10^{-6}$ & $0$
        & $(4.777\pm2.875)\times10^{-5}$
        & $(4.783\pm2.869)\times10^{-5}$
        & $\mathrm{N/A}$ \\
        & $10^{-5}$ & $0$
        & $(4.752\pm2.846)\times10^{-4}$
        & $(4.758\pm2.840)\times10^{-4}$
        & $\mathrm{N/A}$ \\
        & $10^{-4}$ & $0$
        & $(4.527\pm2.584)\times10^{-3}$
        & $(4.533\pm2.579)\times10^{-3}$
        & $\mathrm{N/A}$ \\
        & $10^{-3}$
        & $(3.052\pm9.651)\times10^{-6}$
        & $0.03252\pm0.01330$
        & $0.03256\pm0.01328$
        & $\mathrm{N/A}$ \\
        \hline

        \multirow{4}{*}{18}
        & $10^{-6}$ & $0$
        & $(6.963\pm5.131)\times10^{-5}$
        & $(7.460\pm5.479)\times10^{-5}$
        & $(-8.580\pm3.838)\times10^{-5}$ \\
        & $10^{-5}$ & $0$
        & $(6.878\pm4.981)\times10^{-4}$
        & $(7.369\pm5.320)\times10^{-4}$
        & $(-8.512\pm3.730)\times10^{-4}$ \\
        & $10^{-4}$ & $0$
        & $(6.191\pm3.878)\times10^{-3}$
        & $(6.634\pm4.143)\times10^{-3}$
        & $(-7.967\pm2.957)\times10^{-3}$ \\
        & $10^{-3}$
        & $(2.289\pm2.667)\times10^{-6}$
        & $0.03746\pm0.01270$
        & $0.04015\pm0.01359$
        & $-0.05914\pm0.01119$ \\
        \hline

        \multirow{4}{*}{21}
        & $10^{-6}$ & $0$
        & $(1.384\pm0.450)\times10^{-4}$
        & $(1.808\pm0.590)\times10^{-4}$
        & $(-1.453\pm0.254)\times10^{-4}$ \\
        & $10^{-5}$ & $0$
        & $(1.350\pm0.427)\times10^{-3}$
        & $(1.765\pm0.560)\times10^{-3}$
        & $(-1.433\pm0.241)\times10^{-3}$ \\
        & $10^{-4}$ & $0$
        & $0.01117\pm0.00280$
        & $0.01459\pm0.00367$
        & $-0.01289\pm0.00159$ \\
        & $10^{-3}$
        & $(6.819\pm1.056)\times10^{-6}$
        & $0.05454\pm0.00487$
        & $0.07118\pm0.00639$
        & $-0.08725\pm0.00373$ \\
        \hline

        \multirow{4}{*}{24}
        & $10^{-6}$ & $0$
        & $(6.529\pm3.064)\times10^{-4}$
        & $(1.048\pm0.492)\times10^{-3}$
        & $(-4.293\pm1.024)\times10^{-4}$ \\
        & $10^{-5}$ & $0$
        & $(5.303\pm2.086)\times10^{-3}$
        & $(8.507\pm3.341)\times10^{-3}$
        & $(-3.898\pm0.745)\times10^{-3}$ \\
        & $10^{-4}$
        & $(2.623\pm1.231)\times10^{-7}$
        & $0.02532\pm0.00472$
        & $0.04058\pm0.00746$
        & $-0.02847\pm0.00232$ \\
        & $10^{-3}$
        & $(2.510\pm0.161)\times10^{-5}$
        & $0.08664\pm0.00404$
        & $0.1382\pm0.0062$
        & $-0.1648\pm0.0040$ \\
        \hline

        \multirow{4}{*}{27}
        & $10^{-6}$ & $0$
        & $(2.517\pm1.376)\times10^{-3}$
        & $(4.883\pm2.672)\times10^{-3}$
        & $(-1.362\pm0.270)\times10^{-3}$ \\
        & $10^{-5}$
        & $(8.196\pm6.524)\times10^{-9}$
        & $0.01280\pm0.00271$
        & $0.02484\pm0.00528$
        & $-0.01039\pm0.00077$ \\
        & $10^{-4}$
        & $(8.501\pm0.389)\times10^{-7}$
        & $0.04377\pm0.00217$
        & $0.08471\pm0.00418$
        & $-0.06496\pm0.00146$ \\
        & $10^{-3}$
        & $(8.005\pm0.200)\times10^{-5}$
        & $0.1323\pm0.0016$
        & $0.2516\pm0.0028$
        & $-0.3295\pm0.0024$ \\
        \hline

        \multirow{4}{*}{30}
        & $10^{-6}$
        & $(3.725\pm4.809)\times10^{-10}$
        & $(6.203\pm2.117)\times10^{-3}$
        & $0.01444\pm0.00489$
        & $(-4.030\pm0.742)\times10^{-3}$ \\
        & $10^{-5}$
        & $(2.496\pm0.430)\times10^{-8}$
        & $0.02235\pm0.00312$
        & $0.05198\pm0.00713$
        & $-0.02632\pm0.00140$ \\
        & $10^{-4}$
        & $(2.501\pm0.080)\times10^{-6}$
        & $0.06805\pm0.00178$
        & $0.1569\pm0.0039$
        & $-0.1463\pm0.0020$ \\
        & $10^{-3}$
        & $(2.151\pm0.024)\times10^{-4}$
        & $0.1910\pm0.0013$
        & $0.4175\pm0.0025$
        & $-0.6556\pm0.0035$ \\
        \hline\hline
    \end{tabular}
    \caption{Sensitivity of fluid decoding to the diagnostic regularization $\boldsymbol{u}_\epsilon=\boldsymbol{J}/(\rho+\epsilon\langle\rho\rangle)$.
    Spectral-slope changes are regularized minus unregularized values, fitted over $[15,N/2]$.
    Results are reported as mean $\pm$ sample standard deviation across 10 independent phase realizations at each resolution.
    For $n=15$, the fitting interval contains only two wavenumber shells, so the fitted slopes are undefined.}
    \label{tab:decoding_regularization}
\end{table}

\subsection{Construction and optimization of the quantum circuit for amplitude encoding}
\subsubsection*{Linear ansatz}
We propose an efficient method for preparing an $n$-qubit state $|\varPsi\rangle = \sum_{\vec{k} \in \{0,1\}^n} \sqrt{P(\vec{k})} |\vec{k}\rangle$ characterized by a target Pauli-spinor amplitude scaling $\sqrt{P(\vec{k})} \propto k^{-\beta}$ with a given constant $\beta$.
Crucially, to suppress numerical artifacts, we impose a smooth spectral cutoff at the wavenumber $k_d$.
The ideal power-law scaling is modulated by a super-Gaussian window function, which yields the modified target distribution
\begin{equation}\label{S-eq:target_spectrum}
    \sqrt{P(\vec{k})} \propto k^{-\beta} \ee^{-(k/k_d)^{10}}, \quad \vec{k}\ne\vec{0}.
\end{equation}
This regularization defines an effective bandwidth, ensuring that the probability amplitude decays smoothly to zero.
With the zero-mode probability $P_0$ specified independently, the normalized target distribution is
\begin{equation}
    P(\boldsymbol{k})=
    \begin{cases}
        P_0,
        &\boldsymbol{k}=\boldsymbol{0},\\[4pt]
        \displaystyle
        \frac{1-P_0}{Z_\beta}
        k^{-2\beta}\mathrm{e}^{-2(k/k_d)^{10}},
        &\boldsymbol{k}\ne\boldsymbol{0},
    \end{cases}
\end{equation}
where $Z_\beta=\sum_{\boldsymbol{q}\ne\boldsymbol{0}} q^{-2\beta}\mathrm{e}^{-2(q/k_d)^{10}}$ is evaluated over the represented Fourier grid. 
Here, $P(\boldsymbol{k})$ denotes the normalized target distribution for each component register, and the physical component population is specified separately through $\|\hat\psi_s\|_2^2$.
The present implementation uses $\beta=0$ to encode turbulent fields, whereas the nonzero exponents shown in Fig.~\ref{fig:construct_power_law} illustrate the broader range of amplitude profiles supported by the method.

We represent the discretized probability distribution $P(\vec{k})$ as a joint probability distribution $P(q_0, q_1, \dots, q_{n-1})$ defined over $n$ qubits.
By applying the chain rule of probability, this high-dimensional distribution is factorized into a product of conditional probabilities
\begin{equation}
    P(q_0, q_1, \dots, q_{n-1}) = P(q_0) \cdot P(q_1 \mid q_0) \cdot P(q_2 \mid q_0, q_1) \cdots P(q_{n-1} \mid q_0, \cdots, q_{n-2}).
\end{equation}
Based on this decomposition, the standard approach consists of a sequence of $n$ rotation layers, in which the state of the $j$-th qubit is set by a rotation $R_y(\theta_j)$ conditioned on the states of the preceding $j$ qubits, referred to as the precursor string $\vec{c}_j = q_{j-1}\cdots q_1 q_0$.
Mathematically, the required rotation angle $\theta_j(\vec{c}_j)$ is determined by the conditional probabilities of the target distribution.
For a given precursor configuration $\vec{c}_j$, the rotation angle is expressed as
\begin{equation}\label{eq:exact_angle}
    \theta_j^{\text{target}}(\vec{c}_j)
    = 2 \arccos\sqrt{P(q_j=0 \mid \vec{c}_j)} 
    = 2 \arccos\sqrt{\frac{P(q_j=0, \vec{c}_j)}{P(\vec{c}_j)}},
\end{equation}
where $P(\vec{c}_j) = P(q_j=0, \vec{c}_j) + P(q_j=1, \vec{c}_j)$ denotes the marginal probability of the precursor string.

For a general target amplitude distribution $P(\vec{k})$, the function $\theta_j^{\text{target}}(\vec{c}_j)$ is arbitrary.
Because there are $2^j$ distinct precursor configurations for the $j$-th layer, a faithful implementation requires a lookup table of size $2^j$.
Realizing this dependence in a quantum circuit necessitates either a generic multi-controlled rotation gate or a quantum random access memory (QRAM) implementation~\cite{Giovannetti2008_Quantum}, resulting in a gate count that scales as $O(2^n)$ and is therefore intractable for scalable simulation.

For the target distributions examined here, Gray ordering yields a conditional-angle landscape that is approximately affine over the probability-weighted configurations. 
This is an empirical approximation assessed below, rather than a consequence of one-bit adjacency alone.
Within the probability-weighted region critical for state preparation, this landscape is near-affine and is therefore represented by the linear ansatz below.
The deviation of this landscape from the fitted hyperplane is subsequently quantified following the regression procedure.

We define a linear ansatz in which the rotation angle for qubit $j$ is approximated as a linear combination of the states of the preceding qubits as
\begin{equation}\label{eq:linear_ansatz}
    \theta_j^{\text{target}}(\boldsymbol{c}_j)
    \approx
    \tilde{\theta}_j(\boldsymbol{c}_j)
    =
    b_j
    +
    \sum_{m=0}^{j-1}
    w_{j,m}q_m,
\end{equation}
where $q_m \in \{0, 1\}$ denotes the state of the $m$-th qubit, $b_j$ is a scalar bias term, and $w_{j,m}$ is the weight coefficient coupling qubit $m$ to qubit $j$.
The linear relation in Eq.~\eqref{eq:linear_ansatz} maps directly onto a resource-efficient quantum circuit architecture, in which the bias $b_j$ is implemented as a local rotation $R_y(b_j)$ on qubit $j$, and the weights $w_{j,m}$ are realized through two-qubit controlled rotations $C$-$R_y(w_{j,m})$, with qubit $m$ as the control and qubit $j$ as the target.
By adopting this ansatz, the number of parameters in layer $j$ is reduced from $2^j$ to $j+1$.
Consequently, the total controlled-$R_y$-gate count of the circuit scales as $O(n^2)$.

The parameters $\{b_j, \vec{w}_j\}$ for each layer are determined classically prior to circuit execution.
We cast this task as a supervised learning problem in which the inputs are the precursor bitstrings $\vec{c}_j$ and the targets are the exact rotation angles $\theta_j^{\text{exact}}(\vec{c}_j)$.
Because the physical distribution $P(\vec{k})$ is dominated by high-amplitude modes at low wavenumbers, fitting errors in these regions strongly degrade the state fidelity.
By contrast, errors in the low-probability tail at high wavenumbers have a negligible effect.
Accordingly, we employ an amplitude-weighted ridge regression.

To determine the optimal parameters $\{\vec{w}_j, b_j\}$, we employ amplitude-weighted ridge regression independently for each layer.
The optimization objective is to minimize the weighted mean squared error
\begin{equation}
    \mathcal{L}_j = \sum_{\vec{c}_j \in \mathcal{S}} \sqrt{P(\vec{c}_j)} \left[\theta_j^{\mathrm{target}}(\vec{c}_j) - \tilde{\theta}_j(\vec{c}_j) \right]^2 + \lambda \|\vec{w}_j\|_2^2
\end{equation}
between the predicted angles and the exact target angles.
The set
\begin{equation}
    \mathcal{S}_j := \{\vec{c}_j \in \{0, 1\}^j \mid P(\vec{c}_j) > \epsilon_{\mathrm{cutoff}}\}
\end{equation}
denotes the collection of significant precursor configurations whose marginal probabilities exceed the numerical threshold $\epsilon_{\mathrm{cutoff}} = 10^{-25}$.
The bit string of the $m$-th sample is given by $\vec{c}_j^{(m)} = q_{j-1}^{(m)} \cdots q_1^{(m)} q_0^{(m)}$.
This filtering procedure removes negligible-probability precursor configurations and numerical noise from the regression objective, thereby restricting the ansatz to physically relevant regions of the Hilbert space.
The regularization term $\lambda \|\vec{w}_j\|_2^2$, with fixed regularization parameter $\lambda = 10^{-9}$, stabilizes the regression and suppresses excessively large coefficients.
Sample weights are chosen as $\sqrt{P(\vec{c}_j)}$, ensuring that the regression prioritizes accurate fitting of high-energy modes while remaining tolerant to errors in the low-energy cutoff region.

The optimal parameter vector is determined by a deterministic closed-form solution
\begin{equation}\label{eq:ridge_solution}
    [\vec{w}_j, b_j]\T = \left( \tensor{X}_j\T \tensor{\Omega}_j \tensor{X}_j + \lambda \tensor{I}' \right)^{-1} \tensor{X}_j\T \tensor{\Omega}_j \vec{y}_j,
\end{equation}
thereby eliminating the necessity for iterative optimization.
Here, the augmented design matrix 
\begin{equation}
    \tensor{X}_j 
    = \sum_{l=0}^{j-1} \sum_{m=1}^{|\mathcal{S}_j|} q_l^{(m)}\ket{m}\bra{l} + \sum_{m=1}^{|\mathcal{S}_j|}\ket{m}\bra{j}
\end{equation}
and the target angle vector 
\begin{equation}
    \vec{y}_j = \sum_{m=1}^{|\mathcal{S}_j|} \theta_j^{\mathrm{target}} (\vec{c}_j^{(m)})\ket{m}
\end{equation}
are constructed from the precursor configurations.
These components are weighted by the diagonal matrix 
\begin{equation}
    \tensor{\Omega}_j = \sum_{m=1}^{|\mathcal{S}_j|} \sqrt{P(\vec{c}_j^{(m)})} \ket{m}\bra{m}.
\end{equation}
The regularization term employs a modified identity matrix 
\begin{equation}
    \tensor{I}' = \sum_{l=0}^{j} (1-\delta_{l,j}) \ket{l}\bra{l} 
    = \operatorname{diag} \bigl(\underbrace{1,\cdots,1}_{j\ \mathrm{entries}},0\bigr).
\end{equation}
Here, $\boldsymbol X_j\in\mathbb R^{|\mathcal S_j|\times(j+1)}$, $\boldsymbol y_j\in\mathbb R^{|\mathcal S_j|}$, $\boldsymbol\Omega_j\in\mathbb R^{|\mathcal S_j|\times|\mathcal S_j|}$, and $\boldsymbol I'\in\mathbb R^{(j+1)\times(j+1)}$. 
Sample rows are indexed by $m=1,\cdots,|\mathcal S_j|$, whereas feature columns are indexed by $l=0,\cdots,j$. 
The first $j$ columns correspond to $q_0,\cdots,q_{j-1}$, and the last column contains ones for the bias $b_j$. Thus, $\boldsymbol I'$ penalizes the regression weights but leaves the bias unregularized.

\subsubsection*{Scalable classical compilation}
We design and implement a sparse compiler to bypass the construction of a dense probability array of length $2^n$ and its subsequent use in evaluating the prefix probabilities $P(\boldsymbol{c}_j)$, thereby avoiding the exponential classical time and memory overheads of the circuit compiler.

Given that the target distribution in Eq.~\eqref{S-eq:target_spectrum} is analytic with a super-Gaussian cutoff, this compiler first determines the spectrally sparse set $\mathcal{S}_{\epsilon} = \left\{ \boldsymbol{k}: |\boldsymbol{k}|\leq k_{\epsilon} \right\}$ with the discarded probability $\epsilon_{\mathrm{tail}}=\sum_{\boldsymbol{k}\notin\mathcal{S}_{\epsilon}} P(\boldsymbol{k}) \leq\epsilon$, retaining only the $M_{\epsilon}=|\mathcal{S}_{\epsilon}|$ most significant modes.
For the tenth-order super-Gaussian cutoff, the cutoff wavenumber and the mode count scale as $k_{\epsilon}=O[k_d \log^{1/10} (1/\epsilon)]$ and $M_{\epsilon}=O(k_{\epsilon}^{d})$.
Thus, for fixed $k_d$, spatial dimension $d$, and error tolerance $\epsilon$, $M_{\epsilon}$ remains independent of the full Hilbert-space dimension $2^n$.

Under this approach, each retained mode is converted into a Gray-code bit string and inserted into a sparse prefix tree.
The required marginal probabilities are then accumulated exclusively over occupied tree nodes according to
\begin{equation}
    P(\boldsymbol{c}_j)
    =
    \sum_{\substack{
    \boldsymbol{k}\in\mathcal{S}_{\epsilon}\\
    \operatorname{prefix}[\boldsymbol{g}(\boldsymbol{k}),j]
    =
    \boldsymbol{c}_j
    }}
    P(\boldsymbol{k}),
    \end{equation}
thereby avoiding the construction of a full probability array of length $2^n$.
The target conditional angles $\theta_j^{\mathrm{target}}(\boldsymbol{c}_j)$ are subsequently evaluated solely for prefix nodes with nonzero weight.
Then, the sparse compiler directly accumulates the sufficient statistics required for this weighted regression.
For an integer $r\in\{0,\cdots,n-1\}$, we define the corresponding $r$-local precursor set $\mathcal{N}_r(j):= \{m\in\mathbb{Z} \mid \max(0,j-r)\leq m\leq j-1 \}$.
The corresponding $r$-local affine ansatz reads
\begin{equation}
    \tilde{\theta}_j^{(r)}(\boldsymbol{c}_j)
    =
    b_j
    +
    \sum_{m\in\mathcal{N}_r(j)}
    w_{j,m}q_m,
\end{equation}
where $w_{j,m}=0$ for $j-m>r$.
Here, $r$ represents the maximum backward distance, measured within the chosen qubit ordering, over which a precursor qubit can directly influence the conditional rotation of qubit $j$.
Notably, the limits $r=0$ and $r=n-1$ recover the bias-only and fully connected affine ansatzes, respectively.
For a fixed value of $r$, each layer contains at most $r+1$ regression parameters, and the complete amplitude circuit requires only $O(nr)$ rotation gates.

Consequently, the resulting classical compilation achieves a time complexity of $O(M_{\epsilon}nr^2+nr^3)$ and a memory complexity bounded by $O(M_{\epsilon}n+nr^2)$, without any exponential dependence on $2^n$.
For fixed values of $M_{\epsilon}$ and $r$, as employed in our implementation, the classical compilation cost scales at most linearly with $n$.
In terms of quantum resources, the amplitude circuit contains $O(nr)$ rotation gates.
Meanwhile, the phase circuit comprises at most $O(n^2)$ two-qubit phase gates and exhibits an $O(n)$ logical depth under the standard all-to-all connectivity model.
For fixed precursor radius $r$, the complete state-preparation circuit therefore admits linear logical depth under the stated all-to-all scheduling assumptions.

Additionally, the spectral-truncation and circuit-approximation errors admit distinct analytical bounds.
We define the normalized full target amplitude state
\begin{equation}
    |\varPsi\rangle
    =
    \sum_{\boldsymbol{k}}
    \sqrt{P(\boldsymbol{k})}
    |\boldsymbol{k}\rangle
\end{equation}
and the corresponding normalized sparse target state
\begin{equation}
    |\varPsi_\epsilon\rangle
    :=
    \frac{1}{\sqrt{1-\epsilon_{\mathrm{tail}}}}
    \sum_{\boldsymbol{k}\in\mathcal{S}_\epsilon}
    \sqrt{P(\boldsymbol{k})}
    |\boldsymbol{k}\rangle.
\end{equation}
The overlap between this sparse state and the full target state satisfies
\begin{equation}
    \langle\varPsi|\varPsi_\epsilon\rangle
    =
    \frac{
        \sum_{\boldsymbol{k}\in\mathcal{S}_\epsilon}
        P(\boldsymbol{k})
    }{
        \sqrt{1-\epsilon_{\mathrm{tail}}}
    }
    =
    \sqrt{1-\epsilon_{\mathrm{tail}}}.
\end{equation}
This overlap directly implies the fidelity bound
\begin{equation}
    \left|
        \langle\varPsi|\varPsi_\epsilon\rangle
    \right|^2
    =
    1-\epsilon_{\mathrm{tail}}
    \ge
    1-\epsilon.
\end{equation}

Let $U_j^\epsilon$ denote the exact conditional-rotation layer associated with the sparse target distribution, and let $\tilde{U}_j^{(r)}$ represent its $r$-local affine approximation.
The ideal sparse state and the compiled circuit state are given by $|\varPsi_\epsilon\rangle=U_n^\epsilon\cdots U_2^\epsilon U_1^\epsilon |0\rangle^{\otimes n}$ and $|\varPsi_{\mathrm{circuit}}\rangle = \tilde{U}_n^{(r)} \cdots \tilde{U}_2^{(r)} \tilde{U}_1^{(r)} |0\rangle^{\otimes n}$, respectively.
We then define the maximum angle error in layer $j$ as
\begin{equation}
    \Delta_j
    :=
    \max_{\boldsymbol{c}_j\in\mathcal{S}_j^\epsilon}
    \big|
        \theta_j^\epsilon(\boldsymbol{c}_j)
        -
        \tilde{\theta}_j^{(r)}(\boldsymbol{c}_j)
    \big|,
\end{equation}
where $\mathcal{S}_j^\epsilon$ denotes the set of precursor configurations with nonzero probability under the sparse target distribution.
For precursor configurations with zero probability, the exact conditional angles can be set equal to the compiled angles, as their values do not affect the preparation of $|\varPsi_\epsilon\rangle$.

Since each conditional-rotation layer is block-diagonal in the precursor basis, the operator norm difference satisfies
\begin{equation}
    \big\|
        U_j^\epsilon-\tilde{U}_j^{(r)}
    \big\|
    \leq
    2\sin\frac{\Delta_j}{4}
    \leq
    \frac{\Delta_j}{2}.
\end{equation}
A telescoping expansion of the two circuit products therefore yields
\begin{equation}
    \left\|
        |\varPsi_\epsilon\rangle
        -
        |\varPsi_{\mathrm{circuit}}\rangle
    \right\|_2
    \leq
    2\sum_j
    \sin\left(\frac{\Delta_j}{4}\right)
    \leq
    \frac{1}{2}\sum_j\Delta_j.
\end{equation}
Combining this result with the triangle inequality yields the final distance bound
\begin{equation}\label{S-eq:ini_error}
    \begin{aligned}
    \left\|
        |\varPsi\rangle
        -
        |\varPsi_{\mathrm{circuit}}\rangle
    \right\|_2
    &\leq
    \left\|
        |\varPsi\rangle-|\varPsi_\epsilon\rangle
    \right\|_2
    +
    \left\|
        |\varPsi_\epsilon\rangle
        -
        |\varPsi_{\mathrm{circuit}}\rangle
    \right\|_2
    \\
    &\leq
    \sqrt{
        2-2\sqrt{1-\epsilon_{\mathrm{tail}}}
    }
    +
    \frac{1}{2}\sum_j\Delta_j.
    \end{aligned}
\end{equation}
This bound clearly decouples the error introduced by spectral truncation from that caused by the affine circuit approximation.
Furthermore, applying any subsequent phase-scrambling circuit identically to both states preserves this norm bound due to unitary invariance.

To validate our approach, we benchmark the sparse compiler against the dense reference implementation for $n=30$.
For this evaluation, we set $\epsilon_{\mathrm{tail}}=10^{-12}$ and $r=3$, parameters shown to be sufficient for achieving high numerical accuracy.
Under these settings, the state fidelity between the two prepared configurations is $1.000$, the relative $L_2$ error of the energy spectrum is $3.5\times 10^{-8}$, and the maximum relative $L_2$ error of the resulting velocity, vorticity, and vortex-surface fields is $1.1\times 10^{-6}$.
Specifically, the sparse compiler processes only $30047$ significant modes and $76578$ prefix nodes, whereas the dense implementation must evaluate all $2^{30}$ probability values.
Consequently, the classical compilation time is reduced from $122.5\,\mathrm{s}$ to $0.81\,\mathrm{s}$, yielding an acceleration factor of $151$.
To assist the community, our updated public repository~\cite{code} preserves the dense compiler as a small-scale reference implementation while providing the sparse compiler as the primary, scalable tool.

\begin{figure}[ht!]
    \centering
    \includegraphics{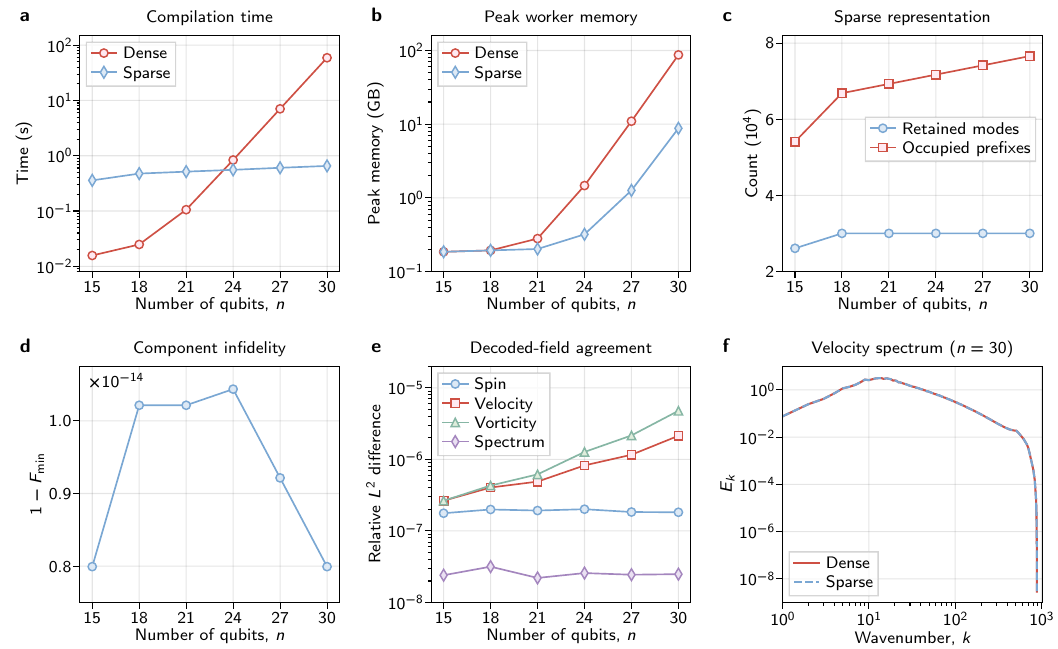}
    \caption{Dense-sparse compiler consistency.
    \textbf{a}, \textbf{b}, Classical compilation time and peak memory of the isolated compilation worker, respectively, for the dense and sparse compilers. Peak memory includes interpreter and import overhead and is reported in decimal GB. Statevector simulation and field reconstruction are excluded from both measurements.
    \textbf{c}, Numbers of retained Fourier modes and occupied prefixes in the sparse representation. The prefix count includes the root and leaves of the retained Gray-code trie.
    \textbf{d}, Component infidelity $1-F_{\min}$, where $F_{\min}$ is the smaller squared overlap between the corresponding normalized component states prepared by the two compilers.
    \textbf{e}, Relative $L^2$ differences in the spin, velocity, vorticity, and velocity energy spectrum, using the dense results as the reference.
    \textbf{f}, Unweighted velocity energy spectra obtained with the two compilers with $n=30$ qubits, shown as solid and dashed curves, respectively.
    One phase realization is used at each qubit count. This comparison assesses consistency between the two compilers rather than preparation fidelity with respect to the analytic target state.}
    \label{fig:dense_sparse_consistency}
\end{figure}

Furthermore, to assess the scalability and numerical consistency of sparse coefficient compilation, we compare the sparse compiler with the dense reference implementation for $n=15,18,21,24,27,30$ at fixed $k_d=15$, using one phase realization at each register size.
Figure~\ref{fig:dense_sparse_consistency} reports compilation time and peak memory for both implementations, retained-mode and occupied-prefix counts, dense-sparse state fidelity, and downstream relative differences.
Both implementations use identical analytic inputs, phase parameters, component populations, locality restrictions, regression settings, and gate-pruning thresholds.
The truncation tolerance is $10^{-12}$, and the locality radius is 3.
Weighted ridge regression uses a regularization coefficient of $10^{-9}$ and a prefix-probability cutoff of $10^{-25}$, with weights proportional to the square root of prefix probability and normalized by their layerwise maximum.
The bias is unpenalized, and controlled rotations retain only weights satisfying $|w_{j,m}|>10^{-3}$.
Compilation resources exclude statevector emulation and full-field reconstruction. 
Peak memory includes interpreter and import overhead.

The retained-mode count remains at 30047 for $n\geq18$, while the occupied-prefix count increases from 54115 at $n=15$ to 76578 at $n=30$.
At $n=30$, sparse compilation requires $8.80$ GB of peak memory, compared with $87.1$ GB for dense compilation.
Across all tested register sizes, component infidelities remain below $1.1\times10^{-14}$, and the largest downstream relative $L^2$ difference among spin, velocity, vorticity, and velocity energy spectrum is $4.78\times10^{-6}$.
These results support numerical consistency between the two compilers for the tested fixed-cutoff family.
The reported fidelity quantifies agreement between corresponding normalized compiled component states.

\subsubsection*{Quantification of the affine conditional-angle landscape}
To quantify the deviation of the conditional-angle landscape from the fitted affine hyperplane, we define the normalized affine-deviation measure
\begin{equation}
    \mathcal{K}_j
    :=
    \frac{
        \displaystyle
        \sum_{\vec{c}_j\in\mathcal{S}_j}
        \sqrt{P(\vec{c}_j)}
        \left[
        \theta_j^{\mathrm{target}}(\vec{c}_j)
        -
        \tilde{\theta}_j(\vec{c}_j)
        \right]^2
    }{
        \displaystyle
        \pi^2
        \sum_{\vec{c}_j\in\mathcal{S}_j}
        \sqrt{P(\vec{c}_j)}
    }.
    \label{eq:conditional_angle_curvature}
\end{equation}
Here, $\sqrt{\mathcal{K}_j}$ represents the amplitude-weighted root-mean-square fitting error normalized by the natural angular scale of $\pi$.
Under this formulation, $\mathcal{K}_j=0$ denotes a perfectly affine landscape, whereas $\mathcal{K}_j\ll1$ implies that the affine ansatz accounts for the vast majority of the conditional-angle variation within the physically occupied region of the Boolean feature space.
Furthermore, we define the accumulated error over the complete preparation circuit as
\begin{equation}
    \mathcal{K}_{\mathrm{global}}
    :=
    \frac{
        \displaystyle
        \sum_j
        \sum_{\boldsymbol{c}_j\in\mathcal{S}_j}
        \sqrt{P(\vec{c}_j)}
        \left[
        \theta_j^{\mathrm{target}}(\boldsymbol{c}_j)
        -
        \tilde{\theta}_j(\boldsymbol{c}_j)
        \right]^2
    }{
        \displaystyle
        \pi^2
        \sum_j
        \sum_{\boldsymbol{c}_j\in\mathcal{S}_j}
        \sqrt{P(\vec{c}_j)}
    }.
    \label{eq:global_conditional_angle_deviation}
\end{equation}

\begin{figure}[ht!]
    \centering
    \includegraphics{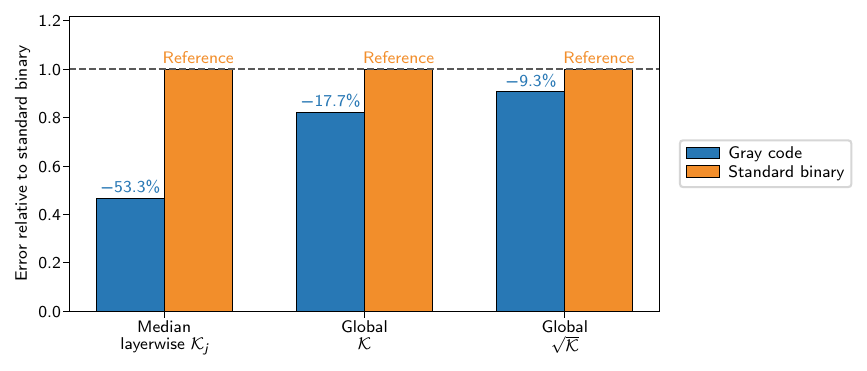}
    \caption{Comparison of affine-approximation errors for Gray-code and standard-binary orderings.
    The three metrics are normalized individually to their respective standard-binary baselines.
    Specifically, Gray-code ordering reduces the median layerwise deviation by $53.3\%$, the global probability-weighted deviation by $17.7\%$, and the corresponding global normalized root-mean-square deviation by $9.25\%$.
    Furthermore, within the reported numerical precision, the Gray-code representation yields an exactly vanishing $\mathcal{K}_j$ across several repeated early-refinement layers of the directional subregisters.}
    \label{fig:gray_binary_error}
\end{figure}

Figure~\ref{fig:gray_binary_error} compares the affine-approximation errors associated with Gray-code and standard-binary orderings.
The accuracy enhancement facilitated by Gray-code ordering is not uniform across all circuit layers.
Instead, it drives a significant portion of the conditional-angle relations into the perfectly affine regime.
Specifically, the numerical evaluation yields $\mathcal{K}_j=0$ for $j\in \{1,2,3,4,11,12,13,14,21,22,23,24\}$.
These layers constitute three recurring groups, namely $j=1$--$4$, $11$--$14$, and $21$--$24$, which correspond to the first four nontrivial conditional refinements within each of the three ten-qubit directional subregisters.
In contrast, standard-binary ordering retains finite residuals across all nontrivial layers.

Across the nontrivial layers $j=1,\cdots,29$, Gray-code ordering reduces the median normalized affine-deviation measure by approximately $53.3\%$ compared to standard-binary ordering.
This reduction reflects the larger proportion of conditional-angle relations that become exactly or nearly affine under the Gray-code representation.
Moreover, the accumulated probability-weighted error decreases from $\mathcal{K}_{\mathrm{global}}=3.51\times10^{-3}$ under standard-binary ordering to $2.89\times10^{-3}$ under Gray-code ordering, representing an overall reduction of approximately $17.7\%$.
Equivalently, the global normalized root-mean-square deviation $\sqrt{\mathcal{K}_{\mathrm{global}}}$ decreases by $9.25\%$, dropping from $5.92\times10^{-2}$ to $5.37\times10^{-2}$.

These findings reveal two complementary effects of the Gray-code representation.
First, the representation renders a recurring subset of early conditional refinements perfectly affine within numerical precision.
Second, it reduces both the median layerwise error and the accumulated probability-weighted error across the entire circuit.
The maximum layerwise errors remain of the same order of magnitude for both representations, indicating that the advantage of Gray-code ordering lies primarily in the overall error distribution rather than in a uniform improvement across all layers.

\subsubsection*{Accuracy and structure of the sparsified amplitude-encoding circuit}
\begin{figure}[t!]
    \centering
    \includegraphics{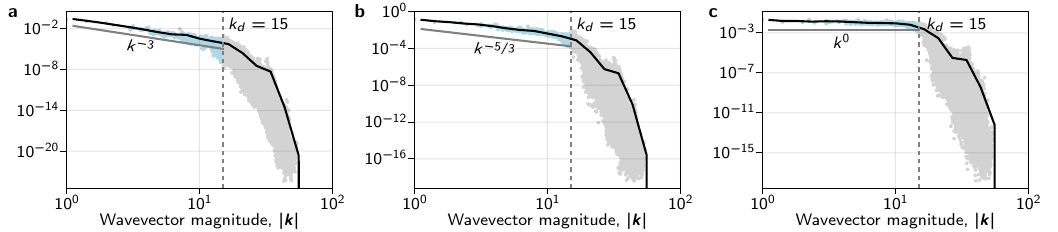}
    \caption{Amplitude encoding across distinct spectral scaling regimes.
    The prepared amplitude distributions are presented for three power-law targets: \textbf{a}, $k^{-3}$; \textbf{b}, $k^{-5/3}$; and \textbf{c}, $k^0$.
    Scatter points represent the discrete amplitudes of individual wavevectors generated by the linear ansatz circuit.
    Blue points denote physically relevant modes within the effective bandwidth defined by the cutoff wavenumber ($k \le k_d$), whereas gray points indicate exponentially suppressed modes in the dissipation region ($k > k_d$).
    The solid black line depicts the radially averaged spectrum $\widebar{|\hat{\psi}_s|}(k) = \sqrt{\frac{1}{|S_k|}\sum_{\vec{k}\in S_k}|\hat{\psi}_s(\vec{k})|^2}$, where $S_k$ denotes the set of discrete wavevectors within the interval $[k, k+\Delta k)$, demonstrating that the statistical properties of the prepared quantum states accurately reproduce theoretical scaling laws. These exponents describe the encoded Pauli-spinor component amplitudes rather than the decoded velocity energy spectrum.}
    \label{fig:construct_power_law}
\end{figure}

\begin{figure}[t!]
    \centering
    \includegraphics{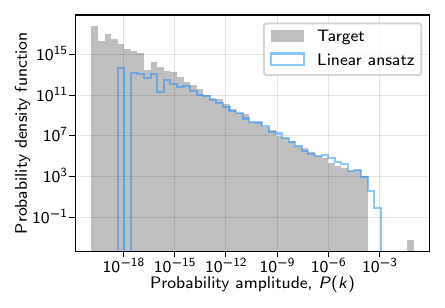}
    \caption{Statistical distributions of the probability amplitude.
    The histogram compares the probability density functions of the target scalar field amplitudes (gray bars) and the quantum-encoded state (blue line).
    Both distributions are visualized on a log-log scale to elucidate their scaling behavior across multiple orders of magnitude.
    The substantial overlap demonstrates that the ridge regression-based circuit accurately captures the statistics of the target distribution throughout the relevant range of amplitudes.}
    \label{fig:pdf_Pk}
\end{figure}

Finally, to further sparsify the circuit, any resulting weight $|w_{j,m}|$ smaller than a threshold $10^{-3}$ is truncated to zero, removing the corresponding $C$-$R_y$ gate from the circuit.
Figure~\ref{fig:construct_power_law} shows the amplitude-encoding results across distinct spectral scaling regimes with a truncated wavenumber $k_d = 15$ obtained using the sparsified circuit.
Figure~\ref{fig:pdf_Pk} compares the probability density functions of the target scalar field amplitudes and the quantum-encoded state.

\begin{figure}[t!]
    \centering
    \includegraphics{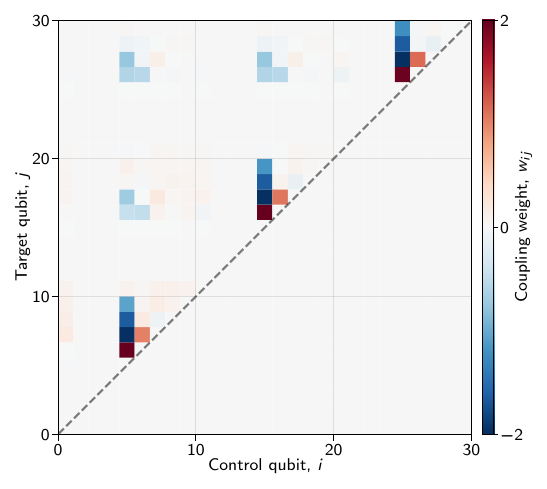}
    \caption{Emergent locality in the quantum circuit parameters.
    The heatmap shows the ridge regression coefficients $w_{ij}$, which quantify the strength of control exerted by precursor qubit $i$ on target qubit $j$ within the layer-wise amplitude encoding scheme.
    The matrix displays a pronounced quasi-diagonal structure, in which significant coupling weights $|w_{ij}| > 10^{-3}$ are concentrated almost exclusively near the main diagonal.
    Within the fitted affine model, nearby precursor bits generally carry larger weights.
    }
    \label{fig:weights_matrix}
\end{figure}

The structural analysis of the circuit parameters illustrated in Fig.~\ref{fig:weights_matrix} elucidates the interplay between the data encoding strategy and the physical properties of the target field.
Although the linear ansatz in Eq.~\eqref{eq:linear_ansatz} allows for a fully connected interaction graph, the solution in Eq.~\eqref{eq:ridge_solution} yields a sparse, quasi-banded weight matrix.
This locality is observed for the present target distribution and Gray ordering.

Physically, the prescribed spinor modal distribution has a regular amplitude envelope, reflecting the hierarchical energy cascade across local wavenumbers.
Unlike binary ordering with Hamming cliffs, Gray ordering assigns one-bit changes to consecutive coordinate indices.
The concentration of weights along the diagonal thus indicates that the refinement of the probability amplitude at scale $j$ primarily necessitates information from the preceding coarse-graining scale $j-1$ rather than the global configuration.
Such observations have critical implications for quantum simulation on near-term hardware.
The weight locality motivates pruning of weak controlled rotations. 
In particular, the subsequent all-pair ZZ phase block changes entanglement without changing computational-basis probabilities. 
The linear-depth construction instead relies on the fixed-radius restriction and scheduling assumptions specified above.

\subsubsection*{Distinction between exact flow-field loading and generative turbulent-state preparation}
The state-preparation task addressed here differs fundamentally from the exact loading of a prescribed classical flow-field snapshot.
Specifically, for a single discretized velocity component $\boldsymbol{u}^{\mathrm{data}} =(u_0,u_1,\ldots,u_{D-1})$ with $D=2^n$, exact amplitude encoding requires preparing the instance-specific state
\begin{equation}
    \left|u_{\mathrm{data}}\right\rangle
    =
    \frac{1}{
        \|\boldsymbol{u}^{\mathrm{data}}\|_2
    }
    \sum_{j=0}^{D-1}
    u_j|j\rangle.
\end{equation}
Such a task treats the $2^n$ amplitudes as independent classical inputs, preserving the prescribed field pointwise.

Within the standard single- and two-qubit gate model, a generic normalized pure state in a $2^n$-dimensional Hilbert space is characterized by $2^{n+1}-2$ independent real parameters.
Its exact preparation therefore necessitates $\Omega(2^n)$ elementary gates in the worst case, corresponding to a circuit depth of $\Omega(2^n/n)$ under optimal parallelization~\cite{Zhang2022_Quantum, Sun2023_Asymptotically}.
Moreover, retrieving all entries of a conventionally stored classical field already requires $\Omega(2^n)$ data-access operations.
A preloaded QRAM reduces the query depth but still requires $\Omega(2^n)$ addressable memory elements and initialization resources~\cite{Giovannetti2008_Quantum}.
Consequently, QRAM shifts this exponential complexity to memory construction and initialization rather than eliminating it.
While specific flow snapshots possessing a compact representation may admit efficient approximate loaders, this scenario constitutes a restricted structured-data problem rather than general, instance-specific loading.

The present method addresses a different state-preparation task.
The classical inputs comprise the register size $n$, the Pauli-spinor power-law scaling exponent $\beta$ defined in Eq.~\eqref{S-eq:target_spectrum}, and the randomly sampled gate-angle vector $\boldsymbol{\xi}$ that determines the fluctuations and intermode correlations.
The prepared state is
\begin{equation}
    \ket{\varPsi_{\beta,\boldsymbol{\xi}}}
    =
    U(\beta,\boldsymbol{\xi})
    |0\rangle^{\otimes n}.
\end{equation}
For fixed $\beta$ and $\boldsymbol{\xi}$, the circuit yields a deterministic, reproducible pure state corresponding to a single realization of the generated flow field.
Varying the parameter vector $\boldsymbol{\xi}$ generates distinct realizations that preserve the prescribed multiscale statistics and geometric features.

The conditional-angle construction in Eqs.~\eqref{eq:exact_angle} to \eqref{eq:ridge_solution} maps this restricted input family to a circuit of depth $\Theta(n)$.
This reduction in circuit depth from $\Omega(2^n/n)$ to $\Omega(n)$ therefore does not imply an acceleration of exact loading for arbitrary classical data.
Rather, it arises from substituting pointwise instance fidelity with a physics-informed generative representation.
The resulting quantum state reproduces the prescribed turbulent statistics and coherent vortex geometry, rather than constituting a quantum copy of a pre-existing classical snapshot.
Table~\ref{tab:method_comparison} summarizes these distinctions among exact amplitude loading, tensor-network compression~\cite{Gourianov2022_A}, classical spectral synthesis~\cite{Kraichnan1970_Diffusion}, and the present method.

The present task should also be distinguished from earlier quantum-fluid representations and vortex-generation methods. Hydrodynamic Schrödinger and spinor formulations~\cite{Meng2023_Quantum, Meng2024_Quantum} specify how fluid variables are represented and evolved after an initial state has been supplied, whereas classical vortex-based synthesis produces an explicitly resolved classical field~\cite{Shen2024_Designing}. The present method instead maps a compact multiscale specification directly to a native turbulent quantum state intended for subsequent quantum evolution and targeted measurement.

We note that the analytic modal profile and compact phase parameters also define a classical generator without requiring full-grid materialization. 
Our comparison with explicit classical synthesis therefore concerns representation and initialization resources, rather than a speedup for an identical output task. 
Tensor-network methods provide an important complementary framework for representing and evolving velocity fields~\cite{Gourianov2022_A, Gourianov2025_Tensor, Esmaeili2026_Tensor}. We quantify the representation requirements of the present initial fields in Sec.~\ref{sec:velocity_mps}.
The preparation ansatz is used only to specify the initial state, and subsequent quantum evolution is not assumed to remain within the same ansatz family.

\begin{table*}[ht!]
	\centering
	\setlength{\tabcolsep}{10pt}
    \renewcommand{\arraystretch}{1.75}
	\begin{tblr}{
        width=\textwidth,
        colspec={
            Q[l,wd=2.25cm]
            Q[l,wd=2.85cm]
            X[l]
            Q[l,wd=2.55cm]
            Q[l,wd=2.75cm]
        },
        colsep=7pt,
        rowsep=6pt,
    }
		\hline\hline
		\textbf{Method}
		&
		\textbf{Classical input}
		&
		\textbf{Output}
		&
		\textbf{Main cost}
		&
		\textbf{Preserved or reproduced information}
		\\
        \hline
		Present method
		&
		Compact multiscale specification
		&
		Quantum state
		&
		$\Theta(n)$
		&
		Prescribed multiscale statistics and geometric features
		\\
		\hline
		Amplitude loading
		&
		Prescribed flow field
		&
		Quantum state
		&
		$\Omega(2^n/n)$
		&
		Exact prescribed flow field
		\\ \hline
		Tensor-network compression~\cite{Gourianov2022_A}
		&
		Prescribed flow field
		&
		Compressed classical tensor representation
		&
		Bond-dimension dependent
		&
		Approximate prescribed flow field
		\\ \hline
		Classical spectral synthesis~\cite{Kraichnan1970_Diffusion}
		&
		Compact multiscale specification
		&
		Classical random field
		&
		$\Omega(2^n)$
		&
		Prescribed spectral statistics
		\\ 
		\hline\hline
	\end{tblr}
    \caption{Comparison of flow-field encoding and generation tasks. The cost column refers to quantum circuit depth for state preparation, representation costs governed by the retained bond dimensions for tensor-network compression, and classical work required to materialize a complete field for spectral synthesis. These quantities describe different computational stages and output formats. The table does not establish a runtime ranking or an end-to-end quantum advantage. A numerical matrix-product-state benchmark for the present initial velocity fields is reported in Sec.~\ref{sec:velocity_mps}.}
    \label{tab:method_comparison}
\end{table*}    

\subsection{Expressive capacity and complexity of random-phase scrambling}
Following amplitude encoding, we apply the random-phase-scrambling circuit 
\begin{equation}
    U_p(\boldsymbol{\xi})
    =
    \prod_{i=1}^{n}
    \prod_{j=i+1}^{n}
    \mathrm{ZZ}^{(i,j)}(\gamma_{ij})
    \bigotimes_{j=1}^{n}
    R_z^{(j)}(\phi_j),
    \label{S-eq:phase_scrambling}
\end{equation}
where the single- and two-qubit phase gates are defined by $R_z^{(j)}(\phi_j) = \mathrm{e}^{-\mathrm{i}\phi_j Z_j/2}$ and $\mathrm{ZZ}^{(i,j)}(\gamma_{ij}) = \mathrm{e}^{-\mathrm{i}\gamma_{ij}Z_iZ_j/2}$, respectively.
Consider a computational basis state $|\boldsymbol{q}\rangle = |q_1\rangle\otimes\cdots\otimes|q_n\rangle$ with $\boldsymbol{q} = (q_1,\cdots,q_n)\in\{0,1\}^n$.
For each qubit, $z_j = (-1)^{q_j} \in\{1,-1\}$ denotes the Pauli-$Z$ eigenvalue associated with the basis value $q_j$, which satisfies $Z_j|\boldsymbol{q}\rangle = z_j|\boldsymbol{q}\rangle$.
Consequently, $z_j=1$ for $q_j=0$ and $z_j=-1$ for $q_j=1$.

Since all gates in $U_p(\boldsymbol{\xi})$ are diagonal in the computational basis, their joint action is given by
\begin{equation}
    U_p(\boldsymbol{\xi})|\boldsymbol{q}\rangle
    =
    \mathrm{e}^{\mathrm{i}
    \varPhi_{\boldsymbol{\xi}}(\boldsymbol{q})}
    |\boldsymbol{q}\rangle,
\end{equation}
where
\begin{equation}
    \varPhi_{\boldsymbol{\xi}}(\boldsymbol{q})
    =
    -\frac{1}{2}\sum_j\phi_j z_j
    -\frac{1}{2}\sum_{i<j}\gamma_{ij}z_i z_j.
\end{equation}
The variables $z_j$ correspond to the Pauli-$Z$ eigenvalues of the computational basis state, while the quadratic term introduces pairwise correlations between these binary variables.

The phase angles $\boldsymbol{\xi} = \bigl(\{\phi_j\},\{\gamma_{ij}\}_{i<j}\bigr)$ are randomly sampled.
Consequently, each sampled parameter set generates a randomized phase distribution across the computational basis.
However, the phases of the different basis states remain correlated because the same $O(n^2)$ gate parameters jointly determine all $2^n$ values of $\varPhi_{\boldsymbol{\xi}}(\boldsymbol{q})$.

The circuit comprises $n$ single-qubit $R_z$ gates and at most $n(n-1)/2$ two-qubit $\mathrm{ZZ}$ gates, yielding a total gate count of $O(n^2)$.
Under all-to-all connectivity, the edges of the complete interaction graph decompose into $n-1$ disjoint-pair layers for even $n$ and $n$ layers for odd $n$.
Consequently, the random-phase-scrambling circuit exhibits an $O(n)$ circuit depth.
For hardware platforms with restricted connectivity, the physical depth additionally incorporates routing overheads dictated by the coupling topology.

\subsection{Physical roles of amplitude encoding and phase encoding}
A smooth amplitude distribution alone does not fully determine the spatial geometry of turbulence.
The amplitude distribution specifies the modal occupation probabilities of the encoded spinor components. 
The decoded velocity spectrum, spatial localization, coherent vortices, and higher-order statistics depend jointly on these amplitudes, the phase correlations, the relative component population, and the nonlinear reconstruction.
The phase-scrambling layer is therefore an essential component of our generative construction.
Importantly, this layer does not load a rugged phase profile pointwise using $2^n$ independent values.
Instead, single-qubit $R_z$ rotations and pairwise $\mathrm{ZZ}$ interactions simultaneously generate correlated phases over the entire computational basis.
Specifically, this operation takes the form
\begin{equation}
    U_{\phi}|\boldsymbol{q}\rangle
    =
    \mathrm{e}^{\mathrm{i}\varPhi(\boldsymbol{q})}
    |\boldsymbol{q}\rangle,
    \quad
    \varPhi(\boldsymbol{q})
    =
    \phi_0
    +
    \sum_{\ell}a_{\ell}q_{\ell}
    +
    \sum_{\ell<m}a_{\ell m}q_{\ell}q_m,
\end{equation}
where $\boldsymbol{q}=(q_1,\cdots,q_n)\in\{0,1\}^n$ represents an $n$-bit computational-basis configuration, with $q_\ell$ denoting the binary state of the $\ell$th qubit and $|\boldsymbol{q}\rangle=\bigotimes_{\ell=1}^{n}|q_\ell\rangle$ being the corresponding basis state.
Thus, at most $O(n^2)$ circuit parameters implicitly generate the phases of all $2^n$ basis states.
Rather than representing an arbitrary $2^n$-parameter profile, these phases belong to a restricted family of correlated configurations.
Varying these phase parameters generates distinct physical realizations that all share the same prescribed amplitude scaling.
The rugged phase structure consequently emerges from the combination of the compact phase generator, the Gray-code mapping, and the nonlinear geometric mapping provided by the generalized Madelung transform and Hopf fibration, rather than being supplied as external classical data.

\begin{figure}[ht!]
    \centering
    \includegraphics{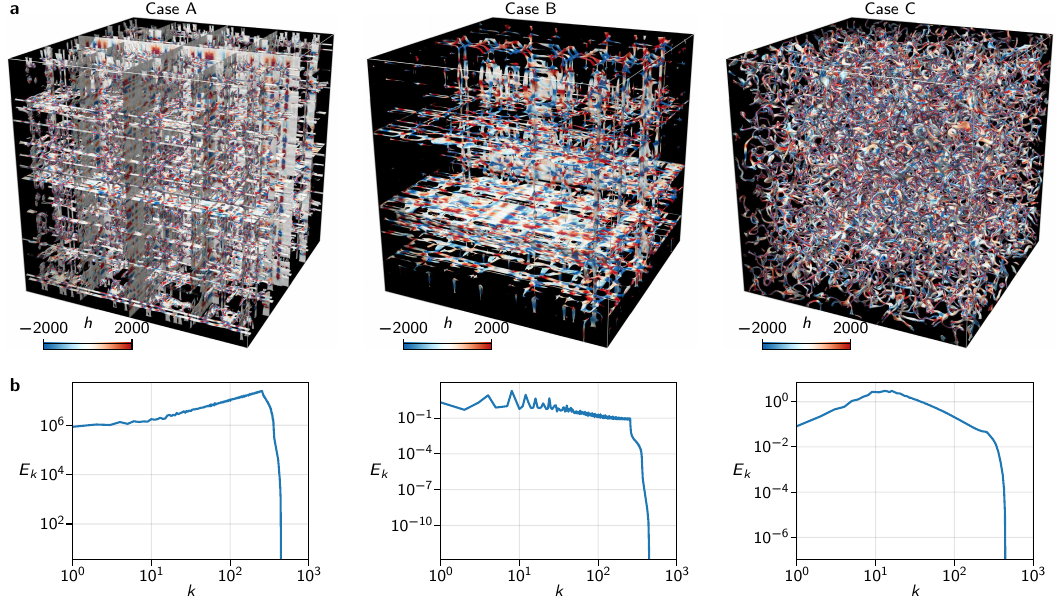}
    \caption{Ablation analysis of the phase-scrambling circuit for turbulent fields generated using $n=27$ qubits.
    \textbf{a}, Vorticity isosurface $|\vec{\omega}|=300$ color-coded by the helicity density.
    \textbf{b}, Corresponding kinetic-energy spectra $E_k(k)$.}
    \label{fig:ablation_phase}
\end{figure}

To quantify the respective contributions of the amplitude and phase layers, we perform an ablation study using the three configurations detailed in Tab.~\ref{tab:phase_ablation}.
All cases employ the same amplitude circuit, relative component population, Gray-code mapping, and physical decoding procedure, differing solely in their phase operations.
Because both $R_z$ and $\mathrm{ZZ}$ are diagonal phase operations, all three circuits preserve an identical Pauli-spinor amplitude distribution.
The differences observed in Fig.~\ref{fig:ablation_phase} therefore arise exclusively from the phase architecture.
Although these phase gates leave the input modal probabilities unaffected, they alter the decoded velocity field because the generalized Madelung transform maps phase gradients and correlations to momentum and vorticity.

\begin{table*}[ht!]
    \centering
    \setlength{\tabcolsep}{10pt}
    \renewcommand{\arraystretch}{1.35}
    \begin{tblr}{
        width=\textwidth,
        colspec={
            Q[l,wd=1cm]
            Q[l,wd=2.4cm]
            Q[l,wd=2.4cm]
            Q[l,wd=1.9cm]
            X[l]
        },
        colsep=7pt,
        rowsep=6pt,
    }
        \hline\hline
        \textbf{Case}
        &
        \textbf{Amplitude circuit}
        &
        \textbf{Single-qubit $R_z$}
        &
        \textbf{Two-qubit $\mathrm{ZZ}$}
        &
        \textbf{Physical question examined}
        \\
        \hline
        A
        &
        Identical
        &
        Absent
        &
        Absent
        &
        Whether the prescribed amplitude scaling and geometric decoding
        alone determine the decoded velocity spectrum and vortex geometry
        \\
        \hline
        B
        &
        Identical
        &
        Present
        &
        Absent
        &
        Whether separable single-qubit phases suppress coherent spectral
        artifacts and promote spatial localization of vorticity
        \\
        \hline
        C
        &
        Identical
        &
        Present
        &
        Present
        &
        Whether nonseparable two-qubit phase correlations produce a
        broadband spectral decay and an organized network of tangled
        vortex tubes
        \\
        \hline\hline
    \end{tblr}
    \caption{Definition of the phase-circuit ablation cases.
    All three cases utilize the same Pauli-spinor amplitude distribution and differ solely in their phase operations.}
    \label{tab:phase_ablation}
\end{table*}

In case A, the absence of phase modulation results in a strong, systematic coherence among the spectral modes.
The resulting velocity spectrum rises toward higher wavenumbers and exhibits a pronounced accumulation of energy near the spectral cutoff, rather than the expected descending broadband range.
Correspondingly, the vorticity isosurfaces are dominated by densely overlapping, extended sheet-like structures, with individual vortex filaments embedded in a largely space-filling background in Fig.~\ref{fig:ablation_phase}a.
Thus, the prescribed amplitude scaling and geometric mapping alone are insufficient to recover the target turbulent spectrum or a well-separated network of vortex tubes.

Introducing single-qubit $R_z$ rotations in case B partially disrupts this coherent modal alignment.
Consequently, the high-wavenumber energy accumulation is substantially reduced, and the vorticity field becomes more localized and filamentary.
Nevertheless, the velocity spectrum still retains pronounced fluctuations and exhibits a relatively shallow decay, while extended sheet-like structures remain visible among the vortical filaments.
Thus, single-qubit phase scrambling introduces spatial complexity but falls short of fully organizing the decoded field across scales.

The complete circuit in case C incorporates nonseparable phase dependence through two-qubit $\mathrm{ZZ}$ interactions.
The corresponding velocity spectrum exhibits a clear peak at large scales, followed by a smoother and broader descending range that shows closer agreement with the Kolmogorov five-thirds scaling law.
Concurrently, the extended sheet-like background is significantly suppressed, with the intense-vorticity regions organizing into a dense network of slender, clearly separated, and tangled vortex tubes.
This progression from case A to case C demonstrates that the characteristic turbulent structures do not arise from the smooth amplitude distribution alone.
While local phase rotations break the initial modal coherence, two-qubit phase correlations are necessary within this circuit architecture to simultaneously recover both the broadband spectral decay and the coherent vortex-tube geometry.
These findings further demonstrate that the phase-scrambling circuit does not merely impose unstructured random phases, but instead generates a compact, correlated phase landscape that structures the velocity spectrum and vortex geometry obtained after convolution and deconvolution.

\subsection{Relative component population}
\begin{figure}[ht!]
    \centering
    \includegraphics{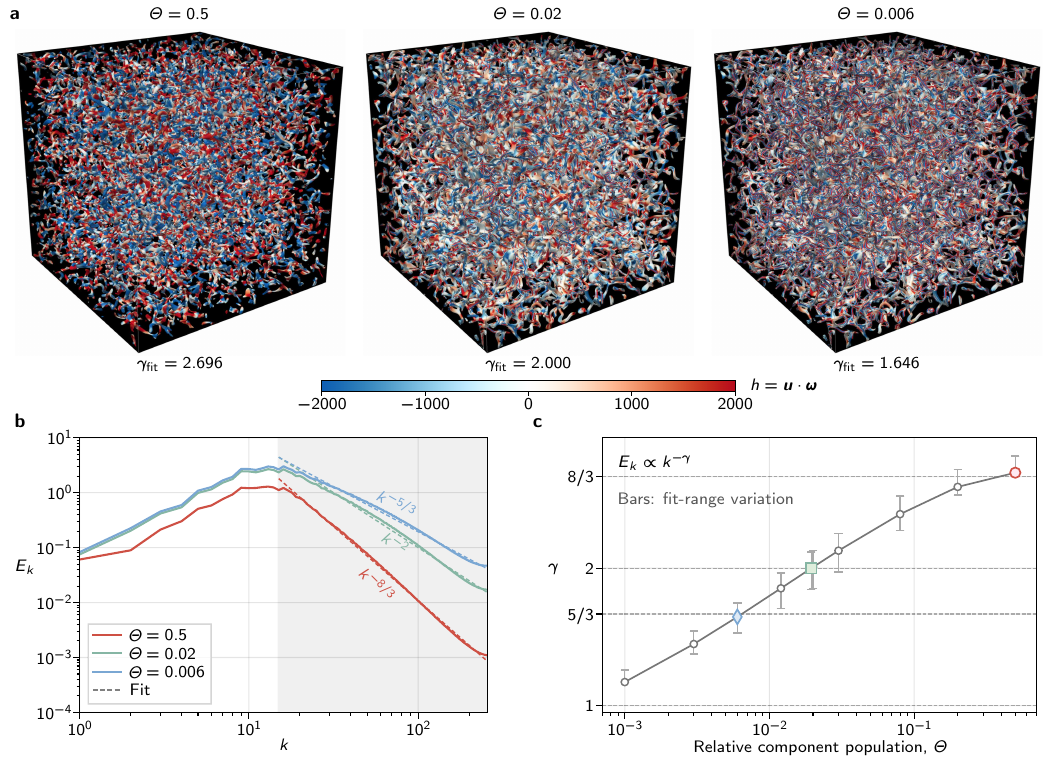}
    \caption{Control of velocity-spectrum scaling through the relative component population.
    All fields are generated with $n=27$ qubits on a $512^3$ grid. The input amplitude exponent $\beta=0$, cutoff $k_d=15$, and all other parameters, including the phase parameters of both components, remain fixed while $\varTheta$ varies.
    \textbf{a}, Vorticity isosurfaces $|\boldsymbol{\omega}|=300$ for $\varTheta=0.5$, $0.02$, and $0.006$, respectively, colored by the helicity density.
    \textbf{b}, Corresponding velocity energy spectra. Dashed curves show least-squares fits of $\log E_k$ against $\log k$ over $15\leq k\leq256$, indicated by the gray background. The fitted exponents in $E_k\propto k^{-\gamma}$ are $\gamma=2.696$, $2.000$, and $1.646$, respectively.
    \textbf{c}, Dependence of the fitted exponent on $\varTheta$ across ten parameter values. Colored symbols identify the cases in \textbf{a}. Gray bars span the exponents obtained over the fitting intervals $[15,256]$, $[15,128]$, $[30,256]$, $[30,200]$, and $[20,200]$, quantifying sensitivity to the fitting interval. Decreasing $\varTheta$ produces progressively shallower spectra within this fixed-phase family.}
    \label{fig:chi_dependence_n27}
\end{figure}

Each component register encodes the normalized profile $\hat{\psi}_s/\|\hat{\psi}_s\|_2$, so that $\hat{\psi}_s(\vec{k}(\vec{j})) = \norm{\hat{\psi}_s}_2 \langle \vec{j}|\hat{\varPsi}_s\rangle$. 
This component norms specify the relative populations in the physical Pauli spinor. 
The population fraction of the first component is
\begin{equation}
    \varTheta=\frac{\|\hat{\psi}_+\|_2^2}{\|\hat{\psi}_+\|_2^2 + \|\hat{\psi}_-\|_2^2}.
\end{equation}
Since a common normalization factor cancels from both the velocity and normalized spin fields, the implementation fixes the total squared component norm to 2, giving $\|\hat{\psi}_+\|_2^2=2\varTheta$ and $\|\hat{\psi}_-\|_2^2=2(1-\varTheta)$. 
At fixed normalized component profiles, varying $\varTheta$ rescales the two physical components without changing their individual spectral shapes.
The population scan in Fig.~\ref{fig:chi_dependence_n27} holds these normalized profiles and their phase parameters fixed.
Changing $\varTheta$ modifies both $\boldsymbol{J}$ and $\rho$, so its effect on $\boldsymbol{u}$ depends on the spatial structures of the two components.

To express this dependence in terms of the spin vector, write $\psi_\pm=|\psi_\pm|\mathrm{e}^{\mathrm{i}\theta_\pm}$, where $\theta_\pm$ denote the local component phases.
Using $s_3 = (|\psi_+|^2 - |\psi_-|^2)/\rho$, the velocity takes the form
\begin{equation}\label{S-eq:vel_theta}
    \boldsymbol{u} = \frac12\boldsymbol{\nabla}(\theta_++\theta_-)
    +\frac{s_3}{2}\boldsymbol{\nabla}(\theta_+-\theta_-),
\end{equation}
where both components are nonzero and their phases are locally smooth. 
Taking the curl of Eq.~\eqref{S-eq:vel_theta} yields
\begin{equation}
    \boldsymbol{\omega} =\frac12\boldsymbol{\nabla}s_3 \times\boldsymbol{\nabla}(\theta_+-\theta_-).
\end{equation}
At fixed normalized component profiles, varying $\varTheta$ leaves the phase gradients unchanged but modifies $s_3$ and its spatial gradient. 
The population thus controls the local weighting of phase gradients in the reconstructed velocity and the resulting vorticity distribution. 
This provides a mechanism for modifying the decoded flow structure and spectral shape while keeping the input amplitude exponent fixed. 
These phase-gradient identities describe the local continuum transform. 

A local estimate clarifies how the weaker component modifies a low-density region. 
Consider $\varTheta\ll1$, so that $\psi_-$ carries most of the total population. 
Suppose that $\psi_-$ vanishes linearly away from a simple zero line, while $\psi_+$ remains nonzero on that line. 
Along a fixed direction normal to the line, let $R$ denote the distance from it. 
For the normalized component profiles, the leading local approximations are $|\psi_-(R)|^2/\norm{\hat{\psi}_-}_2^2\simeq CR^2$ and $|\psi_+(R)|^2/\norm{\hat{\psi}_+}_2^2\simeq B >0$.
Here, the numerators are real-space component densities along the chosen transverse direction, $B$ is the normalized density of the first component on the zero line, and $C>0$ describes the quadratic increase of the normalized second-component density along the chosen direction. 
Using $\norm{\hat{\psi}_+}_2^2=2\varTheta$ and $\norm{\hat{\psi}_-}_2^2=2(1-\varTheta)$, the total density is then
\begin{equation}
    \rho(R)\simeq2\varTheta B+2(1-\varTheta)CR^2.
\end{equation}
The first component therefore supplies a nonzero density $\rho(0)\simeq2\varTheta B$ at the zero of the second component. 
The constant and quadratic contributions become comparable at the transverse transition scale
\begin{equation}
    \ell_\varTheta \sim \left[\frac{\varTheta B}{(1-\varTheta)C}\right]^{1/2}.
\end{equation} 
At fixed normalized component profiles, $B$ and $C$ remain fixed, so varying $\varTheta$ changes both the density at the zero line and the width of this local transition. 
Within the same local approximation, the spin component satisfies
\begin{equation}
    s_3(R) \simeq \frac{\varTheta B - (1-\varTheta)CR^2}{\varTheta B + (1-\varTheta)CR^2}.
\end{equation}
Its transverse gradient scales as $|\p_R s_3| \sim \ell_\varTheta^{-1}$ near $R\sim\ell_\varTheta$, linking the local density balance to the vorticity identity above.
Through $\boldsymbol{u}=\boldsymbol{J}/\rho$, this modifies the reconstruction near the low-density core and its associated velocity gradients. The estimate applies only where the local expansions remain valid. 
The coefficient $C$ generally depends on position along the zero line and on the transverse direction, so $\ell_\varTheta$ characterizes a local transition rather than a universal vortex-core radius.
Figure~\ref{fig:chi_dependence_n27} quantifies this population dependence at $n=27$ on a $512^3$ grid. 

The input exponent $\beta$ and the population $\varTheta$ therefore play distinct roles. 
The former specifies the modal amplitude profile of each component, whereas the latter controls their nonlinear combination in the decoded velocity. 
Choosing $\beta=0$ provides a baseline without additional algebraic decay in the nonzero-mode amplitudes, while retaining the spectral cutoff and circuit-generated phase correlations. 
The resulting velocity-spectrum exponent $\gamma$ is an output statistic that depends on these ingredients.
For different qubit counts $n$, we keep $\beta=0$ and $k_d=15$ fixed and select $\varTheta$ separately to obtain a $k^{-5/3}$ velocity spectra, as listed in Tab.~\ref{tab:n-chi}.
For $n = 21$, 24, 27, and 30, the selected populations approximately follow a geometric sequence with a common ratio of about 0.3, $\varTheta(n+3)/\varTheta(n)\approx 0.3$.
Using the population at $n = 30$ as a reference gives the approximate continuation $\varTheta(n) \approx 0.002\times 0.3^{(n-30)/3}$ for larger $n$.

\begin{table}[ht!]
    \centering
    \setlength{\tabcolsep}{10pt}
    \renewcommand{\arraystretch}{1.35}
    \begin{tabular}{ccccccc}
        \hline\hline
        Qubit count $n$ & 15 & 18 & 21 & 24 & 27 & 30 \\ \hline
        Relative component population $\varTheta$ & 0.5 & 0.5 & 0.08 & 0.02 & 0.006 & 0.002 \\
        \hline\hline
    \end{tabular}
    \caption{Relative component populations $\varTheta$ selected at different qubit counts $n$ to obtain a $k^{-5/3}$ energy spectrum of the decoded velocity field.}
    \label{tab:n-chi}
\end{table}

\subsection{Statistical robustness}
To assess the robustness of the generated initial fields, we sample 10 independent pairs of phase vectors $(\boldsymbol{\xi}_{+},\boldsymbol{\xi}_{-})$ at each qubit count $n=21$, 24, 27, and 30, corresponding to resolutions of $N^3=128^3$, $256^3$, $512^3$, and $1024^3$, respectively.
For reproducibility, the random-seed pairs used to generate $(\boldsymbol{\xi}_{+},\boldsymbol{\xi}_{-})$ in the ten realizations are $(2025,2026)$, $(2027,2028)$, $(2029,2030)$, $(2031,2032)$, $(2033,2034)$, $(2035,2036)$, $(2037,2038)$, $(2039,2040)$, $(2041,2042)$, and $(2043,2044)$.
At each resolution, we keep the amplitude-encoding circuit parameters $\beta=0$, $k_d=15$, and $P_0=0.1$, as well as the component populations specified in Tab.~\ref{tab:n-chi}, fixed across all phase realizations.
Neither the state-preparation parameters nor the fitting intervals are retuned for individual seeds.

For each realization, the velocity-spectrum slope is extracted by ordinary least-squares regression in logarithmic coordinates over $15\leq k\leq N/2$.
The fitting intervals are therefore $[15,64]$, $[15,128]$, $[15,256]$, and $[15,512]$ for $n=21$, 24, 27, and 30, respectively.
These intervals are fixed across phase realizations at each resolution.

\begin{figure}[ht!]
    \centering
    \includegraphics{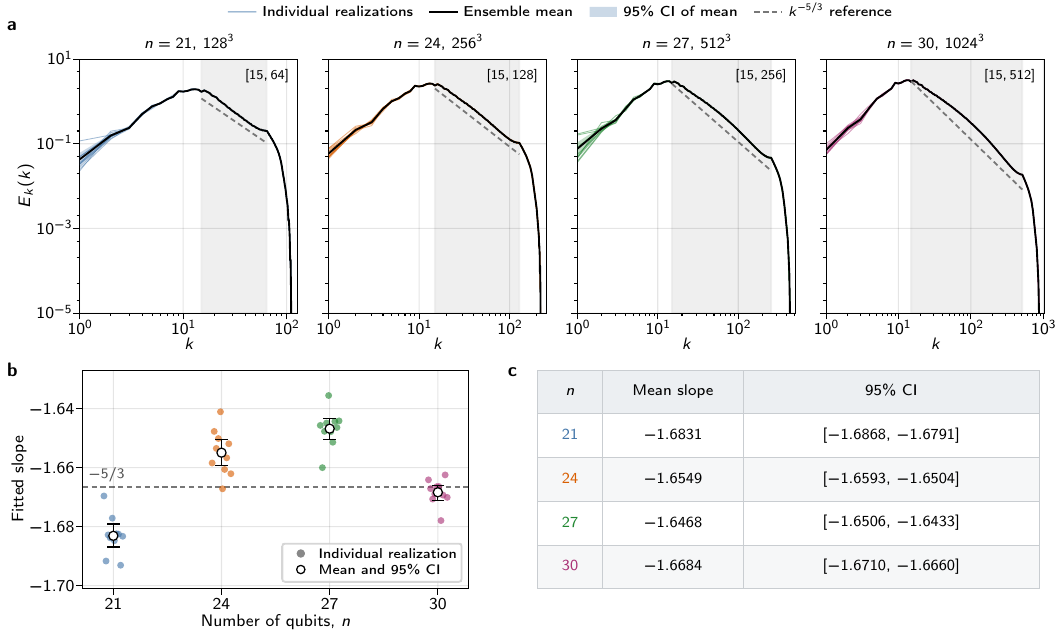}
    \caption{Robustness of the velocity spectra across independent phase realizations and grid resolutions.
    \textbf{a}, Velocity energy spectra for $n=21$, 24, 27, and 30.
    Thin colored lines show the ten individual realizations, black lines show their arithmetic ensemble means, and colored bands indicate pointwise 95\% confidence intervals of the means.
    Gray regions indicate the fitting intervals $[15,N/2]$, and dashed lines show $k^{-5/3}$ references with arbitrary vertical offsets.
    \textbf{b}, Individual fitted slopes and their ensemble means with 95\% confidence intervals, with the dashed line marking $-5/3$.
    \textbf{c}, Numerical summary of the mean slopes and their confidence intervals.
    All confidence intervals use the percentile bootstrap with 50000 resamples of the ten realizations at each resolution.
    The spectral confidence bands are pointwise rather than simultaneous.}
    \label{fig:velocity_spectrum_robustness}
\end{figure}

Figure~\ref{fig:velocity_spectrum_robustness}a presents the ten individual velocity spectra at each resolution, their arithmetic ensemble means, and pointwise 95\% confidence intervals obtained from 50000 bootstrap resamples of the independent phase realizations.
Figures~\ref{fig:velocity_spectrum_robustness}b and c summarize the individual fitted slopes and their ensemble means with 95\% bootstrap confidence intervals.
The mean fitted slopes are $-1.6831$, $-1.6549$, $-1.6468$, and $-1.6684$ for $n=21$, 24, 27, and 30, respectively.
These confidence intervals quantify uncertainty in the ensemble means arising from phase-to-phase variation, rather than the regression uncertainty of a single spectrum.
The mean slopes remain close to $-5/3$ across the tested resolutions, with small resolution-dependent departures.

\subsection{Algorithm summary and division of computational tasks}
The complete workflow is summarized as follows.

\begin{enumerate}
    \item Classical specification. Set qubit count $n$, the fixed target parameters $\beta=0$, $k_d=15$, and $P_0=0.1$, and the prescribed population $\varTheta(n)$. Sample the gate-angle vectors $\vec{\xi}_+$ and $\vec{\xi}_-$ for the two components separately. The component targets determine amplitude compilation, while the sampled phase vectors and physical component weights specify the realization.
    \item Sparse classical compilation. Construct the effective support $\mathcal{S}_\epsilon$, convert the retained modes to Gray-code strings, and accumulate their prefix probabilities in a sparse tree. Evaluate $\theta_j^{\mathrm{target}}(\boldsymbol{c}_j)$ only at occupied prefix nodes.
    \item Parameter regression. Apply probability-weighted ridge regression to obtain $b_j$ and $w_{j,m}$. For fixed $M_\epsilon$ and $r$, the classical time and memory costs scale linearly with $n$.
    \item Quantum state preparation. Apply the $R_y$ and controlled-$R_y$ amplitude circuit, followed by the $R_z$ and ZZ phase circuit and the exact Gray-to-binary transformation. This step produces $|\hat{\varPsi}_s\rangle$, which encodes one physical Pauli-spinor component in an $n$-qubit data register.
    \item Observable extraction. Estimate selected Fourier coefficients through the operators $\hat{\mathcal{O}}_{\mathcal{Q},s}$. Full-field reconstruction and deconvolution are used in the present numerical study for validation; subsequent Hamiltonian evolution instead accepts the prepared spinor state directly as its initial condition.
\end{enumerate}

Steps 1–3 are classical, step 4 is quantum state preparation, and step 5 combines quantum measurement with observable-dependent classical post-processing.

\section{MPS representation of the quantum encoded velocity fields}\label{sec:velocity_mps}
Tensor-network methods provide a structured framework for representing and evolving fluid fields~\cite{Gourianov2022_A, Gourianov2025_Tensor, Esmaeili2026_Tensor}. 
We use the scale-ordered matrix product state (MPS) encoding of that work to examine the representation requirements of the geometric quantum encoded velocity fields generated here. 
The benchmark includes Schmidt spectra, verified reconstruction errors, parameter counts, and storage costs. 
We also report the corresponding generating-circuit resources, with their computational scope specified separately.

We consider $N^3$ grids with $N=32,64,128,256,512$, corresponding to $n=15,18,21,24,27$ qubits. 
Each grid uses one realization, with phase seeds $2025$ and $2026$ for the two spinor components. 
We retain $\beta=0$, $k_d=15$, and $P_0=0.1$. 
Write $N=2^L$, so that $n=3L$. At each scale, the coordinate bits form an octal physical index $x_\ell+2y_\ell+4z_\ell$ giving $L$ sites of physical dimension $8$, ordered from coarse to fine. 
The unfolding at cut $\ell$ has dimensions $8^\ell\times8^{L-\ell}$, with rank bounds $\varGamma_\ell=\min(8^\ell,8^{L-\ell})$ and $\chi_{\max}=8^{\lfloor L/2\rfloor}$ which depend on the tensorization. The staircase in $\chi_{\max}$ therefore reflects the parity of $L$.

\subsection{Error control and MPS construction}
For each velocity component $a\in\{x,y,z\}$, we define
\begin{equation}
    \mathcal{E}_a
    =\frac{\|u_a-\tilde{u}_a\|_2}{\|u_a\|_2},
    \quad
    \mathcal{E}_{\max}=\max_a\mathcal{E}_a,
    \label{eq:mps_error}
\end{equation}
where $\tilde{u}_a$ denotes the velocity component reconstructed from its truncated MPS representation.
We require $\mathcal{E}_{\max}\leq\epsilon$ at $\epsilon=10^{-2}$, $10^{-3}$, and $10^{-4}$. 
Here $\epsilon$ is a velocity-reconstruction tolerance, distinct from the probability-tail tolerance used in sparse circuit compilation. A relative $L^2$ error of $10^{-2}$ corresponds to a relative squared-error budget of $10^{-4}$.

Let $\sigma_{\ell j}^{(a)}$ denote the singular values of the original component unfolding at cut $\ell$. These spectra are computed before MPS truncation. 
The normalized weights and Schmidt entropy are
\begin{equation}
    p_{\ell j}^{(a)}
    =\frac{(\sigma_{\ell j}^{(a)})^2}{\|u_a\|_2^2},
    \quad
    S_\ell^{(a)}
    =-\sum_jp_{\ell j}^{(a)}\log_2p_{\ell j}^{(a)}.
    \label{eq:mps_entropy}
\end{equation}
The entropy characterizes scale separability of the normalized velocity tensor.

The necessary rank at a single cut is
\begin{equation}
    r_{\ell a}(\epsilon)
    =\min\Big\{r:
    \sum_{j>r}(\sigma_{\ell j}^{(a)})^2
    \leq\epsilon^2\|u_a\|_2^2
    \Big\}.
\end{equation}
Consequently, $\chi_{\mathrm{lb}}(\epsilon)=\max_{\ell,a}r_{\ell a}(\epsilon)$ is a necessary lower bound on the maximum bond dimension of any MPS satisfying Eq.~\eqref{eq:mps_error}. 
For a prescribed bond dimension $\chi$, the error lower bound obeys
\begin{equation}
    \mathcal{E}_{\mathrm{lb}}(\chi)
    =\max_{\ell,a}
    \left[
    \frac{\sum_{j>\chi}(\sigma_{\ell j}^{(a)})^2}
    {\|u_a\|_2^2}
    \right]^{1/2},
    \label{eq:mps_error_lower_bound}
\end{equation}
following from optimal low-rank matrix approximation at each cut.

For each component, we select the smallest uniform cap satisfying
\begin{equation}
    \sum_{\ell=1}^{L-1}\sum_{j>\chi}
    (\sigma_{\ell j}^{(a)})^2
    \leq\epsilon^2\|u_a\|_2^2.
    \label{eq:mps_sufficient_budget}
\end{equation}
We then perform fixed-cap TT-SVD in the original coarse-to-fine order and fully reconstruct the field to verify its actual error. 
Equation~\eqref{eq:mps_sufficient_budget} is sufficient and generally more restrictive than the individual-cut conditions. 
We therefore distinguish the necessary bound $\chi_{\mathrm{lb}}$ from the verified construction rank $\chi_{\mathrm{built}}$. 
The construction does not establish a globally minimal parameter count.

\subsection{Representation resources}
Let $r_\ell^{(a)}$ denote the retained bond dimensions, with $r_0^{(a)}=r_L^{(a)}=1$. 
The stored scalar count is
\begin{equation}
    P_{\mathrm{store}}
    =\sum_a\sum_{\ell=1}^{L}
    8r_{\ell-1}^{(a)}r_\ell^{(a)}.
    \label{eq:mps_storage}
\end{equation}
Following the parameter-count convention in Ref.~\cite{Gourianov2022_A}, removing internal gauge freedom gives
\begin{equation}
    P_{\mathrm{NVPS}}
    =P_{\mathrm{store}}
    -\sum_a\sum_{\ell=1}^{L-1}(r_\ell^{(a)})^2.
    \label{eq:mps_nvps}
\end{equation}
The dense reference contains $3N^3$ real scalars. 
All velocity arrays and MPS cores use double precision, so actual core storage is $8P_{\mathrm{store}}$ bytes. 
Gauge-reduced parameter count and stored core bytes are therefore separate resource measures.

\begin{figure}[ht!]
    \centering
    \includegraphics{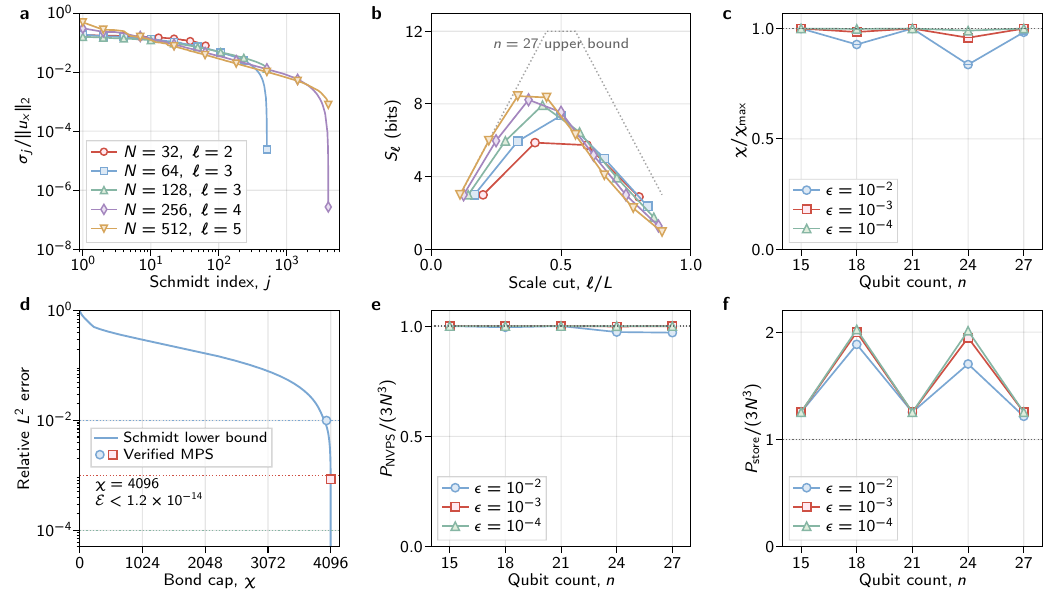}
    \caption{Matrix product state (MPS) representations of geometrically generated velocity fields from a single realization on grids of size $N^3$, with $n=3\log_2 N$ spatial qubits.
    \textbf{a}, Normalized Schmidt spectra of $u_x$ at the cuts requiring the largest single-cut rank for a relative $L^2$ error of $10^{-2}$.
    \textbf{b}, Schmidt entropy $S_\ell$ at the relative scale cut $\ell/L$, where $L=\log_2N$. The gray dotted line gives the entropy bound for $n=27$.
    \textbf{c}, MPS bond dimensions normalized by the octal-encoding bound $\chi_{\max}=8^{\lfloor L/2\rfloor}$.
    Downward bars extend to the necessary ranks inferred from the Schmidt spectra.
    \textbf{d}, Schmidt lower bound on the maximum componentwise relative $L^2$ error for the $512^3$ grid as a function of the bond cap $\chi$.
    Symbols denote measured MPS reconstruction errors. Colored horizontal lines mark the target errors.
    \textbf{e}--\textbf{f}, Gauge-reduced parameter counts and stored MPS scalars, respectively, normalized by the dense velocity representation with $3N^3$ real scalars.}
    \label{fig:mps}
\end{figure}

Figure~\ref{fig:mps} summarizes the spectra, entropies, rank requirements, reconstruction errors, and representation resources. 
At $\epsilon=10^{-2}$, the $512^3$ field requires $\chi_{\mathrm{lb}}=3994$, while a verified construction with $\chi_{\mathrm{built}}=4027$ satisfies the tolerance. 
Both values are close to the encoding bound $\chi_{\max}=4096$. 
A bond cap of $512$ implies an error of at least $40.17\%$ in at least one component.

Across the five grids at this tolerance, the constructed gauge-reduced parameter count is $97.03\%$--$100\%$ of the dense count. 
Stored cores require approximately $1.216$--$1.886$ times the dense-array memory. 
For $512^3$, the MPS cores occupy $3734.96$ MiB, whereas the three dense velocity arrays occupy $3072$ MiB. Table~\ref{tab:velocity_mps_resources} reports the detailed counts. 
These results indicate limited compression for the present initial fields within the selected encoding.

\begin{table*}[!ht]
    \centering
    \setlength{\tabcolsep}{10pt}
    \renewcommand{\arraystretch}{1.35}
    \begin{tabular}{ccccccc}
    \hline\hline
    $n$ & $N$ & $\chi_{\max}$ & $\chi_{\mathrm{lb}}$ & $\chi_{\mathrm{built}}$ & $P_{\mathrm{NVPS}}/(3N^3)$ & Storage (MiB) \\
    \hline
    15 & 32 & 64 & 64 & 64 & 1.0000 & 0.94 \\
    18 & 64 & 512 & 475 & 475 & 0.9947 & 11.32 \\
    21 & 128 & 512 & 512 & 512 & 1.0000 & 60.19 \\
    24 & 256 & 4096 & 3427 & 3427 & 0.9730 & 654.07 \\
    27 & 512 & 4096 & 3994 & 4027 & 0.9703 & 3734.96 \\
    \hline\hline
    \end{tabular}
    \caption{Resources of the velocity MPS at a componentwise relative $L^2$ error tolerance $\epsilon=10^{-2}$.
    The octal scale encoding has $L=\log_2N$ sites and the rank bound $\chi_{\max}=8^{\lfloor L/2\rfloor}$.
    The necessary bound $\chi_{\mathrm{lb}}$ follows from the Schmidt spectra, whereas $\chi_{\mathrm{built}}$ is attained by an explicitly constructed MPS with verified reconstruction error.
    Both ranks are maxima over all cuts and velocity components.
    $P_{\mathrm{NVPS}}$ excludes internal gauge freedom.
    Parameter counts and storage sum the three velocity components.
    Storage uses double precision, with $1\,\mathrm{MiB}=2^{20}$ bytes.
    Each row represents one realization with the manuscript parameters.}
    \label{tab:velocity_mps_resources}
\end{table*}

\subsection{Generating-circuit resources}
Table~\ref{tab:spinor_circuit_resources} reports the corresponding generating-circuit resources from $n=15$ to $n=30$. 
The description count includes shared active amplitude angles counted once, independent one-body and pairwise phase angles for both components, and one component population. 
If $A_{\mathrm{act}}$ is the number of shared active amplitude angles, the total count is $P_{\mathrm{gen}}=A_{\mathrm{act}}+n(n+1)+1$. At $n=30$, this gives $51+930+1=982$ parameters.

\begin{table}[ht!]
    \centering
    \setlength{\tabcolsep}{10pt}
    \renewcommand{\arraystretch}{1.35}
    \begin{tabular}{cccccc}
    \hline\hline
    Qubits & Parameters & \shortstack{CX gates} & \shortstack{$U$ gates} & \shortstack{Maximum depth} \\
    \hline
    15 & 295 & 572 & 400 & 174 \\
    18 & 405 & 784 & 522 & 207 \\
    21 & 520 & 992 & 622 & 223 \\
    24 & 655 & 1244 & 748 & 229 \\
    27 & 808 & 1532 & 896 & 176 \\
    30 & 982 & 1868 & 1070 & 194 \\
    \hline\hline
    \end{tabular}
    \caption{Resources of the geometric quantum encoding circuits.
    The parameter count includes shared amplitude angles, independent phase angles for both spinor components, and one population fraction.
    Gate counts are summed over the two circuits, and the reported depth is their maximum.
    Transpilation uses the $\{U,\mathrm{CX}\}$ basis at optimization level 3 without a hardware coupling constraint.}
    \label{tab:spinor_circuit_resources}
\end{table}

Gate counts are summed over the two component circuits, and the depth is their maximum. 
Transpilation uses Qiskit 2.1.1, the $\{U,\mathrm{CX}\}$ basis, optimization level $3$, seed $2025$, and no hardware coupling constraint. 
These counts cover preparation before Gray decoding and Fourier transforms, while exclude velocity evaluation, measurement, time evolution, routing, and fault-tolerant synthesis.

\subsection{Verification and interpretation}
The numerical implementation follows the public MATLAB scale encoding and fixed-cap construction, using a vectorized Python adapter for the large-grid calculations~\cite{Gourianov2022_A}. 
Production spectra use direct SVD or the smaller Gram eigenproblem. 
An independent audit recomputes all $90$ cut spectra by direct SVD of physical-coordinate unfoldings and reconstructs all $45$ saved MPS approximations using an independent coordinate mapping.

The independent audit gives a maximum normalized spectrum difference of approximately $9.60\times10^{-14}$ and a maximum difference of approximately $3.47\times10^{-17}$ between the reconstruction errors measured by the two procedures.
Regenerated 15- and 27-qubit velocity fields match the saved reference arrays pointwise.
Small-grid tests of the original MATLAB construction, analysis, and reconstruction routines also agree with the Python adapter, including the joint-component error convention.
The resulting compression estimates concern one realization at each resolution, the stated parameter family, and the selected scale ordering.
The MPS implementation is available in the GitHub repository~\cite{code}.

These initial-field representation costs provide a baseline for assessing subsequent MPS evolution, whose efficiency also depends on rank growth and truncation during time integration.
MPS-based solvers offer substantial flexibility for nonlinear and dissipative dynamics without requiring a unitary gate decomposition.
Their compatibility with established numerical methods and present-day classical hardware makes them particularly promising for near-term fluid simulation.
A matched dynamical comparison should assess accumulated error, computational cost, and the extraction of the same physical observables at the same simulation time.
Such a comparison lies beyond the present static representation benchmark.

\section{Measurement of observables in spectral space}
\subsection{Derivation of measurement operators}
The convolution in Eq.~\eqref{S-eq:Q} is quadratic in $\hat{\psi}_s$, thereby defining a linear measurement operator acting on the density matrix $\varrho$. 
We derive the measurement operator for Eq.~\eqref{S-eq:Q}.
For the density $\rho$ and momentum $\vec{J}$, this operator reduces to $\hat{\mathcal{Q}}(\vec{k}) = \sum_{s=\pm} \hat{\mathcal{Q}}_s$, with the component
\begin{equation}\label{S-eq:Q_sim}
    \hat{\mathcal{Q}}_s(\vec{k})
    = \sum_{\vec{k}'\in\mathbb{Z}^d} \hat{c}_{s,s}(\vec{k}, \vec{k}') \hat{\psi}_s(\vec{k}'+\vec{k}) \hat{\psi}_{s}^*(\vec{k}').
\end{equation}
For density and momentum, the kernel is diagonal in the spinor indices.
We measure the normalized component-register expectations separately and restore the physical factors $\|\hat\psi_s\|_2^2$ before summing their contributions.
In contrast, the transverse spin observables $\rho s_1$ and $\rho s_2$ contain $s\neq s'$ contributions and require coherent interference between $\hat\psi_+$ and $\hat\psi_-$. 
Accessing them requires preparing a joint spinor state $|\hat{\boldsymbol{\varPsi}}\rangle = \sqrt{\varTheta}\ket{+}\ket{\hat{\varPsi}_+} + \sqrt{1-\varTheta}\ket{-}\ket{\hat{\varPsi}_-}$~\cite{Meng2023_Quantum} and measuring the spin-label qubit in the Pauli-$X$ or Pauli-$Y$ basis together with the corresponding spatial operator. These cross-component observables are not included in the present measurement-resource analysis.
Physical quadratic observables obtained from this normalized joint state include the total squared component norm, which equals 2 under our convention.

The finite grid size restricts the wavenumber $k_\alpha$ in each dimension $\alpha$ to the interval $[-2^{n_\alpha-1}, 2^{n_\alpha-1}-1]$, which truncates the summation in Eq.~\eqref{S-eq:Q_sim}.
Consequently, the summation domain is reduced to the set
\begin{equation}\label{S-eq:set_K}
    \mathbb{K}(k_\alpha)=\left\{k_\alpha' \mid k_\alpha'\in\mathbb{Z} \cap \left[-2^{n_\alpha-1}, 2^{n_\alpha-1}-1\right] \cap \left[-2^{n_\alpha-1} - k_\alpha, 2^{n_\alpha-1}-1-k_\alpha \right] \right\}.
\end{equation}
By taking the sum over the Cartesian product of the admissible per-axis wavenumber sets $\mathcal K(\boldsymbol{k}):=\mathbb{K}(k_0)\times \mathbb{K}(k_1)\times\cdots\times \mathbb{K}(k_{d-1})$, Eq.~\eqref{S-eq:Q_sim} is accordingly modified as
\begin{equation}\label{S-eq:Qs_sim}
    \hat{\mathcal{Q}}_s(\vec{k})
    = \sum_{\vec{k}'\in\mathcal{K}(\vec{k})} \hat{c}_{s,s}(\vec{k}, \vec{k}') \hat{\psi}_s(\vec{k}+\vec{k}') \hat{\psi}_s^*(\vec{k}').
\end{equation}

To derive the measurement operator, we re-express this summation in terms of the state vector index $j_\alpha$.
For each dimension $\alpha$, the indices $j_\alpha$ and $k_\alpha$ are related through the bijection $k_\alpha = \mathrm{mod}(j_\alpha+2^{n_\alpha-1}, 2^{n_\alpha}) - 2^{n_\alpha-1}$ and its inverse $j_\alpha = k_\alpha + 2^{n_\alpha} [1 - H(k_\alpha)]$, where $\mathrm{mod}(\cdot, \cdot)$ is the modulo operation and $H(\cdot)$ is the Heaviside step function.
Under this transformation, the summation set $\mathbb{K}(k_\alpha)$ is mapped to its counterpart in the index space
\begin{equation}\label{S-eq:J_jalpha}
    \mathbb{J}(j_\alpha) 
    = \begin{cases}
        \mathbb{Z} \cap \left( \left[0, 2^{n_\alpha-1}-1-j_\alpha \right] \cup \left[ 2^{n_\alpha-1}, 2^{n_\alpha}-1\right] \right), & j_\alpha<2^{n_\alpha-1}, \\
        \mathbb{Z} \cap \left( \left[0, 2^{n_\alpha-1}-1 \right] \cup \left[ 3\cdot 2^{n_\alpha-1}-j_\alpha, 2^{n_\alpha}-1\right] \right), & j_\alpha\ge 2^{n_\alpha-1}.
    \end{cases}
\end{equation}
Defining the set
\begin{equation}
    \mathcal J(\boldsymbol{j})
    :=
    \prod_{\alpha=0}^{d-1}\mathbb{J}(j_\alpha)
    =
    \left\{
    \boldsymbol{j}'
    \,\middle|\,
    j'_\alpha\in \mathbb{J}(j_\alpha)
    \ \text{for all } \alpha
    \right\}
\end{equation}
and summing over the multidimensional summation index $\boldsymbol{j}'$ over $\mathcal J(\boldsymbol{j})$ yield
\begin{equation}
    \hat{\mathcal{Q}}_{s}(\vec{j}) = \sum_{\vec{j}'\in\mathcal{J}(\vec{j})} \mathcal{C}(\vec{j}, \vec{j}') \hat{\psi}_{s, m(\vec{j}, \vec{j}')} \hat{\psi}_{s, n(\vec{j}')}^*,
\end{equation}
with the state indices
\begin{equation}\label{S-eq:m}
    m(\boldsymbol{j},\boldsymbol{j}')
    =\sum_{\alpha=0}^{d-1} 2^{\sum_{i=0}^{\alpha-1}n_i} \operatorname{mod}(j_\alpha+j'_\alpha,2^{n_\alpha}),
\end{equation}
and
\begin{equation}\label{S-eq:n}
    n(\boldsymbol{j}') = \sum_{\alpha=0}^{d-1} 2^{\sum_{i=0}^{\alpha-1}n_i}j'_\alpha,
\end{equation}
given by coupled wavenumber pairs.
The modulo operation in $m(\boldsymbol{j},\boldsymbol{j}')$ maps each signed wavenumber back to the computational-basis range $0,\cdots,2^{n_\alpha}-1$. 
The restriction $\boldsymbol{j}'\in\mathcal J(\boldsymbol{j})$ excludes physical aliasing across the spectral boundary.
This procedure yields measurement operators
\begin{equation}\label{S-eq:rho_meas_opt}
    \hat{\mathcal{O}}_{\rho,s} = \sum_{\vec{j}'\in\mathcal{J}(\vec{j})} \ket{n(\vec{j}')}\bra{m(\vec{j}, \vec{j}')}
\end{equation}
for the density, and
\begin{equation}\label{S-eq:momentum_meas_opt}
    \hat{\mathcal{O}}_{J_\alpha,s} = \sum_{\vec{j}'\in\mathcal{J}(\vec{j})} \bigg[ \frac12\mathrm{mod}(j_\alpha+2^{n_\alpha-1}, 2^{n_\alpha}) + \mathrm{mod}(j_\alpha'+2^{n_\alpha-1}, 2^{n_\alpha}) - 3\cdot 2^{n_\alpha-2}\bigg] \ket{n(\vec{j}')}\bra{m(\vec{j}, \vec{j}')}
\end{equation}
for the momentum. 

Consequently, the measurement operators for $\rho$ and $\vec{J}$ in spectral space are expressed in the unified form
\begin{equation}\label{S-eq:observable_fluid}
    \hat{\mathcal{O}}_{\mathcal{Q},s}(\vec{j}) = \sum_{\vec{j}'\in \mathcal{J}(\vec{j})} \mathcal{C}(\vec{j}, \vec{j}') \ket{n(\vec{j}')}\bra{m(\vec{j}, \vec{j}')}.
\end{equation}
The strength of the two-wave interaction is determined by the matrix elements $\mathcal{C}$ in Eq.~\eqref{S-eq:observable_fluid}, which are unity for density and $\mathrm{mod}(j_\alpha+2^{n_\alpha-1}, 2^{n_\alpha})/2 + \mathrm{mod}(j_\alpha'+2^{n_\alpha-1}, 2^{n_\alpha}) - 3\cdot 2^{n_\alpha-2}$ for momentum in the $\alpha$-th direction. 
Since the matrix $\hat{\mathcal{O}}_{\mathcal{Q},s}$ in Eq.~\eqref{S-eq:observable_fluid} is non-Hermitian, the corresponding Hermitian observables $(\hat{\mathcal{O}}_{\mathcal{Q},s} + \hat{\mathcal{O}}_{\mathcal{Q},s}^\dagger)/2$ and $(\hat{\mathcal{O}}_{\mathcal{Q},s} - \hat{\mathcal{O}}_{\mathcal{Q},s}^\dagger)/(2\ii)$ are measured separately.

We verify the multidimensional index mapping on two- and three-dimensional grids with $n_\alpha=2$ using random complex spinor amplitudes. 
For every allowed output wavenumber, the direct Cartesian-product convolution agrees with the expectation value obtained from Eq.~\eqref{S-eq:observable_fluid}. 
The comparison includes output modes at the positive and negative spectral boundaries.

\subsection{Measurement of spectral observables}
To estimate the expectation value of the observable in Eq.~\eqref{S-eq:observable_fluid}, we propose a measurement procedure based on the linear combination of unitaries (LCU) method~\cite{Childs2012_Hamiltonian}.
For a given $\vec{j}$, this observable is expressed as
\begin{equation}\label{S-eq:meas_opt_sim}
    \hat{\mathcal{O}} = \sum_{\ell} c_\ell \ket{m_\ell}\bra{n_\ell},
\end{equation}
where $\ell$ labels the complete multidimensional index $\boldsymbol{j}'\in\mathcal J(\boldsymbol{j})$ in Eq.~\eqref{S-eq:observable_fluid}.
The number of terms in the explicit expansion is $|\mathcal J(\boldsymbol{j})|=\prod_{\alpha=0}^{d-1}|\mathbb{J}(j_\alpha)|$.
The coefficients $c_\ell$ are unity for $\hat{\mathcal{O}}_{\rho,s}$ and are the real numbers defined in Eq.~\eqref{S-eq:momentum_meas_opt} for $\hat{\mathcal{O}}_{J_\alpha,s}$.
Since a direct, term-by-term measurement of $\ket{m_\ell}\bra{n_\ell}$ is infeasible, we employ the LCU to construct a block-encoding for estimating the total expectation value.
We then analyze the expectation $\langle\hat\varPsi_s|\hat{\mathcal{O}}|\hat\varPsi_s\rangle$ in a normalized component register. 
The known physical component norms are restored after estimation.

First, we decompose each term in Eq.~\eqref{S-eq:observable_fluid} into a linear combination of Hermitian operators
\begin{equation}
    \hat{X}_\ell := \ket{m_\ell}\bra{n_\ell} + \ket{n_\ell}\bra{m_\ell}
\end{equation}
and
\begin{equation}
    \hat{Y}_\ell := \ii\ket{m_\ell}\bra{n_\ell} -\ii\ket{n_\ell}\bra{m_\ell},
\end{equation}
which allows the non-Hermitian term $\ket{m_\ell}\bra{n_\ell}$ to be expressed as
\begin{equation}\label{S-eq:decompose_element_meas}
    \ket{m_\ell}\bra{n_\ell} = \frac{1}{2}(\hat{X}_\ell - \ii\hat{Y}_\ell).
\end{equation}
Substituting Eq.~\eqref{S-eq:decompose_element_meas} into Eq.~\eqref{S-eq:meas_opt_sim} yields
\begin{equation}\label{S-eq:meas_Q_decompose}
    \hat{\mathcal{O}} = \sum_\ell \frac{c_\ell}{2}(\hat{X}_\ell - \ii\hat{Y}_\ell).
\end{equation}

Then, we construct a unitary representation for each Hermitian operator, $\hat{X}_\ell$ and $\hat{Y}_\ell$. 
We define the projection operator $\hat{P}_\ell := \ket{m_\ell}\bra{m_\ell} + \ket{n_\ell}\bra{n_\ell}$, which projects onto the subspace spanned by $\ket{m_\ell}$ and $\ket{n_\ell}$.
The projector onto the orthogonal complement is then $\hat{P}_\ell^\perp = I-\hat{P}_\ell$, with the identity operator $I$ on the full Hilbert space. 
Subsequently, we define the unitary extension of $\hat{X}_\ell$ as
\begin{equation}\label{S-eq:U_ell_X}
    \hat{U}_{\ell, X} 
    := \hat{X}_\ell + \hat{P}_\ell^\perp
    = (\ket{m_\ell}\bra{n_\ell} + \ket{n_\ell}\bra{m_\ell}) + (I - \ket{m_\ell}\bra{m_\ell} - \ket{n_\ell}\bra{n_\ell}).
\end{equation}
This operator is both Hermitian and unitary, satisfying $\hat{U}_{\ell, X}^2=I$.
The operator $\hat{U}_{\ell, X}$ acts as a Pauli-$X$ gate on the subspace spanned by $\{\ket{m_\ell}, \ket{n_\ell}\}$ and as the identity on its orthogonal complement.
Similarly, we define the unitary extension of $\hat{Y}_\ell$ as
\begin{equation}\label{S-eq:U_ell_Y}
    \hat{U}_{\ell, Y} 
    := \hat{Y}_\ell + \ii \hat{P}_\ell^\perp
    = (\ii\ket{m_\ell}\bra{n_\ell} - \ii\ket{n_\ell}\bra{m_\ell}) + \ii(I - \ket{m_\ell}\bra{m_\ell} - \ket{n_\ell}\bra{n_\ell}).
\end{equation}
Substituting Eqs.~\eqref{S-eq:U_ell_X} and \eqref{S-eq:U_ell_Y} into Eq.~\eqref{S-eq:meas_Q_decompose} yields
\begin{equation}\label{S-eq:meas_Q_decompose_1}
    \hat{\mathcal{O}} = \sum_\ell \frac{c_\ell}{2}(\hat{U}_{\ell, X} - \ii\hat{U}_{\ell, Y}) - \sum_{\ell} c_\ell \hat{P}_\ell^\perp.
\end{equation}
We then decompose this operator as $\hat{\mathcal{O}}=\hat{H}_1 - \hat{H}_2$, with $\hat{H}_1=\sum_\ell \frac{c_\ell}{2}(\hat{U}_{\ell, X} - \ii\hat{U}_{\ell, Y})$ and $\hat{H}_2=\sum_{\ell} c_\ell \hat{P}_\ell^\perp$.

We introduce two oracles and auxiliary quantum registers for a LCU procedure to separately measure $\langle \hat{\varPsi}_s|\hat{H}_1|\hat{\varPsi}_s\rangle$ and $\langle \hat{\varPsi}_s|\hat{H}_2|\hat{\varPsi}_s\rangle$, thereby determining $\langle \hat{\varPsi}_s|\hat{\mathcal{O}}|\hat{\varPsi}_s\rangle$ via Eq.~\eqref{S-eq:meas_Q_decompose_1}.
Beyond the $n$-qubit grid register, this procedure requires a control register (with subscript c) with $n_c=\lceil \log_2 |\mathcal J(\boldsymbol{j})|\rceil + 1$ qubits for indexing the terms in the linear combination, and an ancilla qubit (with subscript a) to extract the real and imaginary parts of the expectation value through interference.
We define a PREPARE oracle $\hat{O}_P$. 
It encodes only the coefficient magnitudes in the control-register amplitudes via the mapping
\begin{equation}
    \hat{O}_P: \ket{0}_c^{\otimes n_c} \mapsto \frac{1}{\norm{\vec{c}}_1^{1/2}}\sum_\ell \bigg(\sqrt{\frac{|c_\ell|}{2}}\ket{\ell,0}_c + \sqrt{\frac{|c_\ell|}{2}}\ket{\ell,1}_c \bigg).
\end{equation}
The coefficient phases are included in the selection oracle. The states $\lvert\ell,0\rangle_c$ and $\lvert\ell,1\rangle_c$ select $s_\ell\hat{U}_{\ell,X}$ and $-\mathrm{i}s_\ell\hat{U}_{\ell,Y}$, respectively, where $s_\ell := c_\ell/|c_\ell|$ for $c_\ell\neq0$, while terms with $c_\ell=0$ are omitted.
We then define the corresponding SELECT oracle
\begin{equation}
    \hat{O}_S = \sum_{\ell} \left[\ket{\ell,0}_c\bra{\ell,0}_c \otimes s_\ell \hat{U}_{\ell, X} + \ket{\ell,1}_c\bra{\ell,1}_c \otimes (-\ii s_\ell \hat{U}_{\ell, Y}) \right].
\end{equation}
Since $|s_\ell|=|-\ii s_\ell|=1$, every selected operator remains unitary, and hence $\hat{O}_S$ is unitary.

The PREPARE oracle is implemented by first preparing a uniform superposition over the $|\mathcal J(\boldsymbol{j})|$ valid indices, reversibly evaluating $|c_\ell|/\norm{\vec{c}}_\infty$, and applying a controlled rotation followed by amplitude amplification. Its cost satisfies
\begin{equation}
    C_P(n,\eta)
    =
    O\bigg[
    \sqrt{\frac{|\mathcal J(\boldsymbol{j})| \norm{\vec{c}}_\infty}{\norm{\vec{c}}_1}}\,
    \operatorname{poly}\left(
    n,\log\frac{1}{\eta}
    \right)
    \bigg],
\end{equation}
where $\eta$ denotes the oracle-synthesis error. 
For density, $c_\ell=1$ and $|\mathcal J(\boldsymbol{j})| \norm{\vec{c}}_\infty/\norm{\vec{c}}_1=1$. 
For momentum, the coefficients are piecewise affine over the valid index intervals, which gives $|\mathcal J(\boldsymbol{j})| \norm{\vec{c}}_\infty/\norm{\vec{c}}_1=O(1)$. 
The SELECT oracle reversibly evaluates $m_\ell$, $n_\ell$, and $s_\ell$, and then applies the corresponding two-level unitary. Standard reversible arithmetic and multi-controlled-gate decompositions give the conservative bound $C_S(n)=O(n^2)$ for fixed spatial dimension. Choosing $\eta=O(\varepsilon/\norm{\vec{c}}_1)$ controls the oracle contribution to the absolute error at $O(\varepsilon)$. Therefore, a single block-encoding query has polynomial gate cost, while the complete estimation has cost $O[\norm{\vec{c}}_1 (C_P+C_S) / \varepsilon]$.

We now detail the algorithm for measuring $\real(\langle \hat{\varPsi}_s|\hat{H}_1|\hat{\varPsi}_s\rangle)$.
The procedure starts from the initial state $\ket{0}_a\ket{0}^{\otimes n_c}\ket{\hat{\varPsi}_s}$. A Hadamard gate is first applied to the ancilla qubit as
\begin{equation}
    H\otimes I^{\otimes (n_c+n)}: \ket{0}_a\ket{0}^{\otimes n_c}\ket{\hat{\varPsi}_s} \mapsto \frac{1}{\sqrt{2}}(\ket{0}_a + \ket{1}_a) \ket{0}^{\otimes n_c}\ket{\hat{\varPsi}_s}.
\end{equation}
Then, applying the preparation oracle $\hat{O}_P$ to the control register yields
\begin{equation}\label{S-eq:result_O_P}
    I\otimes \hat{O}_P \otimes I^{\otimes n}: \frac{1}{\sqrt{2}}(\ket{0}_a + \ket{1}_a) \ket{0}^{\otimes n_c}\ket{\hat{\varPsi}_s} \mapsto \frac{1}{\sqrt{2}}(\ket{0}_a + \ket{1}_a) (\hat{O}_P\ket{0}^{\otimes n_c}) \ket{\hat{\varPsi}_s}.
\end{equation}
Next, the controlled-select gate, defined as
\begin{equation}\label{S-eq:C-O_S}
    C_a(\hat{O}_S) = \ket{0}_a\bra{0}_a\otimes I^{\otimes (n_c+n)} + \ket{1}_a\bra{1}_a\otimes \hat{O}_S,
\end{equation}
is applied, transforming the state to
\begin{equation}
    C_a(\hat{O}_S): \frac{1}{\sqrt{2}}(\ket{0}_a + \ket{1}_a) (\hat{O}_P\ket{0}^{\otimes n_c}) \ket{\hat{\varPsi}_s} \mapsto \frac{1}{\sqrt{2}}\left(\ket{0}_a (\hat{O}_P\ket{0}^{\otimes n_c}) \ket{\hat{\varPsi}_s} + \ket{1}_a \hat{O}_S (\hat{O}_P\ket{0}^{\otimes n_c}) \ket{\hat{\varPsi}_s} \right).
\end{equation}
Applying the inverse preparation oracle $\hat{O}_P^\dagger$ to the control register then yields
\begin{equation}\label{S-eq:O_P_dagger}
    I\otimes \hat{O}_P^\dagger\otimes I^{\otimes n}:
    \frac{1}{\sqrt{2}}\left(\ket{0}_a (\hat{O}_P\ket{0}^{\otimes n_c}) \ket{\hat{\varPsi}_s} + \ket{1}_a \hat{O}_S (\hat{O}_P\ket{0}^{\otimes n_c}) \ket{\hat{\varPsi}_s} \right) \mapsto \frac{1}{\sqrt{2}}\left(\ket{0}_a\ket{0}^{\otimes n_c}\ket{\hat{\varPsi}_s} + \ket{1}_a \hat{O}_P^\dagger \hat{O}_S \hat{O}_P \ket{0}_c\ket{\hat{\varPsi}_s} \right).
\end{equation}
By using the identity
\begin{equation}
    \langle 0|_c^{\otimes n_c} \hat{O}_P^\dagger \hat{O}_S \hat{O}_P \ket{0}_c^{\otimes n_c} \ket{\hat{\varPsi}_s}
    = \frac{1}{\norm{\vec{c}}_1}\hat{H}_1\ket{\hat{\varPsi}_s}
\end{equation}
and projecting the control register onto $\ket{0}^{\otimes n_c}$, the state simplifies to
\begin{equation}
    \frac{1}{\sqrt{2}}\left(\ket{0}_a\ket{0}_c^{\otimes n_c}\ket{\hat{\varPsi}_s} + \ket{1}_a\ket{0}_c^{\otimes n_c}\frac{1}{\norm{\vec{c}}_1}\hat{H}_1\ket{\hat{\varPsi}_s} \right).
\end{equation}
A final Hadamard gate on the ancilla register produces the state
\begin{equation}\label{S-eq:Hadamard_final}
    \begin{aligned}
        H\otimes I^{\otimes (n_c+n)}: 
        \frac{1}{\sqrt{2}}\left(\ket{0}_a\ket{0}_c^{\otimes n_c}\ket{\hat{\varPsi}_s} + \ket{1}_a\ket{0}_c^{\otimes n_c}\frac{1}{\norm{\vec{c}}}\hat{H}_1\ket{\hat{\varPsi}_s} \right)
        \mapsto \ &
        \frac{1}{2}\ket{0}_a\ket{0}_c^{\otimes n_c} \otimes \bigg(I^{\otimes n} + \frac{1}{\norm{\vec{c}}_1}\hat{H}_1 \bigg)\ket{\hat{\varPsi}_s} 
        \\
        &+ \frac{1}{2}\ket{1}_a\ket{0}_c^{\otimes n_c} \otimes \bigg(I^{\otimes n} - \frac{1}{\norm{\vec{c}}_1}\hat{H}_1 \bigg)\ket{\hat{\varPsi}_s}.
    \end{aligned}
\end{equation}

As a direct verification, consider a two-qubit grid register with $\lvert m\rangle=\lvert00\rangle$, $\lvert n\rangle=\lvert01\rangle$, $\hat{P}^{\perp}=\lvert10\rangle\langle10\rvert+\lvert11\rangle\langle11\rvert$, and one nonzero real coefficient $c$. 
Equations \eqref{S-eq:U_ell_X} and \eqref{S-eq:U_ell_Y} give
\begin{equation}
    \frac{1}{2}
    \left(
    \hat{U}_X-\mathrm{i}\hat{U}_Y
    \right)
    =
    \lvert00\rangle\langle01\rvert
    +
    \hat{P}^{\perp}.
\end{equation}
For this instance, $s=c/|c|$, and the corrected oracles satisfy
\begin{equation}
    |c|
    \langle0|
    \hat{O}_P^\dagger
    \hat{O}_S
    \hat{O}_P
    |0\rangle
    =
    \frac{c}{2}
    \left(
    \hat{U}_X-\mathrm{i}\hat{U}_Y
    \right)
    =
    c\lvert00\rangle\langle01\rvert
    +
    c\hat{P}^{\perp}.
\end{equation}
Subtracting $\hat{H}_2=c\hat{P}^{\perp}$ therefore gives $\hat{H}_1-\hat{H}_2=c\lvert00\rangle\langle01\rvert=\hat{\mathcal{O}}$, which verifies both the LCU identity and the selected block for this small instance.

A measurement of the ancilla register is then performed.
Let $P_0$ denote the joint probability of measuring the ancilla in $\lvert0\rangle_a$ and the control register in $\lvert0\rangle_c^{\otimes n_c}$. 
After some algebra, we obtain
\begin{equation}
    \begin{aligned}
        P_0
        &=
        \frac14
        \left\|
        \left(
        I+\frac{\hat H_1}{\norm{\vec{c}}_1}
        \right)
        |\hat\varPsi_s\rangle
        \right\|_2^2\\
        &=
        \frac14
        +
        \frac{1}{2\norm{\vec{c}}_1}
        \operatorname{Re}
        \langle\hat\varPsi_s|\hat H_1|\hat\varPsi_s\rangle
        +
        \frac{1}{4\norm{\vec{c}}_1^2}
        \langle\hat\varPsi_s|
        \hat H_1^\dagger\hat H_1
        |\hat\varPsi_s\rangle.
    \end{aligned}
\end{equation}
Similarly, the joint probability associated with the ancilla outcome $\lvert1\rangle_a$ is
\begin{equation}
    \begin{aligned}
        P_1
        &=
        \frac14
        \left\|
        \left(
        I-\frac{\hat H_1}{\norm{\vec{c}}_1}
        \right)
        |\hat\varPsi_s\rangle
        \right\|_2^2\\
        &=
        \frac14
        -
        \frac{1}{2\norm{\vec{c}}_1}
        \operatorname{Re}
        \langle\hat\varPsi_s|\hat H_1|\hat\varPsi_s\rangle
        +
        \frac{1}{4\norm{\vec{c}}_1^2}
        \langle\hat\varPsi_s|
        \hat H_1^\dagger\hat H_1
        |\hat\varPsi_s\rangle.
    \end{aligned}
\end{equation}
The quadratic terms cancel exactly in the probability difference, which gives
\begin{equation}\label{S-eq:P0-P1_normc}
    \operatorname{Re}
    \langle\hat\varPsi_s|\hat H_1|\hat\varPsi_s\rangle
    =
    \norm{\vec{c}}_1(P_0-P_1).
\end{equation}
In contrast, the quadratic terms add in the postselection probability, yielding
\begin{equation}
    P_{\mathrm{success}}
    =
    P_0+P_1
    =
    \frac12+
    \frac{1}{2\norm{\vec{c}}_1^2}
    \langle\hat\varPsi_s|
    \hat H_1^\dagger\hat H_1
    |\hat\varPsi_s\rangle
    \geq\frac12.
\end{equation}
Thus, postselection introduces at most a constant repetition overhead, independently of the precision required for $P_0-P_1$.

Equation \eqref{S-eq:P0-P1_normc} shows that the measured probability difference is normalized by $\norm{\vec{c}}_1$. Let $\hat D$ estimate $D=P_0-P_1=\operatorname{Re}\langle\hat\varPsi_s|\hat H_1|\hat\varPsi_s\rangle/\|\vec{c}\|_1$ with $|\hat D-D|\leq\delta$. 
The resulting error in the register expectation value satisfies
\begin{equation}
    \left|
    \|\vec{c}\|_1\hat D
    -
    \operatorname{Re}
    \langle\hat\varPsi_s|\hat H_1|\hat\varPsi_s\rangle
    \right|
    \leq
    \|\vec{c}\|_1\delta.
\end{equation}
Therefore, an absolute error $\varepsilon$ requires $\delta\leq\varepsilon/\|\vec{c}\|_1$, corresponding to $O(\|\vec{c}\|_1/\varepsilon)$ quantum-amplitude-estimation queries. Since $\hat{\mathcal{O}}=\hat H_1-\hat H_2$, assigning an error of at most $\varepsilon/2$ to each expectation value gives an error of at most $\varepsilon$ in $\langle\hat{\mathcal{O}}\rangle$, with the same $O(\|\vec{c}\|_1/\varepsilon)$ scaling up to constant factors. 

To measure the imaginary part $\imag(\langle\hat{\varPsi}_s|\hat{H}_1|\hat{\varPsi}_s\rangle)$, we apply an additional $S^\dagger$ gate to the ancilla before the final Hadamard gate in Eq.~\eqref{S-eq:Hadamard_final}.
The real and imaginary measurements together determine the complex expectation value of $\hat{\mathcal{O}}$. 
A single block-encoding query has depth $O[C_P(n)+C_S(n)]$, while a fixed absolute error $\varepsilon$ requires $O(\norm{\vec{c}}_1/\varepsilon)$ queries.

Next, we detail the algorithm for measuring $\langle\hat{\varPsi}_s|\hat{H}_2|\hat{\varPsi}_s\rangle$.
We expand $\hat{H}_2$ as
\begin{equation}
    \hat{H}_2 = \bigg(\sum_\ell c_\ell \bigg)I - \sum_{\ell} c_\ell \ket{m_\ell}\bra{m_\ell} - \sum_{\ell} c_\ell \ket{n_\ell}\bra{n_\ell}.
\end{equation}
The first of these three terms has an expectation value $S_c=\sum_\ell c_\ell$, which can be computed analytically using the definitions of $c_\ell$ in Eqs.~\eqref{S-eq:rho_meas_opt} and \eqref{S-eq:momentum_meas_opt}.
The remaining two terms, $\hat{H}_{2,m}=\sum_{\ell} c_\ell \ket{m_\ell}\bra{m_\ell}$ and $\hat{H}_{2,n}=\sum_{\ell} c_\ell \ket{n_\ell}\bra{n_\ell}$, are both Hermitian and possess similar structures.
We therefore detail the measurement method for $\langle\hat{\varPsi}_s|\hat{H}_{2,m}|\hat{\varPsi}_s\rangle$ as a representative example.
The key element of $\hat{H}_{2,m}$ is the operator $\ket{m_\ell}\bra{m_\ell}$.
To express this term as a linear combination of unitaries, we define the Pauli-$Z$-class operator
\begin{equation}
    \hat{U}_{\ell, Z_m} := I - 2\ket{m_\ell}\bra{m_\ell},
\end{equation}
which applies a $-1$ phase to the basis state $\ket{m_\ell}$ while leaving orthogonal states invariant.
Consequently, from the decomposition $\ket{m_\ell}\bra{m_\ell} = \frac{1}{2}(I - \hat{U}_{\ell, Z_m})$, the Hamiltonian $\hat{H}_{2,m}$ becomes
\begin{equation}
    \hat{H}_{2,m} = \sum_{\ell} c_\ell\bigg(\frac{1}{2}(I-\hat{U}_{\ell, Z_m})\bigg)
    = \frac{S_c}{2}I - \frac{1}{2}\sum_{\ell} c_\ell \hat{U}_{\ell, Z_m}.
\end{equation}
Hence, its expectation value with respect to the state $|\hat{\varPsi}_s\rangle$ is
\begin{equation}
    \langle\hat{\varPsi}_s|\hat{H}_{2,m}|\hat{\varPsi}_s\rangle
    = \frac{S_c}{2} - \frac{1}{2}\langle\hat{\varPsi}_s|\sum_{\ell} c_\ell \hat{U}_{\ell, Z_m}|\hat{\varPsi}_s\rangle.
\end{equation}
The second term on the right-hand side is evaluated using the LCU construction analogous to that for $\hat{H}_1$.
The preparation and selection oracles are defined as $\hat{O}_{P,2}\ket{0}_c^{\otimes n_c}=\frac{1}{\norm{\vec{c}}_1^{1/2}}\sum_\ell \sqrt{|c_\ell|} \ket{\ell}_c$ and $\hat{O}_{S,2}=\sum_{\ell}\ket{\ell}_c\bra{\ell}_c \otimes s_\ell \hat{U}_{\ell, Z_m}$, respectively.
These definitions give $\langle0|_c^{\otimes n_c}\hat O_{P,2}^{\dagger}\hat O_{S,2}\hat O_{P,2}\lvert0\rangle_c^{\otimes n_c}=\frac{1}{\norm{\vec{c}}_1}\sum_\ell c_\ell\hat U_{\ell,Z_m}$.

The same construction applies to $\hat H_{2,n}$.
Because the operator $\sum_\ell c_\ell\hat U_{\ell,Z_m}$ is Hermitian, its expectation value is real. Its block encoding is normalized by $\norm{\vec{c}}_1$, and hence an absolute error $\varepsilon_2$ requires $O(\norm{\vec{c}}_1/\varepsilon_2)$ quantum-amplitude-estimation queries. The oracle costs remain explicit, as in the analysis of $\hat H_1$.

In summary, for a fixed wavenumber $\boldsymbol{k}$, a single block-encoding query has depth $O[C_P(n)+C_S(n)]$. Estimating the corresponding normalized physical expectation to an absolute error $\varepsilon$ requires $O(\norm{\vec{c}}_1/\varepsilon)$ quantum-amplitude-estimation queries, giving a block-encoding gate cost
\begin{equation}
    O\left[
    \frac{
    \norm{\vec{c}}_1\left(C_P(n)+C_S(n)\right)
    }{\varepsilon}
    \right].
\end{equation}
To obtain a physical density or momentum coefficient with absolute error at most $\varepsilon$, it suffices to estimate each normalized component expectation with error at most $\varepsilon/2$. Indeed, restoring the component norms gives
\begin{equation}
    |\Delta\hat{\mathcal{Q}}|
    \leq
    \sum_{s=\pm}
    \|\hat\psi_s\|_2^2
    |
    \Delta\langle\hat{\mathcal{O}}_{\mathcal{Q},s}\rangle
    |
    \leq\varepsilon,
\end{equation}
because $\|\hat\psi_+\|_2^2+\|\hat\psi_-\|_2^2=2$. This allocation changes only constant factors in the query and gate costs.

A complete amplitude-estimation implementation also includes repeated calls to the input-state preparation circuit and its inverse.
The postselection probability is bounded below by $1/2$, but this bound does not remove the normalization-dependent precision cost. Thus, the present term-by-term LCU construction gives polynomial total readout complexity only when $\norm{\vec{c}}_1$ is polynomial in $n$.
The method provides a structured measurement circuit for selected spectral observables without reconstructing all $2^n$ state amplitudes.
Here, $\varepsilon$ specifies the estimation error for a selected convolution coefficient of the given input state, separately from the initialization error in Eq.~\eqref{S-eq:ini_error}. 
Note that measuring a fixed number of density and momentum coefficients does not generally suffice to recover a velocity mode, because density inversion couples Fourier modes.

\vspace{1em}
\subsection{Measurement of mean momentum on computational basis}
Mean flow observables are characterized by diagonal matrices.
This property permits their direct evaluation via measurements in the computational basis, thereby obviating the requirement for the LCU framework.

We first derive the diagonal representation of the mean momentum observables.
The mean flow is associated with the wavevector condition $\vec{k} = \vec{0}$, where $k_\alpha = 0$ implies $j_\alpha = k_\alpha + 2^{n_\alpha}[1 - H(k_\alpha)] = 0$.
This relationship identifies the set of indices
\begin{equation}
    \mathbb{J}(j_\alpha) = \mathbb{Z} \cap \left( \left[0, 2^{n_\alpha-1} - 1\right] \cup \left[2^{n_\alpha-1}, 2^{n_\alpha} - 1\right] \right) = \{0, 1, 2, \cdots, 2^{n_\alpha} - 1\}.
\end{equation}
The row and column indices of the operator matrix are then expressed as
\begin{equation}\label{S-eq:rc_position}
    \begin{aligned}
        m(\vec{0}, \vec{j}') &= \sum_{\alpha=0}^{d-1} 2^{\sum_{0 \leqslant i < \alpha} n_i} \left( \bmod(2^{n_\alpha-1}, 2^{n_\alpha}) + \bmod(j'_\alpha + 2^{n_\alpha-1}, 2^{n_\alpha}) - 2^{n_\alpha} \right) \\
        &= \sum_{\alpha=0}^{d-1} 2^{\sum_{0 \leqslant i < \alpha} n_i} \left( \bmod(j'_\alpha + 2^{n_\alpha-1}, 2^{n_\alpha}) - 2^{n_\alpha-1} \right) = n(\vec{j}').
    \end{aligned}
\end{equation}
Because the observable is a diagonal matrix, it permits direct measurement in the computational basis.
Under the assumption of an identical number of qubits across all spatial directions, $n_\alpha = n_0 = n_1 = n_2$, the expansion of Eq.~\eqref{S-eq:rc_position} yields
\begin{equation}
    m(\vec{0}, \vec{j}') = n(\vec{j}') = \bmod(j'_0 + 2^{n_0-1}, 2^{n_0}) - 2^{n_0-1} + 2^{n_0} (\bmod(j'_1 + 2^{n_0-1}, 2^{n_0}) - 2^{n_0-1}) + 2^{2n_0} (\bmod(j'_2 + 2^{n_0-1}, 2^{n_0}) - 2^{n_0-1}).
\end{equation}

The measurement operator for mean momentum with spin $s$ in the $\alpha$-direction is defined as
\begin{equation}
    \hat{J}_{s,\alpha} = \sum_{j'_2=0}^{2^{n_0}-1} \sum_{j'_1=0}^{2^{n_0}-1} \sum_{j'_0=0}^{2^{n_0}-1} \left[ \bmod(j'_\alpha + 2^{n_\alpha-1}, 2^{n_\alpha}) - 2^{n_\alpha-1} \right] |m(\vec{0}, \vec{j}')\rangle \langle n(\vec{j}')|.
\end{equation}
Consequently, the matrix element at the $r$-th row and $c$-th column for the mean momentum in the $x$-direction is given by
\begin{equation}
    \hat{p}_x(r, c) = [\bmod(j'_0 + 2^{n_0-1}, 2^{n_0}) - 2^{n_0-1}]\delta_{rc} = [\bmod(r + 2^{n_0-1}, 2^{n_0}) - 2^{n_0-1}]\delta_{rc}.
\end{equation}
Similarly, the matrix element for the mean momentum in the $y$-direction is
\begin{equation}
    \hat{p}_y(r, c) = [\bmod(j'_1 + 2^{n_0-1}, 2^{n_0}) - 2^{n_0-1}]\delta_{rc} 
    = \bigg[\bmod\left( \left\lfloor \frac{r + 2^{n_0-1}}{2^{n_0}} \right\rfloor + 2^{n_0-1}, 2^{n_0} \right) - 2^{n_0-1}\bigg]\delta_{rc},
\end{equation}
The corresponding element for the mean momentum in the $z$-direction is
\begin{equation}
    \hat{p}_z(r, c) = [\bmod(j'_2 + 2^{n_0-1}, 2^{n_0}) - 2^{n_0-1}]\delta_{rc} = \left\lfloor \frac{r + 2^{n_0-1} + 2^{2n_0-1}}{2^{2n_0}} \right\rfloor \delta_{rc},
\end{equation}
where $\lfloor \cdot \rfloor$ denotes the floor function.

A projective measurement of each normalized state $\ket{\hat{\varPsi}_s}$ is performed in the computational basis $\{\ket{j}\}$.
The probability of obtaining outcome $j$ is $c_{s,j}=|\langle j|\hat{\varPsi}_s\rangle|^2$, with $\sum_j c_{s,j}=1$.
Because $\hat{p}_x$ is diagonal in this basis, the component-register expectation is $\bar{J}_{s,x}=\langle\hat{\varPsi}_s|\hat{p}_x|\hat{\varPsi}_s\rangle=\sum_j c_{s,j}\hat{p}_x(j,j)$, without requiring a basis transformation.
The physical mean momentum is recovered by restoring the component norms as $\bar{J}_x=\sum_{s=\pm}\|\hat{\psi}_s\|_2^2\bar{J}_{s,x}$.

For each component, we collect $T$ independent and identically distributed outcomes $j_{s,1},j_{s,2},\cdots,j_{s,T}$ and construct the estimator $\tilde{J}_{s,x}=\frac{1}{T}\sum_{t=1}^{T}\hat{p}_x(j_{s,t},j_{s,t})$.
The two components use independent sample sets.
Each single-shot momentum has expectation $\mathbb{E}[\hat{p}_x(j_{s,t},j_{s,t})]=\bar{J}_{s,x}$, so $\mathbb{E}[\tilde{J}_{s,x}]=\bar{J}_{s,x}$.
The physical estimator $\tilde{J}_x=\sum_{s=\pm}\|\hat{\psi}_s\|_2^2\tilde{J}_{s,x}$ is therefore unbiased, with $\mathbb{E}[\tilde{J}_x]=\bar{J}_x$.
Independence of the $T$ trials gives
\begin{equation}\label{S-eq:Var_J}
    \mathrm{Var}(\tilde{J}_{s,x})
    =
    \frac{1}{T}
    \bigg(
    \sum_j c_{s,j}\,\hat{p}_x^2(j,j)
    -
    \bar{J}_{s,x}^2
    \bigg).
\end{equation}

The momentum eigenvalues are signed Fourier wavenumbers.
For an $n_0=n/3$ qubit directional subregister, they satisfy $-2^{n/3-1}\leq\hat{p}_x(j,j)\leq2^{n/3-1}-1$.
Popoviciu's inequality and Eq.~\eqref{S-eq:Var_J} therefore give
\begin{equation}
    \operatorname{Var}(\tilde{J}_{s,x})
    \leq
    \frac{(2^{n/3}-1)^2}{4T}.
\end{equation}
For independent component measurements, the variance of the physical estimator satisfies
\begin{equation}
    \operatorname{Var}(\tilde{J}_x)
    =
    \sum_{s=\pm}
    \|\hat{\psi}_s\|_2^4
    \operatorname{Var}(\tilde{J}_{s,x})
    \leq
    \frac{(2^{n/3}-1)^2}{4T}
    \sum_{s=\pm}\|\hat{\psi}_s\|_2^4.
\end{equation}
The same bounds apply to the $y$- and $z$-momentum estimators.
These are worst-case bounds for fixed component norms and scale as $T^{-1}$, where $T$ denotes the number of trials per component.
The actual variances depend on the component momentum distributions.

\section{Theoretical lower bound of quantum encoding for a turbulent field}
Classical simulations of 3D turbulence are computationally demanding, with the required number of degrees of freedom scaling as $\Rey^{9/4}$.
We establish an information-theoretic lower bound on the number of qubits $n$ required for a quantum representation of a turbulent state. 
We prove that, under the operational-discrimination and spectral-packing assumptions, representing a flow at Reynolds number $\Rey$ requires $n = \Omega(\log_2\Rey)$ qubits.
Thus, our geometric quantum encoding using $n \approx 3[\log_2(\Rey^{3/4}/5)+1]$ qubits has the same logarithmic scaling as this conditional lower bound.
This comparison concerns data-register size at fixed spectral tolerance, under the discrimination and packing assumptions stated below.

We consider the set $\mathcal{S}_{\Rey}$ of all statistically stationary turbulent states at a given $\Rey$.
Each state $s \in \mathcal{S}_{\Rey}$ is characterized by its energy spectrum $E_k^{(s)}(k)$, which describes the distribution of kinetic energy over the wavenumber $k$.
To quantify the dissimilarity between two states $s_1$ and $s_2$, we introduce a metric
\begin{equation}
    d(s_1, s_2) = \sup_{k \in [k_L, k_\eta]} |E_k^{(s_1)}(k) - E_k^{(s_2)}(k)|
\end{equation}
on $\mathcal{S}_{\Rey}$.
The supremum is taken over the wavenumber range $[k_L, k_\eta]$, spanning from the integral to the Kolmogorov scales.

Consider a $n$-qubit quantum encoding $\mathcal{E}: s \mapsto \varrho_s$, where $\varrho_s$ is a density operator on an $n$-qubit Hilbert space $\mathcal{H}_n$.
The mapping operator $\mathcal{E}$ is considered $\epsilon$-faithful if for any two states $s_1$ and $s_2$ with $d(s_1, s_2) > \epsilon$, their quantum representations are distinguishable. 
This condition requires that for a fixed constant $\delta \in (0, 1]$, their trace distance satisfies
\begin{equation}
    D(\varrho_{s_1}, \varrho_{s_2}) \equiv \frac{1}{2} \text{Tr}|\varrho_{s_1} - \varrho_{s_2}| \ge \delta.
\end{equation}
To prove the lower bound, we first construct a large set of mutually distinguishable turbulent states and then apply an information-theoretic argument to determine the requisite number of qubits.

We employ metric space packing to construct a set of mutually distinguishable turbulent states.
An $\epsilon$-packing of the metric space $(\mathcal{S}_{\Rey}, d)$ is a subset $\mathcal{C} = \{s_1, \dots, s_M\} \subseteq \mathcal{S}_{\Rey}$ such that $d(s_i, s_j) > \epsilon$ for all $i \neq j$.
The packing number $M(\mathcal{S}_{\Rey}, d, \epsilon)$ denotes the maximum size of such a set. 
Here, we make the additional operational assumption that the $M$ states in an $\epsilon$-packing occur with equal prior probabilities and are perfectly distinguishable by a single measurement, so identifying the encoded state yields $\log_2M$ bits, whereas Holevo’s bound~\cite{Holevo1998_Quantum} limits the accessible information $I_{\mathrm{acc}}$ to $n$ bits.
This implies the relation
\begin{equation}
    I_{\text{acc}} =  \log_2 M \le n.
\end{equation}
Therefore, the number of qubits is lower-bounded by the logarithm of the packing number as
\begin{equation}\label{eq:main_inequality}
    n \ge \log_2 M(\mathcal{S}_{\Rey}, d, \epsilon).
\end{equation}
 
The packing number $M(\mathcal{S}_{\Rey}, d, \epsilon)$ is estimated from the physical properties of turbulence.
We determine the effective degrees of freedom by discretizing the energy spectrum $E_k(k)$ in both wavenumber and energy.
The inertial range spans from the integral scale $L$ to the Kolmogorov scale $\eta$, corresponding to a wavenumber ratio $k_\eta/k_L \sim \Rey^{3/4}$.
As an additional spectral-packing assumption, we treat the logarithmically spaced modes as mutually independent and assume that every mode supports at least $m_{\min}=\Theta(1)>1$ $\epsilon$-distinguishable energy levels, independently of $\Rey$.
The number of independent modes
\begin{equation}
    N_{\text{modes}} \approx C_1'\log_2(k_\eta/k_L) 
    \approx \frac{3}{4}C_1'\log_2 \Rey
    = C_1 \log_2 \Rey
\end{equation}is therefore proportional to the logarithmic range of wavenumbers, with a constant $C_1=\Theta(1)$. 
For each of these modes, indexed by $i$, the spectral energy $E_k(k_i)$ varies over a physical range $\Delta E_i$.
The number of $\epsilon$-distinguishable levels for the $i$-th mode is thus
\begin{equation}\label{eq:mi_def}
    m_i \approx \frac{\Delta E_i}{\epsilon}.
\end{equation}

Assuming the modes are independent, the total number of $\epsilon$-distinguishable states 
\begin{equation}
M(\mathcal{S}_{\Rey}, d, \epsilon) \approx \prod_{i=1}^{N_{\text{modes}}} m_i \approx \prod_{i=1}^{C_1 \log_2 \Rey} \frac{\Delta E_i}{\epsilon}
\end{equation}
is the product of the levels in each mode.
Substituting this estimate into Eq.~\eqref{eq:main_inequality} gives
\begin{equation}
    n \ge \log_2 M 
    \approx \sum_{i=1}^{C_1 \log_2 \Rey} \log_2 \left( \frac{\Delta E_i}{\epsilon} \right).
\end{equation}
By replacing each $\Delta E_i$ with a minimum variation $\Delta E_{\min}$, we obtain a conservative lower bound
\begin{equation}
    n \gtrsim C_1 \log_2 \Rey \cdot \log_2 \left( \frac{\Delta E_{\min}}{\epsilon} \right).
\end{equation}
Given that the minimum number of distinguishable levels is $m_{\min}=\Theta(1)$, Eq.~\eqref{eq:mi_def} implies $\Delta E_{\min}/\epsilon=\Theta(1)$.
The number of qubits $n$ required for a faithful encoding is therefore lower-bounded by
\begin{equation}
    n = \Omega(\log_2\Rey).
\end{equation}
The required number of qubits is consistent with the classical lower bound on the number of grid points, $N=\Omega(\Rey^{9/4})$ \cite{Pope2000_Turbulent}, through the correspondence $N=2^n$.

We next relate the register size to the scale-inferred Reynolds number used in this work. 
For equal directional registers, the represented Fourier domain is a cube with $k_\alpha\in[-2^{n/3-1},2^{n/3-1}-1]$, and its coordinate Nyquist wavenumber is $k_{\max}=2^{n/3-1}$. 
We use $k_{\mathrm{peak}}\approx k_{\max}$ as a bandwidth-dependent characteristic scale and adopt $k_\eta\approx5k_{\mathrm{peak}}$ with $k_L=1$. 
This convention gives $\Rey\approx(5\cdot2^{n/3-1})^{4/3}$. Inverting this model relation gives the register-size estimate
\begin{equation}
    n=3n_\alpha\approx 3[\log_2(\Rey^{3/4}/5)+1].
\end{equation}

Consequently, our geometric encoding, requiring only $n\approx 3[\log_2(\Rey^{3/4}/5)+1]$ qubits, has the same logarithmic dependence on $\Rey$ as the conditional lower bound under the stated assumptions.
This encoding captures the generative rules for multiscale vortex structures via the Hopf fibration~\cite{Irvine2008_Linked}.
This approach thereby circumvents the direct encoding of high-entropy, small-scale stochastic details. 

\section{A brief review of fluid dynamics and turbulence}
Fluid dynamics is the branch of physics concerned with the motion of fluids (e.g., liquids and gases) and the forces acting on them. 
A quantitative description of fluid motion is founded upon the principles of conservation of mass, momentum, and energy. 
For a vast range of problems, particularly those involving incompressible flows at constant temperature, the fluid motion is governed by a set of coupled, nonlinear partial differential equations known as the Navier-Stokes equations. 
These equations form the cornerstone of fluid dynamics. 
For an incompressible flow with constant density $\rho$ and kinematic viscosity $\nu$, the conservation of momentum reads
\begin{equation}
    \frac{\partial \vec{u}}{\partial t} + \vec{u} \cdot \bn \vec{u} = -\frac{1}{\rho} \bn p + \nu \nabla^2 \vec{u} + \vec{f},
\end{equation}
where $\vec{u}(\vec{x}, t)$ is the fluid velocity, $p(\vec{x}, t)$ the pressure, and $\vec{f}$ the external body forces per unit mass, such as gravity. 
The term on the left, $\partial \vec{u}/\partial t$, represents the local or unsteady acceleration of a fluid element, while the term $\vec{u} \cdot \bn \vec{u}$ is the convective acceleration. 
This convective term is nonlinear and is the primary source of the rich and complex behavior observed in fluid flows, including turbulence. 
The terms on the right represent the forces driving the flow: the pressure gradient force $-\bn p / \rho$, the viscous force $\nu \nabla^2 \vec{u}$ which accounts for internal friction, and external body forces. 
This equation is supplemented by the continuity equation, which for an incompressible fluid simplifies to the constraint $\bn \cdot \vec{u} = 0$ that the velocity field must be divergence-free.

Fluid flows can be broadly categorized into two distinct regimes: laminar and turbulent. Laminar flow is characterized by smooth, orderly fluid motion, where fluid layers slide past one another with little to no mixing. 
It is highly predictable and typically occurs at low velocities or in highly viscous fluids. 
In contrast, turbulent flow is chaotic and disordered. 
It is characterized by the presence of swirling, three-dimensional structures known as vortices, which span a wide range of sizes and lead to rapid fluctuations in velocity and pressure. 
The transition between these two regimes is governed by a dimensionless parameter called the Reynolds number, $\Rey$, defined as the ratio of inertial forces to viscous forces:
\begin{equation}
    \Rey := \frac{UL}{\nu}.
\end{equation}
Here, $U$ is a characteristic velocity scale of the flow, $L$ is a characteristic length scale, and $\nu$ is the kinematic viscosity. 
When $\Rey$ is low, viscous forces dominate, damping out any disturbances and maintaining a laminar state. 
As $\Rey$ increases beyond a critical value, inertial forces become dominant, leading to instabilities that amplify disturbances and cause the flow to transition to a turbulent state.
Note that this dynamical definition is distinct from the scale-inferred Reynolds numbers assigned to the generated fields in Tab.~\ref{tab:Rey_lam}. 
Identifying the two requires an additional specification of viscosity and consistency with the dissipation statistics.

\begin{figure}[ht!]
    \centering
    \includegraphics{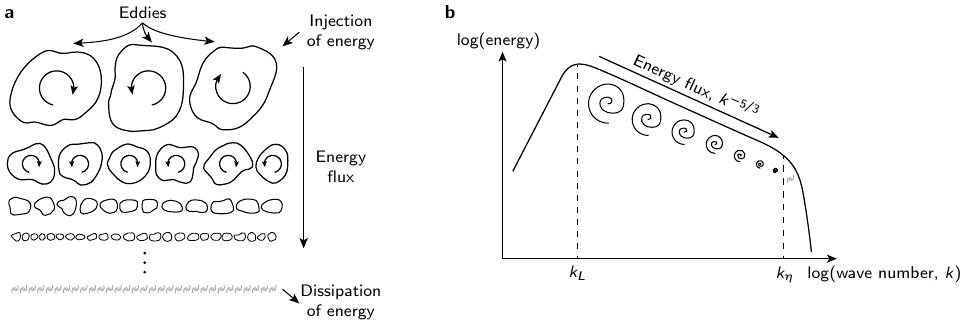}
    \caption{Schematic of the Kolmogorov energy cascade.
    \textbf{a}, Energy injected at large scales cascades to smaller eddies and is ultimately dissipated by viscosity.
    \textbf{b}, The corresponding energy spectrum consists of an injection range near the integral wavenumber $k_L$, an inertial range characterized by a constant energy flux and the Kolmogorov $k^{-5/3}$ scaling, and a dissipation range near the Kolmogorov wavenumber $k_\eta$.}
    \label{fig:cascade_schematic}
\end{figure}

A fundamental tenet of turbulence theory is the concept of the energy cascade, first quantified by Kolmogorov~\cite{Kolmogorov1991_The}, as illustrated in Fig.~\ref{fig:cascade_schematic}. 
In a turbulent flow, energy is typically injected into the system at the largest scales of motion (large eddies), comparable to the characteristic length scale $L$. 
These large, energy-containing eddies are inherently unstable and break down, transferring their kinetic energy to slightly smaller eddies, as shown in Fig.~\ref{fig:cascade_schematic}a. 
This process continues, creating a cascade of energy from larger to smaller scales through the inertial range, where viscous effects are negligible. 
Finally, at the smallest scales, known as the Kolmogorov length scale $\eta$, the eddies are small enough that viscous forces become significant again, and the kinetic energy is dissipated into heat. 
This multiscale nature, where a vast continuum of spatial and temporal scales interacts simultaneously, makes turbulence extremely difficult to model and simulate. 
The chaotic dynamics and the immense range of scales present a formidable computational challenge, motivating the exploration of novel computational paradigms, such as the quantum encoding method discussed in the main text, to represent and analyze these intricate systems more efficiently.

\bibliographystyle{apsrev4-2.bst}
\bibliography{supp.bib}

%% file: setup.tex
\usepackage{newtxtext, newtxmath}
\usepackage[usenames, dvipsnames]{xcolor}
\usepackage{graphicx}
\usepackage{dcolumn} 
\usepackage{xpatch}
    \makeatletter \xpretocmd \start@align{\linenomathWithnumbers}{}{\fail}
\usepackage{mathtools, url}
\usepackage{multirow}
\usepackage{CJKutf8}
\usepackage[toc]{appendix}
\usepackage[ruled, linesnumbered]{algorithm2e}
\usepackage[unicode=true, bookmarks=true, bookmarksnumbered=false, bookmarksopen=false, breaklinks=false, pdfborder={0 0 1}, backref=false, colorlinks=false, hidelinks]{hyperref}
    \hypersetup{linkcolor=black, urlcolor=black, citecolor=black, pdfstartview={FitH}}

\usepackage{xr}
\usepackage{calc}
\usepackage{bbm}
\usepackage{tabularray}

\usepackage[math]{cellspace}
\DeclareSymbolFont{CMlargesymbols}{OMX}{cmex}{m}{n}
\let\sumop\relax\let\prodop\relax
\DeclareMathSymbol{\sumop}{\mathop}{CMlargesymbols}{"50}
\DeclareMathSymbol{\prodop}{\mathop}{CMlargesymbols}{"51}

\SetSymbolFont{operators}{normal}{OT1}{ntxtlf}{m}{n}
\SetSymbolFont{operators}{bold}{OT1}{ntxtlf}{b}{n}

\renewcommand{\vec}[1]{\boldsymbol{#1}}

\newcommand{\bn}{\vec{\nabla}}

\newcommand{\ii}{\mathrm{i}}

\renewcommand{\le}{\leqslant}
\DeclareMathOperator*{\real}{Re}

\DeclareMathOperator*{\poly}{poly}
\newcommand{\ket}[1]{| #1 \rangle}
\newcommand{\bra}[1]{\langle #1 |}
\newcommand{\T}{^{\mathrm{T}}}

\newcommand{\Rey}{\mathit{Re}}

\newcommand{\firtitle}[1]{\vspace{1em} \noindent \textbf{\large #1} \newline \noindent}
\newcommand{\sndtitle}[1]{\noindent \textbf{#1} \newline \noindent}

\allowdisplaybreaks